\documentclass[twoside,preprint,11pt]{elsarticle}

\usepackage{lineno}

\newenvironment{pbitemize}{
\begin{itemize}
  \setlength{\itemsep}{1pt}
  \setlength{\parskip}{0pt}
  \setlength{\parsep}{0pt}}
{\end{itemize}}
\newenvironment{pbenumerate}{
\begin{enumerate}
  \setlength{\itemsep}{1pt}
  \setlength{\parskip}{0pt}
  \setlength{\parsep}{0pt}}
{\end{enumerate}}

\newcommand{\n}[1]{\mathrm{#1}}

\begin{document}

\begin{frontmatter}

\title{Studies of a three-stage dark matter and neutrino observatory
  based on multi-ton combinations of liquid xenon and liquid argon
  detectors}

\author{K.~Arisaka}
\author{P.~Beltrame\corref{cor}}
\cortext[cor]{Corresponding author}
\ead{pbeltrame@physics.ucla.edu}
\author{C.~W.~Lam}
\author{P.~F.~Smith}
\author{C.~Ghag} 
\author{D.~B.~Cline}
\author{K.~Lung}
\author{Y.~Meng}
\author{E.~Pantic}
\author{P.~R.~Scovell}
\author{A.~Teymourian}

\address{Department of Physics and Astronomy, University of California Los Angeles}

\begin{abstract}
  
  We study a three stage dark matter and neutrino observatory based on
  multi-ton two-phase liquid Xe and Ar detectors with sufficiently low
  backgrounds to be sensitive to WIMP dark matter interaction cross
  sections down to $10^{-47}~\n{cm^2}$, and to provide both
  identification and two independent measurements of the WIMP mass
  through the use of the two target elements in a 5:1 mass ratio,
  giving an expected similarity of event numbers. The same detection
  systems will also allow measurement of the pp solar neutrino
  spectrum, the neutrino flux and temperature from a Galactic
  supernova, and neutrinoless double beta decay of $^{136}$Xe to the
  lifetime level of $10^{27} - 10^{28}$ y corresponding to the
  Majorana mass predicted from current neutrino oscillation data. The
  proposed scheme would be operated in three \textbf{G}eneric stages G2, G3, G4,
  beginning with fiducial masses 1-ton Xe + 5-ton Ar (G2),
  progressing to 10-ton Xe + 50-ton Ar (G3)
  then, dependent on results and performance of the latter, expandable
  to 100-ton Xe + 500-ton Ar (G4). This method of scale-up
  offers the advantage of utilizing the Ar vessel and ancillary
  systems of one stage for the
  Xe detector of the succeeding stage, requiring only one new detector
  vessel at each stage. Simulations show the feasibility of reducing
  or rejecting all external and internal background levels to a level
  \textless 1 events per year for each succeeding mass level, by
  utilizing an increasing outer thickness of target material as
  self-shielding. The system would, with increasing mass scale, become
  increasingly sensitive to annual signal modulation, the agreement of
  Xe and Ar results confirming the Galactic origin of the signal. Dark
  matter sensitivities for spin-dependent and inelastic interactions
  are also included, and we conclude with a discussion of possible
  further gains from the use of Xe/Ar mixtures.
\end{abstract}


\end{frontmatter}

% \linenumbers

\clearpage
\newpage
\tableofcontents

\section{Objectives and overview}
\label{sec:overview}

Liquid noble gas detectors have proven capability to identify and
distinguish both nuclear recoil and electron recoil events~\cite{Aprile:2010},
and are shown in this paper to have the potential of achieving the
ultra-low backgrounds needed to detect weakly interacting dark matter
particles (WIMPs) with nucleon interaction cross sections down to
$10^{-47}\,\n{cm^{2}}$ and with sufficient spectral precision to
estimate their mass. With multi-ton target masses, these detectors
could at the same time measure the spectra of pp solar and galactic
supernova neutrinos, and detect neutrinoless double beta decay in
$^{\mathrm{136}}$Xe at the lifetime level ($10^{27} -
10^{28}\,\n{y}$) predicted for Majorana neutrinos from neutrino
oscillation data.

It is shown in Sec.\ref{sec:sensitivity}, that for unambiguous
identification and mass measurement of dark matter particles, it is
essential to have detectors observing signals in targets with two
different atomic numbers, in particular to exploit the
$A^2$ dependence of the spin-independent cross section. For
this, liquid Xe and liquid Ar provide an ideal pair of targets, having
similar operating principles and construction, but having spectral and
$A^2$ differences giving approximately a factor 5
difference in event rates. Hence similar signal rates should be
achievable for a liquid argon detector having five times the mass of a
corresponding liquid xenon detector.

We study a dark matter and neutrino observatory of this type
constructed in three progressively larger stages, with \textbf{G}eneric designs
referred to as G2,
G3 and G4. The sizes and masses of each stage are shown in
Tab.\ref{tab:g2g3g4_1}.
\begin{table}[!htbp]
  \centering
  \begin{tabular}{|c|c|c|c|c|c|c|}
      \hline
      & \multicolumn{2}{c|}{G2} & \multicolumn{2}{c|}{G3} & \multicolumn{2}{c|}{G4 \par } \\
      \hline
      & Xe & Ar & Xe & Ar & Xe & Ar \\
      \hline
      \multicolumn{1}{|l|}{Target dimensions, masses } & & & & & & \\
      \hline
      \multicolumn{1}{|r|}{ diameter $\times$ height (m)} & $1 \times 1$ & $2 \times 2$ & $2 \times 2$ &
      $4 \times 4$& $4 \times 4$ & $8 \times 8$ \\
      \hline
      \multicolumn{1}{|r|}{ total target mass (t)} & 2.2 & 9 & 18 & 73 & 146 & 580 \\
      \hline
      \multicolumn{1}{|r|}{ nominal fiducial target mass (t)} & 1& 5 & 10 & 50 & 100 & 500 \\
      \hline
      \multicolumn{1}{|l|}{No. of photodetectors } & & & & & & \\
      \hline
      \multicolumn{1}{|r|}{ top} & 120 (3'') & 600 (3'') & 600 (3'') & 670 (6'') & 670 (6'') & 2000 (6'') \\
      \hline
      \multicolumn{1}{|r|}{ sides (if instrumented)} & 520 (3'') & 670 (6'') & 670 (6'') & 2400 (6'') & 2400 (6'') & 8000 (6'') \\
      \hline
      \multicolumn{1}{|r|}{ bottom} & 120 (3'') & 160 (6'') & 160 (6'') & 670 (6'') & 670
      (6'') & 2000 (6'') \\
      \hline
    \end{tabular}
  \caption{Summary of sizes, masses and photodetector numbers for G2, G3 and G4 detectors}
  \label{tab:g2g3g4_1}
\end{table}
\begin{figure}[!htbp]
  \centering
  \includegraphics[width=0.9 \columnwidth]{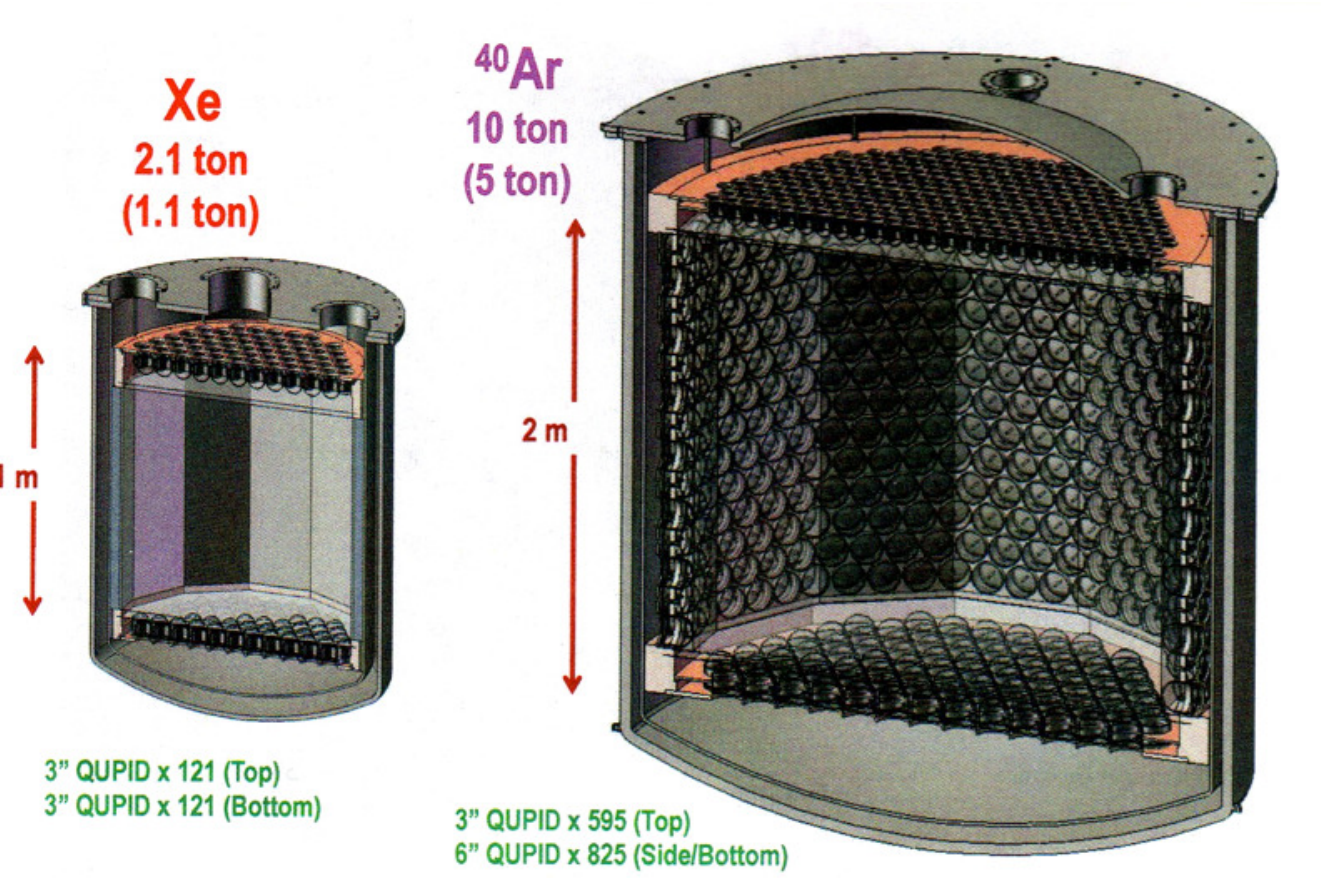}
  \caption{Main parameters of the $1~\n{ton}/5~\n{ton}$ (fiducial) G2 system}
  \label{fig:g2inner}
\end{figure}
\begin{figure}[!htbp]
  \centering
  \includegraphics[width=0.9 \columnwidth]{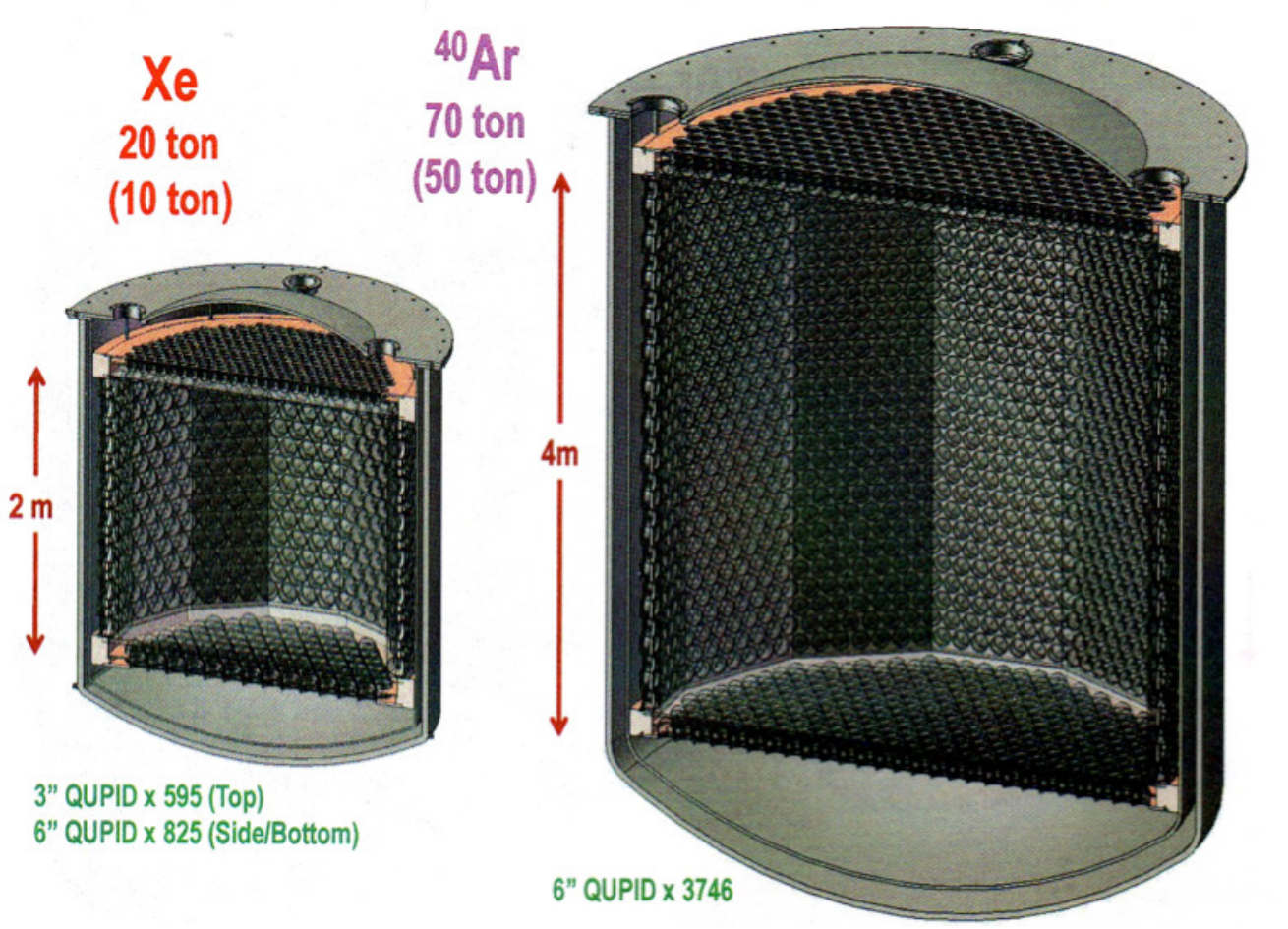}
  \caption{Main parameters of the $10~\n{ton}/50~\n{ton}$ (fiducial) G3 system}
  \label{fig:g3inner}
\end{figure}
\begin{figure}[!htbp]
  \centering
  \includegraphics[width=0.9 \columnwidth]{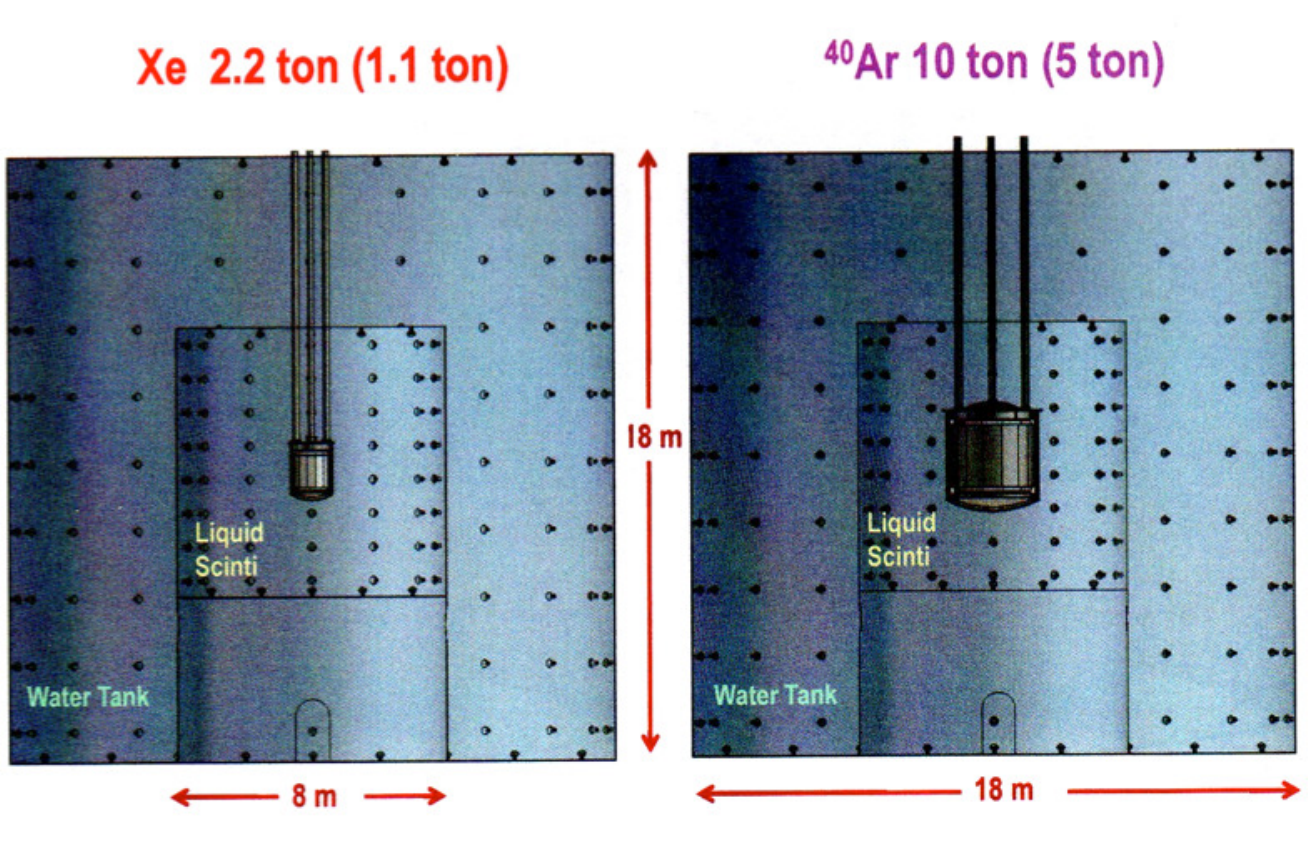}
  \caption{G2 system (1t Xe/5t Ar) in water and liquid scintillator shields}
  \label{fig:g2water}
\end{figure}
\begin{figure}[!htbp]
  \centering
  \includegraphics[width=0.9 \columnwidth]{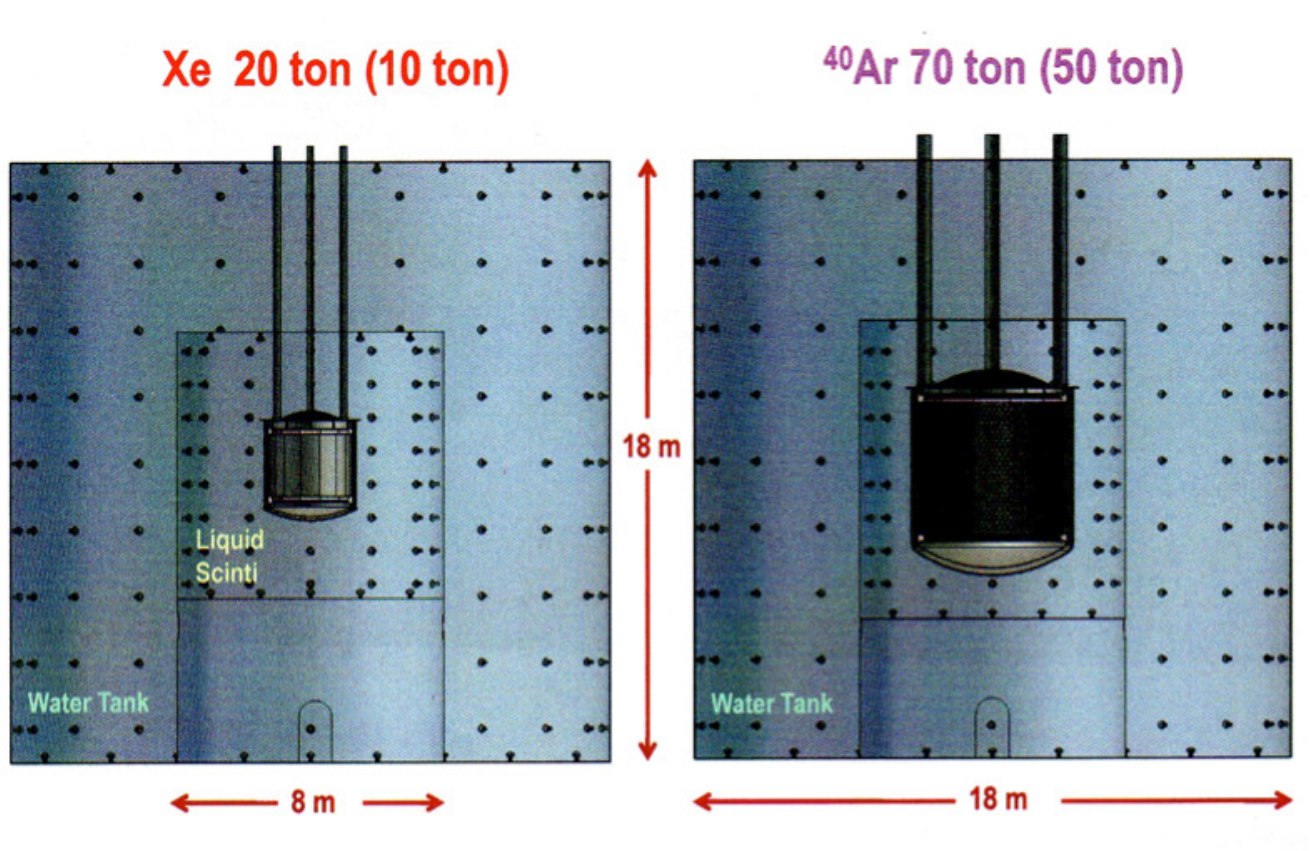}
  \caption{G3 system (10t Xe/50t Ar) in water and liquid scintillator shields}
  \label{fig:g3water}
\end{figure}

Each stage has the required 5:1 ratio of the Ar and Xe fiducial
masses, beginning with 1 ton of Xe and 5 tons of Ar. After a period of
data-taking with the G2 system, subsequent scale up can then be
achieved by the following steps:
\begin{pbenumerate}
\item Replacing the 5-ton Ar with an equal volume of Xe, taking
  advantage of the factor 2 density ratio to give a G3 Xe target an
  order of magnitude larger in mass than G2 Xe.
\item Construction of one new vessel - a matching G3 Ar target,
  which would again have a mass 5 times that of the G3 Xe mass. 
\item If warranted by the performance of the G3 system, replacing the
  G3 argon by 100 tons xenon to give the order of magnitude G4 scale up. 
\item A final option of constructing a G4 Ar target with 5 times the
  G4 Xe fiducial mass.
\end{pbenumerate}

The detectors will operate using the two-phase principle~\cite{Dolgoshein:1970},
yielding both scintillation and ionization signals whose ratio
provides a factor $\sim 10^{2} - 10^{3}$
discrimination between nuclear recoils and background electron
recoils. This has been demonstrated at the target mass level of $10-20$
kg, both technically and operationally, by three collaborations
ZEPLIN~\cite{Alner:2007,Lebedenko:2009,Lebedenko_1:2009,Akimov:2011},
XENON~\cite{Angle:2008,Sorensen:2010,Aprile_1:2010} and WARP~\cite{Benetti:2008},
the latter using liquid Ar and the others liquid Xe. Such devices have
demonstrated stable operation for periods in excess of a year. 
A 60 kg fiducial mass (48 kg fiducial mass)
detector in the XENON series is now in successful operation~\cite{Aprile_1:2010}. In this paper we study the achievable
backgrounds and potential physics sensitivity of the detection system
in Tab.\ref{tab:g2g3g4_1}, without further consideration of hardware
design. 

Illustrations of the basic internal structure of the G2 and G3 systems
are shown in Fig.\ref{fig:g2inner} and
Fig.\ref{fig:g3inner}. Fig.\ref{fig:g2water} and
Fig.\ref{fig:g3water} show the detectors immersed in water and liquid
scintillator shields (to be discussed further in \ref{app:externalbkg}). A constant
water-shielding diameter is used to illustrate both systems, although
smaller water shield thicknesses would be acceptable for the G2
system. The G4 Xe detector would be identical in configuration to the
G3 Ar detector of Fig.\ref{fig:g3inner}. 
The photodetectors shown in
Fig.\ref{fig:g2inner} and Fig.\ref{fig:g3inner}
register both the direct scintillation light from each event (referred
to as S1) and also the simultaneous ionization which is drifted to the
liquid surface by an electric field and then extracted and further
drifted to produce proportional scintillation in the gas (referred to as
S2). The electric drift field will be typically 0.5 - 1 kV/cm.
XENON100~\cite{Aprile:2012} operated at 0.5 kV/cm, while XENON1T~\cite{Aprile:dm2012}
is designed at 1 kV/cm with a drift length of 100 cm, for which the
total voltage 100 kV has been achieved in the R\&D phase of that
project. Thus the voltage requirements of the G2 stage are already
demonstrated. The G3 stage would need a voltage of 200 kV for the
xenon detector, which will also suffice for the argon detector at a
field of 0.5 kV/cm, and a liquid argon TPC with this maximum voltage
is already under development for the LBNE experiment~\cite{LBNE}. A G4
xenon detector would require either a (foreseeable) factor 2
improvement over the LBNE value or a compensating reduction in the
drift field. 

The ratio S2/S1 provides a parameter which separates nuclear
recoil and electron recoil events into two populations with less than 1\%
overlap. For the photodetectors we envisage the use of 3 inch or 6
inch QUPIDs (avalanche photodiodes within a quartz enclosure~\cite{Arisaka:2009}),
which are designed to have a factor 100 lower U/Th activity, per unit
area of coverage, than the best photomultipliers. The required numbers
of these are summarised in the lower half of
Tab.\ref{tab:g2g3g4_1}. The QUPIDs have a quantum efficiency
$\sim 35-37\%$~\cite{Teymourian:2011} for the 180 nm Xe wavelength, while for the liquid
argon detectors, a TPB coating~\cite{TPBref} would be used to wavelength-shift
the argon scintillation light into the QUPID range, again with a QE
$\sim 35-37\%$~\cite{TPBAr:UCLA}. Development studies~\cite{Teymourian:2011} show that the QUPID
dynamic range is sufficient to cover both the low energy dark matter
searches and the MeV-range neutrinoless double beta decay search.

Ideally the QUPIDs would fully surround the
detector in a $4\pi$ array, but a lower-cost option (at some sacrifice
of light collection, energy threshold, and position resolution) would
be to locate the QUPIDs at top and bottom only, using
high-reflectivity ($>95\%$) PTFE on an array of $10-12$ flat panels around the
sides of the target volume (illustrated for the G2 Xe detector in
Fig.\ref{fig:g2inner}). Simulations below will show that, despite the
greater simplicity and lower cost of these reflecting panels, the use
of QUPIDs on the side walls would improve the light collection, and
hence the energy threshold and detector sensitivity, by a factor 2
throughout most of the target volume. 

In Sec.\ref{sec:sensitivity} we use the basic expressions for WIMP
nuclear recoil spectrum (on the assumption of a canonical velocity
distribution), and spin-independence of the WIMP-nucleon cross
section, to determine the sensitivity
of each of these detector systems to WIMP-nucleus cross-section and
WIMP mass, on the assumption that backgrounds (from neutron, gamma,
and electron recoil events) can be reduced to the level $\sim~0.1-0.5$
background events per year -- i.e. signals or limits not background
limited. We also show the sensitivity to the expected annual
modulation arising from the Earth's orbital and Galactic motion, and
the sensitivity to other types of WIMP-nucleus interaction, in
particular spin-dependent and hypothetical inelastic interactions. 

In Sec.\ref{sec:backgrounds} we discuss in detail simulation
results of both external water shielding and the reduction of detector
backgrounds from local radioactivity, in the photodetectors and
detector vessel, showing that a combination of a liquid scintillator
veto and self-shielding by an outer thickness of target material
provides the key to achieving an essentially background-free fiducial
target volume in all three of the G2, G3 and G4 systems. In addition
we discuss the reduction of backgrounds from radioactive contaminants
of liquid Xe and Ar.

In Sec.\ref{sec:nulowenergy} we calculate the sensitivity of
these detectors to neutrinos from the solar pp chain and from a
Galactic supernova. Finally in Sec.\ref{sec:nulessdoublebeta} we show that
sufficiently low backgrounds could be achieved in the G3 or G4 systems
for the observation of neutrinoless double beta decay from
$^{136}$Xe at the lifetime level 10$^{27}$
$-10^{28}~\n{y}$ corresponding to the Majorana mass
estimated from neutrino mixing data. Other possible configurations for this
include a concentric target of enriched $^{136}$Xe shielded
by $^{136}$Xe-depleted Xe (the ``XAX'' scheme, described in ~\cite{Arisaka:2009}) and the use of Xe/Ar mixtures.

\section{Sensitivity to WIMP cross-section and mass}
\label{sec:sensitivity}

\subsection{Nuclear recoil spectra from WIMP-nucleus collisions}
\label{subsec:recoil_spectra}

The basic detector requirements, expected counting rates, and the
dependence on target material and WIMP mass, can be seen from the
expected form of the nuclear recoil energy spectrum. Adopting the
``canonical'' model of a Galactic dark matter velocity distribution
similar in general shape to a Maxwell-Boltzman distribution, the
resulting differential counting rate (events/keV/kg/day) with respect to recoil energy
$E_R$ (keV) is given by~\cite{Smith:1990,Lewin:1996}
\begin{equation}
  dR/dE_R~\n{(events/keV/kg/d)} = c_1[R_{0}/E_{0}r]e^{- c_{2} E_{R}/ E_{0}r} F^{2} (E_R)
  \label{eq:rate}
\end{equation}
where $r = 4M_DA/(M_D+A)^2$ for target element $A$ and incident
particle mass $M_D$ (both expressed in GeV), $E_0 = 0.5\times
10^6 M_D (v_0/c)^2$ ($v_0$ = Galactic velocity dispersion
$\approx 0.0007c$, $E_0$ in keV) and $F$ is a nuclear form factor
correction. The coefficients $0.5 < c_1$, $c_2 <1 $
provide an approximation for the motion of the Earth relative to the
Galaxy. For a stationary Earth $c_1 = c_2 = 1$, giving a total
event rate $R_0$. The orbital motion of the sun modifies this to
$c_1 = 0.78$, $c_2 = 0.58$ with a further $\pm 4\%$ seasonal
variation arising from the orbital motion of the
Earth~\cite{Smith:1990,Lewin:1996}.

Thus an experimental limit on, or measurement of, the differential
rate Eq.\ref{eq:rate} for a given range of $E_R$ leads to a
corresponding limit or value of $R_0$ (events/kg/d) for each assumed value
of $M_D$. This in turn can be converted~\cite{Roszkowski:2007} into a limit or
value for the WIMP-nucleon cross section $\sigma_{W-N}$
using the following relation, assuming a coherent spin-independent
interaction and a detection efficiency and energy threshold capable of
registering a fraction $f$~(typically $\sim 0.3 - 0.5$) of the total
events in the spectrum of Eq.\ref{eq:rate}:
\begin{equation}
  \sigma _{W-N}~\n{(pb)}= 0.0091\frac{R_0}{A^2\,r\,f}
  \label{eq:xsection}
\end{equation}
The above basic Eq.\ref{eq:rate} and Eq.\ref{eq:xsection} reveal two key features:
\begin{pbenumerate}
\item Eq.\ref{eq:xsection} shows that the total spin-independent normalized
  event rate $R_0/r$ is proportional both to the interaction
  cross-section $\sigma_{W-N}$ and to $A^2$,
  the square of the target atomic mass number. 
\item Eq.\ref{eq:rate} shows a dependence on the dark matter
  particle mass $M_D$ through the quantity $E_0$, so that a
  larger dark matter mass gives a more slowly falling spectrum. Thus,
  if a spectrum of events could be observed, the mass of particle
  responsible could be estimated as the maximum likelihood fit of
  $M_D$ to Eq.\ref{eq:rate}.
\end{pbenumerate}

The above properties enable us to resolve the concern that it would be
difficult to distinguish a genuine dark matter signal from a residual
background of neutrons or spurious instrumental noise effects. A
unique identification of a genuine dark matter signal could be
obtained by two methods:
\begin{itemize}
\item[a)] The use of two target elements, to verify the rate dependence on 
  $A^2$. 
\item[b)] Confirmation that the same value of $M_D$ results, within 
  experimental error, from the recoil spectrum of each target element. 
\end{itemize}

It is of particular note that while method {\it a)} is based on comparing
absolute counting rates for two different elements, method {\it b)} depends
on the shape of the spectrum only, and not on the absolute counting
rate. Thus the nuclear coherence of the interaction, and the mass
determination, provide two mathematically independent methods of
signal identification.

\begin{figure}[!htbp]
  \centering
  \includegraphics[width=0.9 \columnwidth]{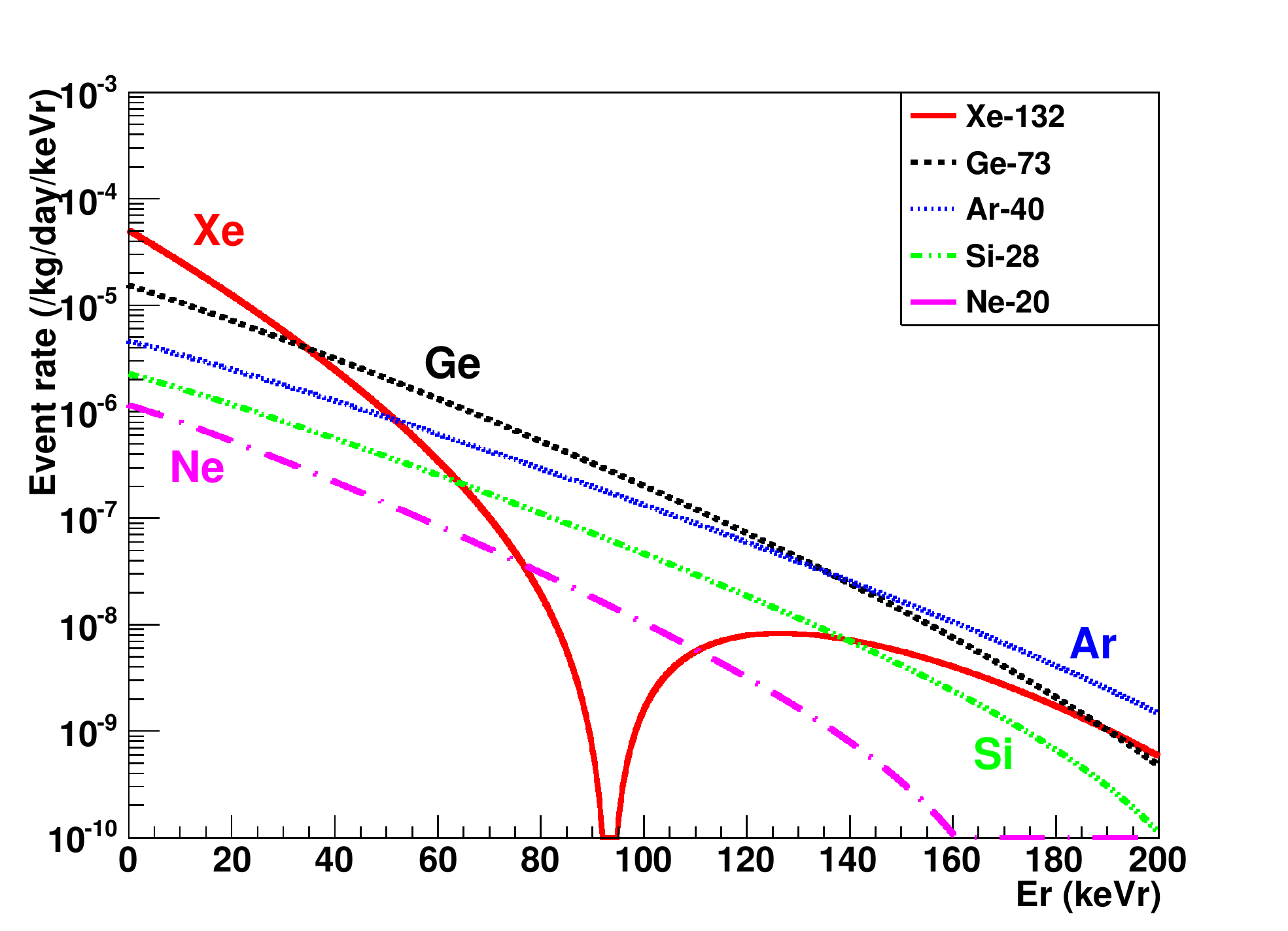}
  \caption{Differential nuclear recoil spectra for Xe and Ar and other
    nuclei, calculated from Eq.\ref{eq:rate} for a WIMP mass of 100
    GeV.}
  \label{fig:spectrum}
\end{figure}

These comparisons ideally require that the two elements should be used
in detectors with similar physical principles and technology. To
achieve this, we can take advantage of the fact that detectors using
similar principles have already been developed for the noble liquids
Xe and Ar. Fig.\ref{fig:spectrum} compares the normalized nuclear
recoil energy spectra for Xe and Ar (with other nuclei for comparison)
for a dark matter particle mass of 100 GeV and interaction
cross-section of $10^{-44}~\n{cm^2}$. This shows that a Xe target
has a greater sensitivity than an Ar target at low energy, but that Ar
is less affected by the form factor correction in Eq.\ref{eq:rate}, so
higher energy recoils can usefully contribute (up to a limit set by
the Galactic escape velocity -- see below). It is thus evident that
the design and operation of two such detectors of similar design, one
using Xe and the other Ar as the target element, and with similar
operating conditions, provides a powerful combination for dark matter
identification and particle mass estimation. It also follows by
integrating the curves in Fig.\ref{fig:spectrum} that to achieve
similar dark matter counting rates, and assuming experimental recoil
energy thresholds 8 keV for Xe and 20 keV for Ar (discussed in
detailed below) the Ar
detector needs to have about 5 times the fiducial mass of the Xe
detector, which gives rise to the choices of target mass values
summarized above in Tab.\ref{tab:g2g3g4_1}.

\subsection{Sensitivity to spin-independent cross section}
\label{subsec:spin_independent}

We next use the above Eq.\ref{eq:xsection} to estimate the cross
section sensitivity for the systems described in
Sec.\ref{sec:overview}. The term ``sensitivity'' has two meanings in
common use. The ``limiting sensitivity'' is the cross section
corresponding to zero observed events (in a pre-selected
background-free ``event box'') and hence a statistical limit of $2-3$
events. For larger detectors envisioning a non-zero signal, we need to
define the ``useful sensitivity'' as the cross section for which the
observed number of events would be sufficient to provide a fit to a
recoil spectrum to allow an estimate of the particle mass. As an
example, for typical WIMP-nucleon cross sections expected from
supersymmetry ($10^{-44}~\n{cm}^2$ to $10^{-46}~\n{cm}^2$ or
$10^{-8\,}$ to $10^{-10}~\n{pb}$), the corresponding interaction rates
from Eq.\ref{eq:rate} and Eq.\ref{eq:xsection} are expected to be in the
range $0.1-10$ per day per ton (fiducial) of Xe and a factor $\sim 5$ 
lower for Ar. Net experimental efficiencies of $\sim 30\% - 50\%$ are
assumed (discussed further below) and already known to be achievable
in detectors with target masses $\sim 10-100$
kg~\cite{Arisaka:2009,Diehl:1995}. Thus a target mass of 1- ton (fiducial) Xe would provide a quantitatively useful signal of
$\sim 20 - 30$ events/y at $10^{-46}~\n{cm^2}$, or an upper limit
of $2-3$ events/y at $10^{-47}~\n{cm^2}$. Comparable signal rates would be
achieved in 5 tons fiducial mass of Ar.

\begin{figure}[!htbp]
  \centering 
  \includegraphics[width=0.9 \columnwidth]{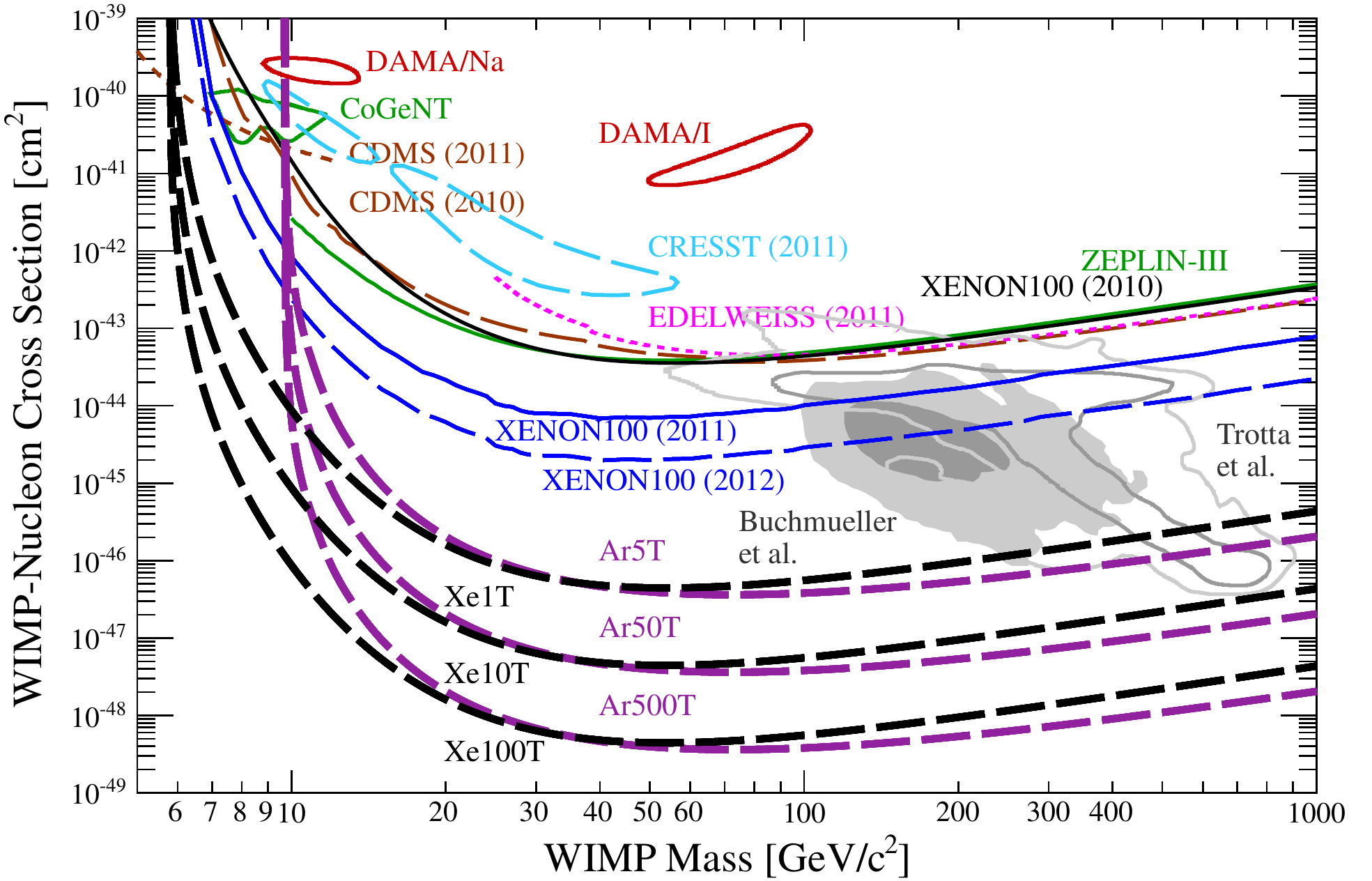}
  \caption{Spin-independent sensitivity plots for Xe and Ar detectors
    in G2, G3 and G4 systems, assuming future nuclear recoil energy
    thresholds $\sim 8$ keVnr for Xe and $\sim 20$ keVnr for Ar, as
    discussed in \ref{app:detmaterial}. Curves show 90\% confidence limits for
    WIMP-nucleus cross section ($10^{-46}~\n{ cm}^2 =
    10^{-10}~\n{pb}$) for live time of 1 year, for zero background
    events in a signal region $50\%$ of total nuclear recoil band and
    $80\%$ cut efficiency. Some recent limits and hypothetical signal
    region are shown~\cite{Bernabei:2009,Savage:2009,Aalseth:2011,Angloher:2011,Ahmed:2009,Akerib:2010,Ahmed:2011,Armengaud:2011,Lebedenko:2009,Akimov:2011},
    together. The shaded region
    shows the most favoured regions of parameter space from
    supersymmetry theory. above the $\sim 40$ GeV mass limit from
    accelerator
    searches~\cite{Trotta:2008,Buchmueller:2011}, but
    could extend to lower cross sections in some variations of the
    theory.} 
  \label{fig:sensitivity}
\end{figure}

Fig.\ref{fig:sensitivity} shows the spin-independent cross section
limits for the G2, G3, G4, Xe and Ar detectors in one year of
operation, compared with some existing limits. These are calculated
for the case of zero observed signal (using 2.4 event $90\%$ upper
limit) and assuming all fiducial backgrounds reduced to
$< 0.2$ events/y. It is the purpose of this paper to show that such
background levels can be achieved for each of the detector systems of
Tab.\ref{tab:g2g3g4_1}, and will be discussed in Sec.\ref{sec:backgrounds}.

\subsection{Number of signal events}
\label{subsec:nevents}

\begin{figure}[!htbp]
  \centering
  \includegraphics[width=0.9 \columnwidth]{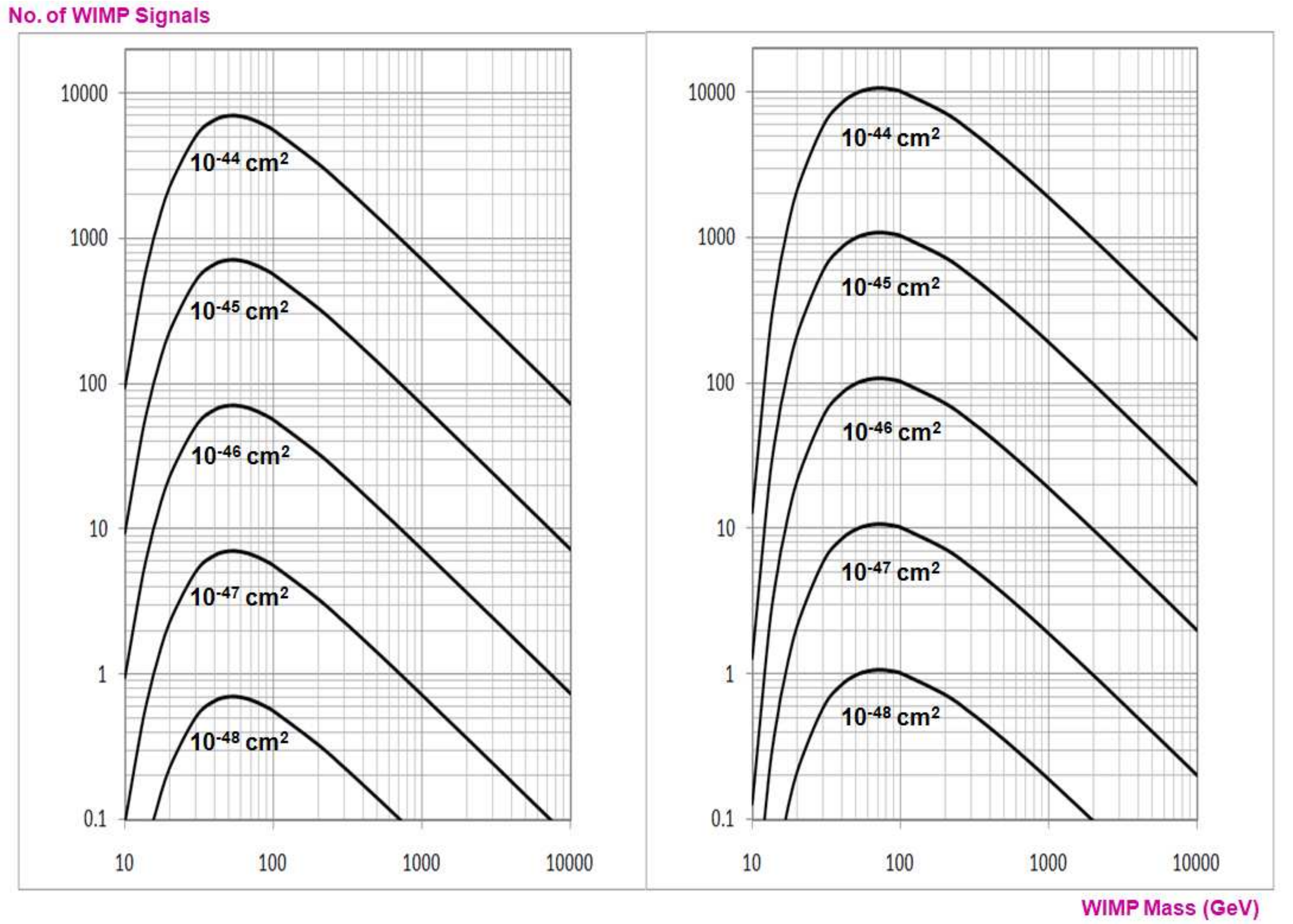}
  \caption{Number of events (vertical axis) in signal range versus
    WIMP mass for different WIMP-nucleon cross-sections, assuming the
    same efficiency factors used for Fig.\ref{fig:sensitivity} and nuclear recoil energy
    thresholds $\sim 8$ keVnr for Xe and $\sim 20$ keVnr for Ar, as
    discussed in \ref{app:detmaterial}. \protect\\
    (\textit{Left plot}) 10 ton-y (fiducial) liquid Xe in range $8-45$
    keVr. (\textit{Right plot}) 50 ton-y (fiducial) liquid Ar in range
    20-200 keVr.}
  \label{fig:nevents}
\end{figure}

If a non-zero dark matter nuclear recoil signal exists at a cross
section larger than the minimum sensitivity plotted in
Fig.\ref{fig:sensitivity}, then the number of signal events available to
form a recoil spectrum is calculable from Eq.\ref{eq:rate} of
Sec.\ref{subsec:recoil_spectra} and plotted in
Fig.\ref{fig:nevents} against WIMP mass for 10 ton-y Xe and 50 ton-y
Ar, the numbers for other running times and target masses being
proportional.

\subsection{Dependence on recoil energy threshold and Galactic escape velocity}
\label{sec:vesc}

The nuclear recoil energy scale is based on the LXe scintillation
signal and thus requires knowledge of the relative scintillation
efficiency of nuclear recoils, with recent direct measurements down to
energies as low as 3 keV indicating a slow decrease from $~15$ keV
~\cite{Plante:2011}, with Monte Carlo simulation techniques
providing consistent results~\cite{Horn:2011}.  Here we
have assumed a recoil threshold of 8.5 keV~\cite{Aprile_1:2010} for the
purpose of calculating the Xe sensitivity estimates of Fig.\ref{fig:sensitivity},
although improved light collection (for example, from full
photodetector coverage or higher quantum efficiency) would result in a
lower effective threshold. In the case of Ar, the recoil energy threshold is dependent on the
degree of success in reducing the low energy background from $^{39}$Ar, which restricted
the energy threshold to $\sim 40 - 55$ keVnr in the initial WARP experiment with
natural Ar~\cite{Benetti:2008}. From the discussion and references
in~\ref{subapp:bkg_ar5t}, we adopt an expectation
of 20 keV threshold for the purpose of calculating the limit curves in
Fig.\ref{fig:sensitivity}. In addition we note, from the flatter
Ar spectrum in Fig.\ref{fig:spectrum}, that the Ar limits will be
less sensitive to energy threshold than the faster-falling spectrum
for Xe.

To estimate the expected total count rate, the spectra of
Fig.\ref{fig:spectrum} have to be integrated between a lower limit
given by the detector nuclear recoil energy threshold, and an upper
limit given by the Galactic escape velocity at the position of the sun
(hence giving an upper limit on the energy of a dark matter particle
in our vicinity). This escape velocity, although governed largely by
the amount and distribution of the dark matter, can be estimated from
the observed velocity distribution of the stellar population, the most
recent detailed study concluding that the escape velocity lies in the
range $550 \pm 50$ km/s~\cite{Smith:2007}. 

From this, and for simplicity treating the escape velocity ($v_{esc}$)
as a ``cut-off'' ($E_{cut}$) rather than an additional attenuation of the velocity
distribution, we can calculate from Eq.\ref{eq:rate} the observable
percentage of the total events in the spectrum as function of the
recoil energy threshold ($E_{thr}$) and the escape velocity. The results are shown
in Tab.\ref{tab:recoil_xe} (for a Xe target) and
Tab.\ref{tab:recoil_ar} (for an Ar target) with those for the most
probable escape velocity shown in bold.

These reduced collection figures, arising from the lower and upper
cut-off in the spectrum, are in addition to the basic factor 5
difference in the total counts resulting from the form factor and
$A^2$ differences in Eq.\ref{eq:rate}
and Eq.\ref{eq:xsection}. Tab.\ref{tab:recoil_xe} and Tab.\ref{tab:recoil_ar}
also show that the Xe detector threshold needs to be about a factor 2
lower than the Ar detector threshold to achieve a similar
event-collection efficiency. Assumed energy threshold values of 8.5
keVnr for Xe and 20 keVnr for Ar are used to calculate the G2, G3, and G4
sensitivity curves in Fig.\ref{fig:sensitivity}. 
\begin{table}[!htbp]
  \begin{center}
    \begin{tabular*}{.75\textwidth}{ @{\extracolsep{\fill}} | c || c | c
        | c | c |}
     \hline
     & \multicolumn{4}{c|}{Detector recoil} \\
     & \multicolumn{4}{c|}{threshold keV} \\
     \hline
     \hline
     Escape Velocity km/s & 5 & 10 & 15 & 20 \\
     (energy cut-off)  & & & & \\
     \hline
      600 (120 keV) & 74 & 51 & 32 & 18 \\
      \hline
      \textbf{550} (\textbf{100} keV) & \textbf{74} & \textbf{51} & \textbf{32} & \textbf{18} \\
      \hline
      500 (83 keV) & 74 & 51 & 32 & 18 \\
      \hline
      450 (67 keV) & 73 & 50 & 31 & 17 \\
      \hline
    \end{tabular*}
    \caption{Percentage of total recoil spectrum registered in Xe
      target for $M_D = 60$ GeV as a function of recoil
      energy threshold and escape velocity cut-off.}
    \label{tab:recoil_xe}
  \end{center}
\end{table}
\begin{table}[!htbp]
  \begin{center}
    \begin{tabular*}{0.75\textwidth}{ @{\extracolsep{\fill}} | c || c | c
        | c | c |}
      \hline
      & \multicolumn{4}{c|}{Detector recoil} \\
      & \multicolumn{4}{c|}{threshold keV} \\
      \hline
      \hline
      Escape Velocity km/s & 10 & 20 & 30 & 40 \\
      (energy cut-off)  & & & & \\
      \hline
      600 (120 keV) & 68 & 53 & 37 & 26 \\
      \hline
      \textbf{550} (\textbf{100} keV) & \textbf{66} & \textbf{51} & \textbf{35} & \textbf{24} \\
      \hline
      500 (83 keV) & 62 & 47 & 32 & 21 \\
      \hline
      450 (67 keV) & 57 & 42 & 27 & 16 \\
      \hline
    \end{tabular*}
    \caption{Percentage total recoil spectrum registered in Ar target
      for $M_D = 60$ GeV as a function of recoil energy
      threshold and escape velocity cut-off.}
    \label{tab:recoil_ar}
  \end{center}
\end{table}

\begin{figure}[!htbp]
  \centering
\includegraphics[width=0.75 \columnwidth]{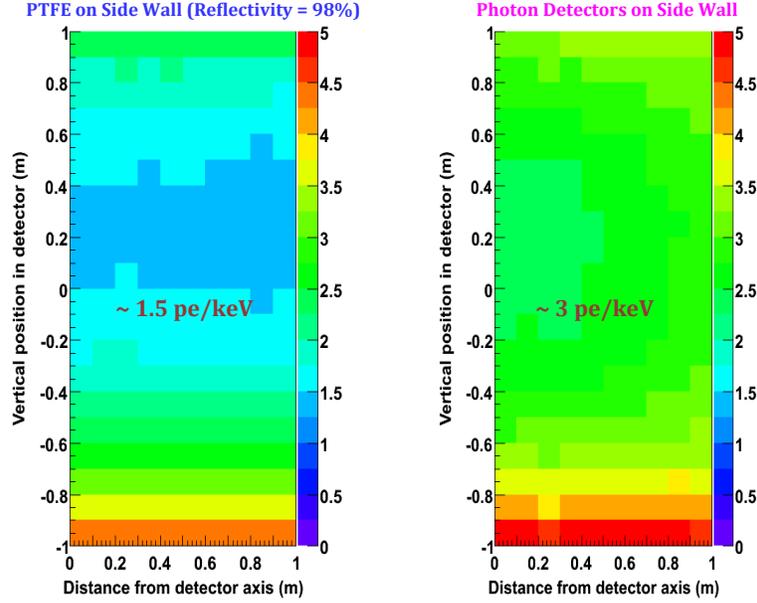}
\caption{Comparison of light collection for sidewalls covered in QUPID
  photodetectors or 98\% reflective PTFE panels for a $2~\n{m} \times 2~\n{m}$ Xe
  target. The simulations used typical input parameters of Xe light
  output of 60 uv photons/keV, quenched to 30 photons by an assumed
  electric field of $\sim 50-100$ kV/m, with a 30\% photocathode
  coverage, and 35\% QUPID quantum efficiency.}
\label{fig:light_coll}
\end{figure}

It is of interest also to simulate the difference in light collection
between the use of high reflectivity panels on the detector side-walls
(see Xe detector illustration in Fig.\ref{fig:g2inner}) and the
replacement of these by photodetectors. For Xe several mm PTFE inner
walls, comparisons of simulated and observed detector performance are
consistent with reflectivities 95-98\%~\cite{Yamashuta:2004,Chepel:2005} and for Ar comparable
reflectivities appear to be achievable by coating inner wall with TPB
wavelength shifter~\cite{Boccone:2009}. Fig.\ref{fig:light_coll}
shows, for simulations in liquid Xe, that the (lower cost) reflecting panels give {\it i)} a less uniform
light collection and {\it ii)} up to a factor 2 lower light output in
photoelectrons/keV, creating a higher and position-dependent energy
threshold.

\subsection{WIMP mass determination -- single detector}
\label{subsec:wimpmass_single}

As discussed in Sec.\ref{subsec:recoil_spectra}, a best fit of an
observed spectrum of events to Eq.\ref{eq:rate} can yield a value for
the WIMP mass with an accuracy dependent on target mass and running
time. 

\begin{figure}[!htbp]
  \centering
\includegraphics[width=0.75 \columnwidth]{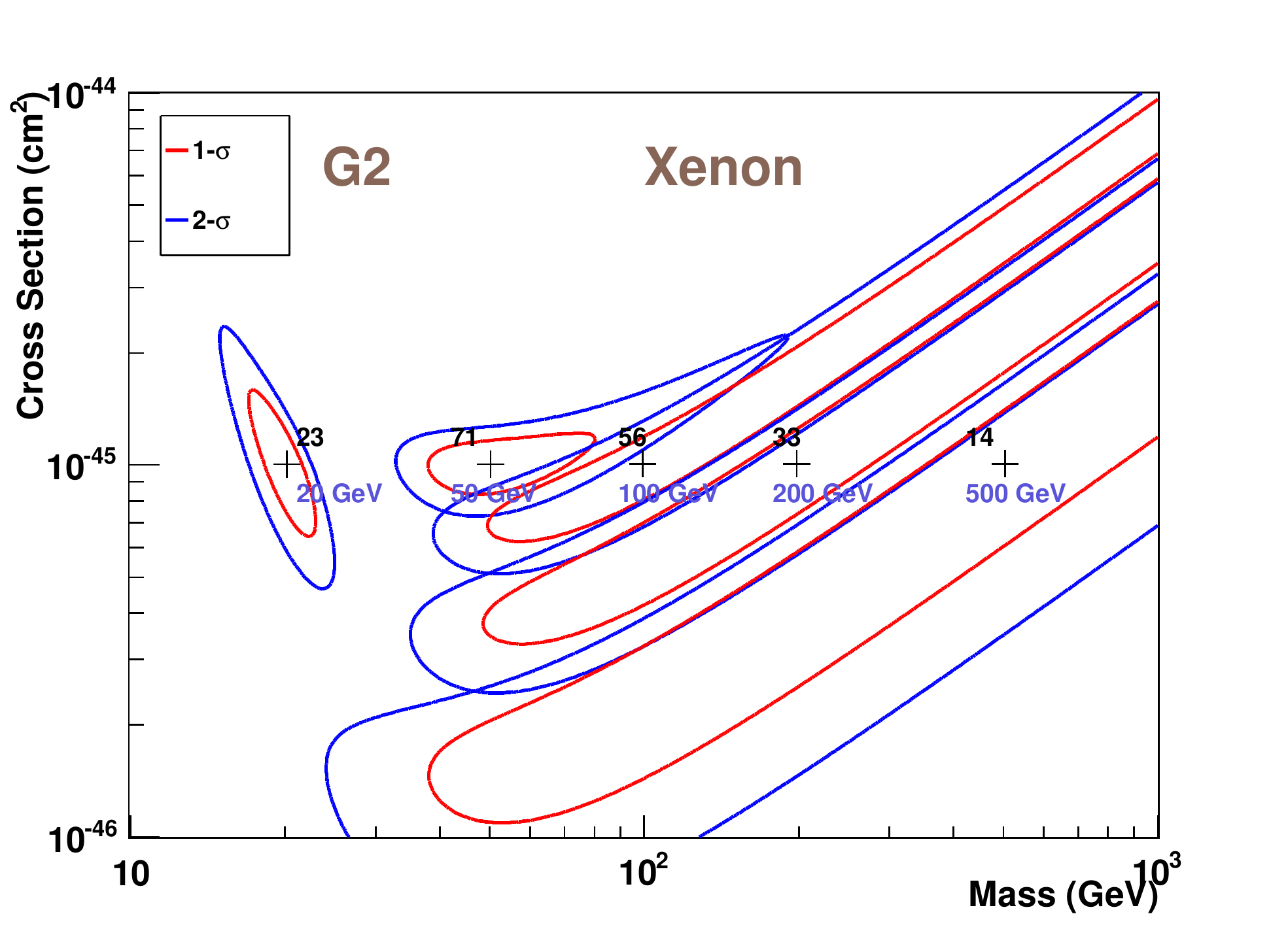}
\includegraphics[width=0.75 \columnwidth]{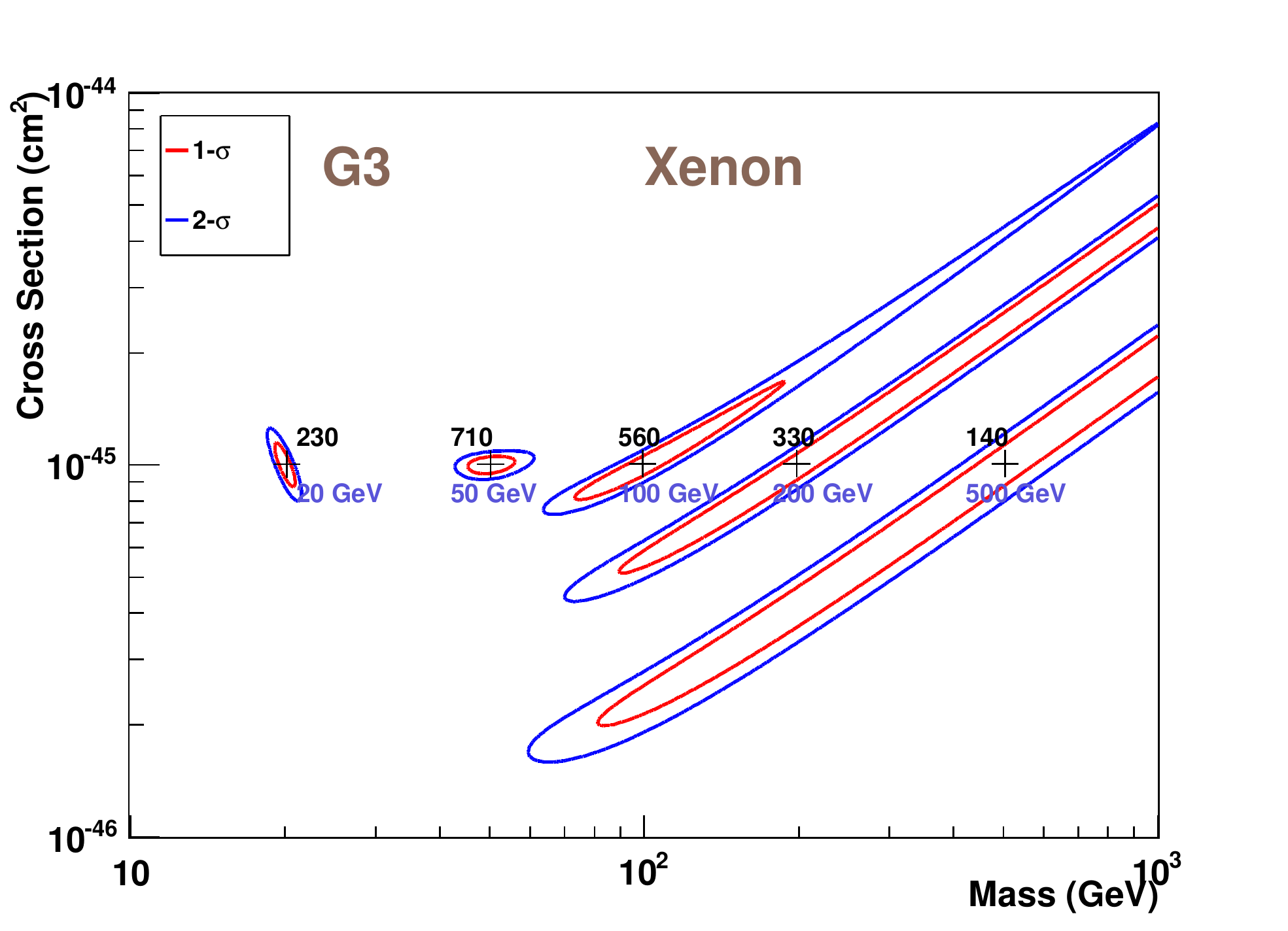}
\caption{Examples of estimation of WIMP mass from spectrum of events in xenon detectors running for 1 year:\protect\\
  \textit{(upper plot)} G2 1-ton,\protect\\
  \textit{(lower plot)} G3 10-ton\protect\\
  Red and blue contours show $1\sigma$ and $2\sigma $ confidence spread
  for cross section $10^{-45}~\n{cm}^2$ and mass values labelled in
  blue, event numbers in black.\protect\\
  The plots apply equally to G3 (upper) and G4 (lower) for cross section
  $10^{-46}~\n{cm}^2$.}
\label{fig:xs_mass_xe}
\end{figure}

Fig.\ref{fig:xs_mass_xe} (upper) illustrates this for the case of an
interaction cross section $10^{-45}~\n{cm}^2$ and the G2 1-ton Xe
detector running for 1 year, showing $1\sigma$ and $2\sigma$ error
contours for different WIMP mass values. Fig.\ref{fig:xs_mass_xe} (lower)
shows the smaller errors resulting from the G3 10-ton detector running
for the same time period. The same plots would apply for G3 and G4
detectors for an interaction cross-section $10^{-46}~\n{cm}^2$, the
order of magnitude smaller cross section, with an order of magnitude
greater mass, giving the same event numbers.

\begin{figure}[!htbp]
  \centering
\includegraphics[width=0.9 \columnwidth]{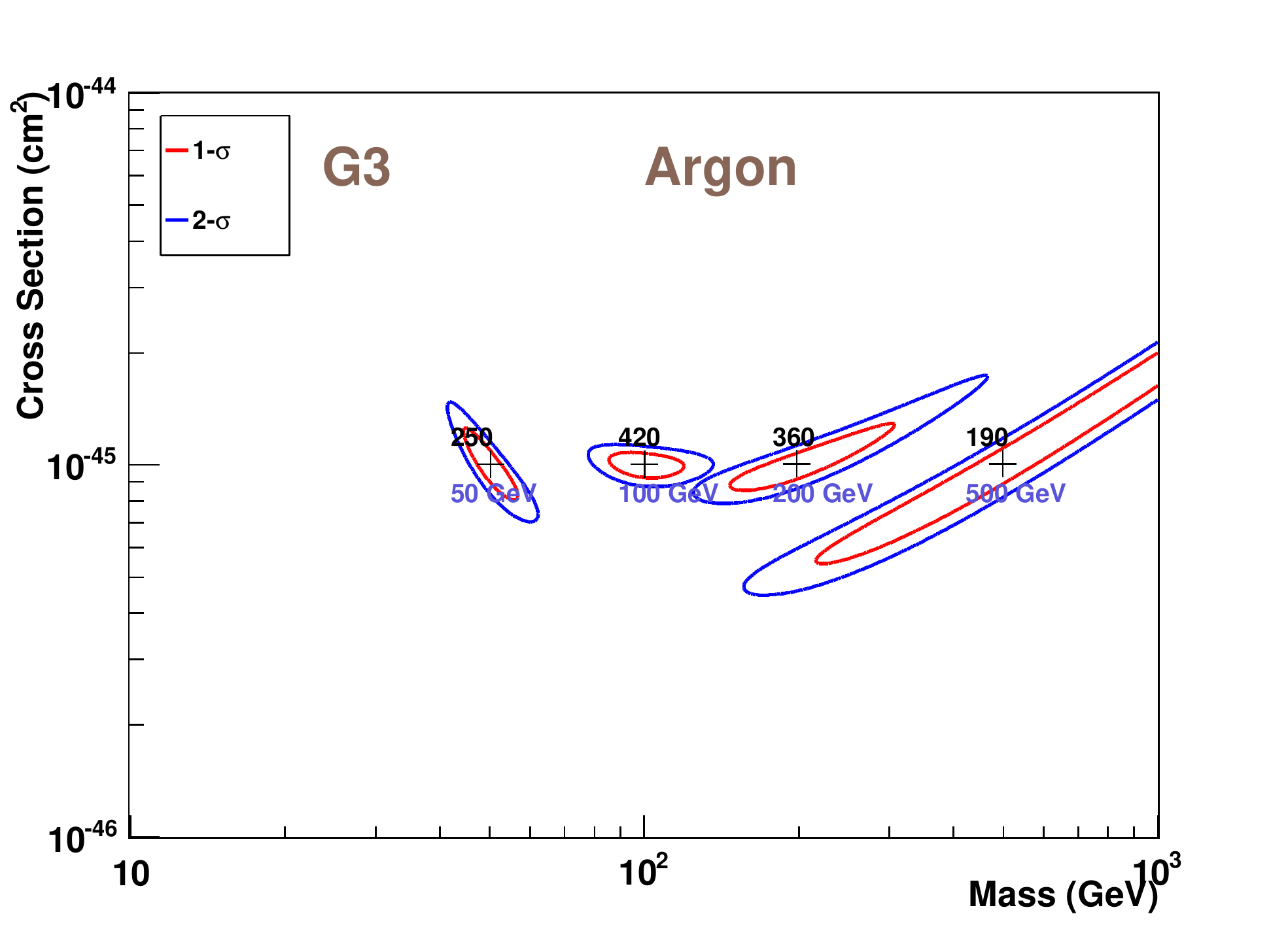}
\caption{$1\sigma$ and $2\sigma$ contours of mass precision for
  cross-section $10^{-45}~\n{cm}^2$ as in Fig.\ref{fig:xs_mass_xe},
  but for 50 ton-y G3 Ar detector. Numbers of events (in black) for
  each WIMP mass value are similar in order of magnitude to those in
  Fig.\ref{fig:xs_mass_xe}, consistent with the requirement for an
  Ar/Xe target mass ratio $\sim 5$ to obtain similar signals in
  each. The same plot, with the same event numbers, would apply to the
  G4 Ar detector for a cross-section $10^{-46}~\n{cm}^2$.}
\label{fig:xs_mass_ar}
\end{figure}

Fig.\ref{fig:xs_mass_ar} shows corresponding error contours for the
G3 50-ton Ar detector, the higher mass of Ar providing event numbers
comparable to those from the Xe detector, in accordance with the
factor $\sim 5$ lower yield of events/ton for Ar, as calculated
above. Again the G4 Ar detector would provide a similar yield of
events for a cross-section $10^{-46}~\n{cm}^2$. The one and
two-sigma errors shown in these plots, and also in
Figs. \ref{fig:xs_mass_xear} and \ref{fig:xs_mass_xear_g2g3},
are those arising specifically from statistical uncertainty in the
numbers of events, and assume no major departures from the
``canonical'' velocity distribution discussed in Sec.\ref{subsec:recoil_spectra} and the associated ``standard'' value of $v_o$.

\subsection{WIMP mass determination using combined Xe and Ar signals}
\label{sec:mass_xear}

\begin{figure}[!htbp]
  \centering
\includegraphics[width=0.9 \columnwidth]{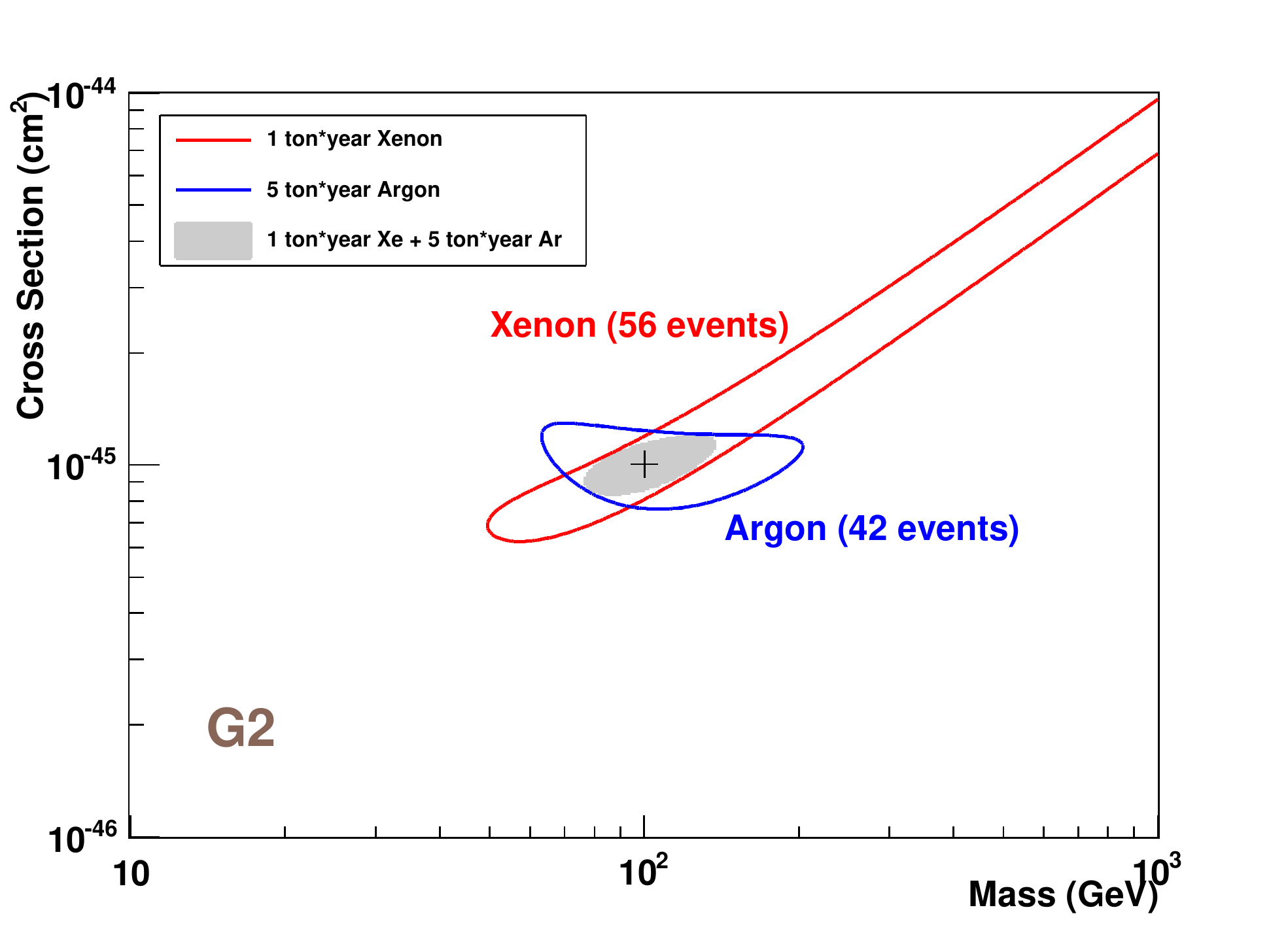}
\caption{Overlaying Xe and Ar signals. Example of overlap of $1\sigma$
  contours from 1 ton-y Xe and 5 ton-y Ar for a specific WIMP Mass 100
  GeV and WIMP-nucleus cross-section $10^{-45}~\n{cm}^2$, showing
  agreement of WIMP mass values and similar event numbers from Xe and
  Ar detectors with target mass ratio $\sim 5$. This plot applies also
  to the larger G3 system with a lower cross section
  $10^{-46}~\n{cm}^2$ and to the G4 system with a cross section
  $10^{-47}~\n{cm}^2$.}
\label{fig:xs_mass_xear}
\end{figure}

As discussed in Sec.\ref{subsec:recoil_spectra}, an important aspect
of the use of two different elements is that they should provide
independent signals that agree in the derived WIMP
mass. Fig.\ref{fig:xs_mass_xear} gives an example of this, for the G2
system with a cross section $10^{-45}~\n{cm}^2$ and a specific dark
matter mass of 100 GeV, showing that the error contours for each will
overlap and surround the same mass and cross-section values. The same
plot applies to the G3 system for a cross-section $10^{-46}~\n{cm}^2$
, and to the G4 system for a cross-section $10^{-47}~\n{cm}^2$.
\begin{figure}[!htbp]
  \centering
\includegraphics[width=0.75 \columnwidth]{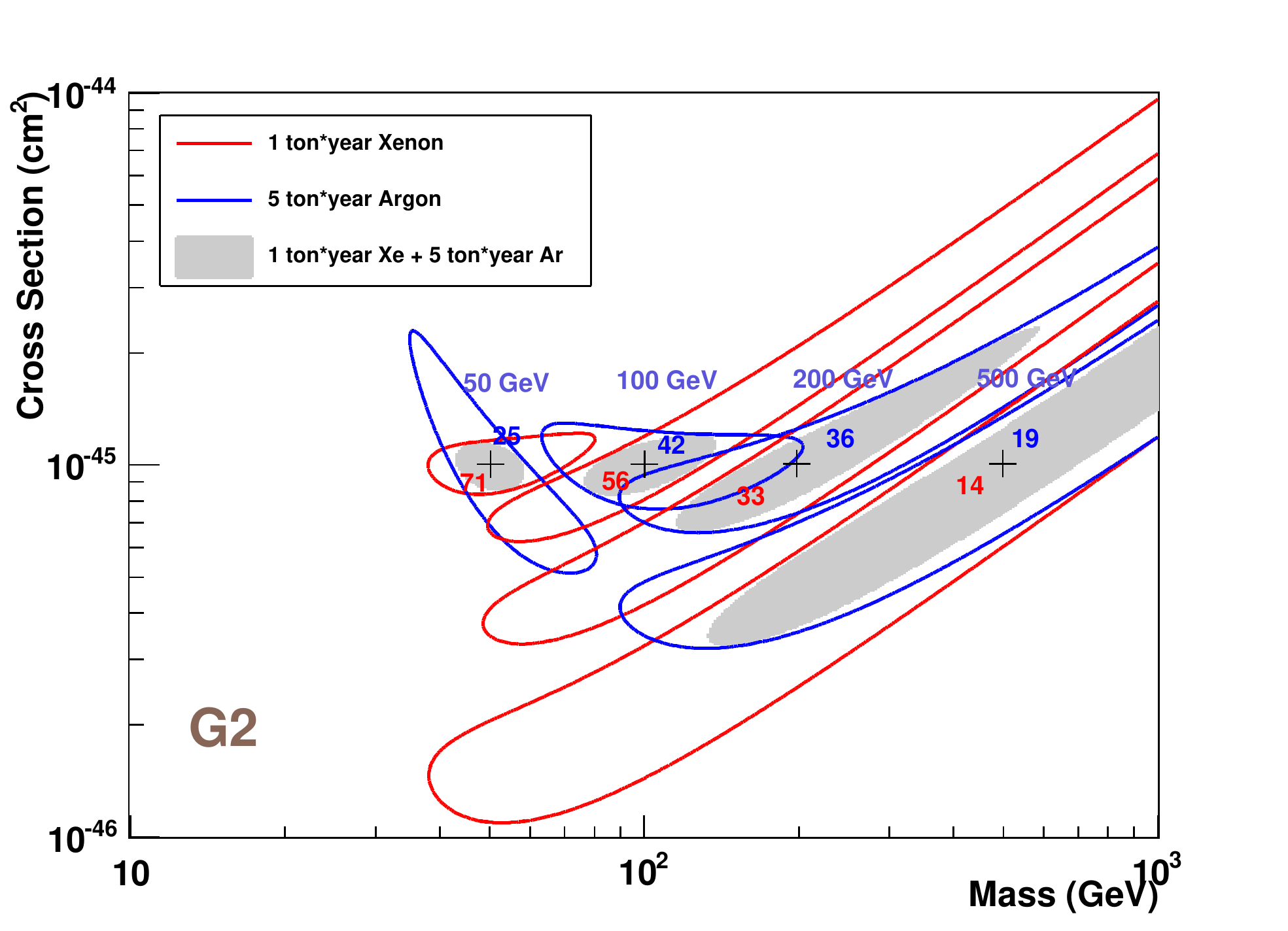}
\includegraphics[width=0.75 \columnwidth]{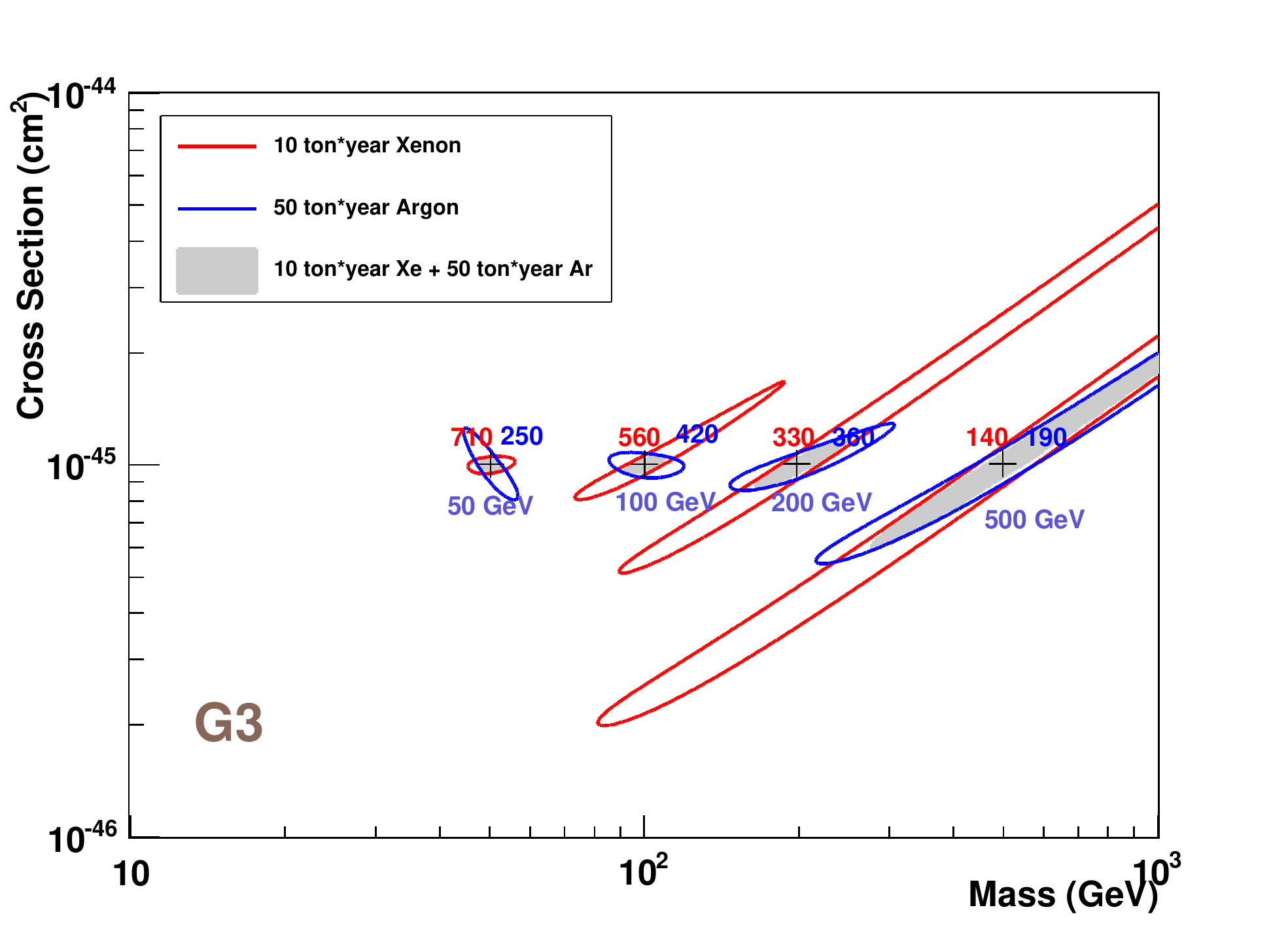}
\caption{Examples of overlap of $1\sigma$ Xe and Ar contours for WIMP masses 50 - 500 GeV\protect\\
\textit{(upper) }cross-section $10^{-45}~\n{cm}^2$ with G2 system: 1 ton-y Xe {\&} 5 ton-y Ar \protect\\
or cross-section $10^{-46}~\n{cm}^2$ with G3 system: 10 ton-y Xe {\&} 50 ton-y Ar\protect\\
or cross-section $10^{-47}~\n{cm}^2$ with G4 system: 100 ton-y Xe {\&} 500 ton-y Ar. \protect\\
\textit{(lower)} cross-section $10^{-45}~\n{cm}^2$ with G3 system: 10 ton-y Xe {\&} 50 ton-y Ar\protect\\
or cross-section $10^{-46}~\n{cm}^2$ with G4 system: 100 ton-y Xe {\&} 500 ton-y Ar.}
\label{fig:xs_mass_xear_g2g3}
\end{figure}

Further examples of this overlap, illustrating the level of precision
obtainable (for a one year exposure) by combining Xe and Ar signals, are
shown in Fig.\ref{fig:xs_mass_xear_g2g3} for the following cases:
\begin{pbenumerate}
\item \textit{(upper plot)} G2 and \textit{(lower plot)} G3, showing
  WIMP-nucleon mass measurement at cross-section $10^{-45}~\n{cm}^2$
  for WIMP masses $50 - 500$ GeV. The plots also show the number of
  signal events from Xe and Ar in each case.
\item \textit{(upper plot)} G3 and \textit{(lower plot)} G4, allowing
  WIMP mass measurement at cross section $10^{-46}~\n{cm}^2$. The event
  numbers remain unaltered by this scaling.
\item \textit{(upper plot)} G4, allowing the lower precision WIMP mass
  measurement at a cross section $10^{-47}~\n{cm}^2$.
\end{pbenumerate}

\subsection{Annual signal modulation}
\label{sec:ann_modulation}

\begin{figure}[!htbp]
  \centering
  \includegraphics[width=0.9 \columnwidth]{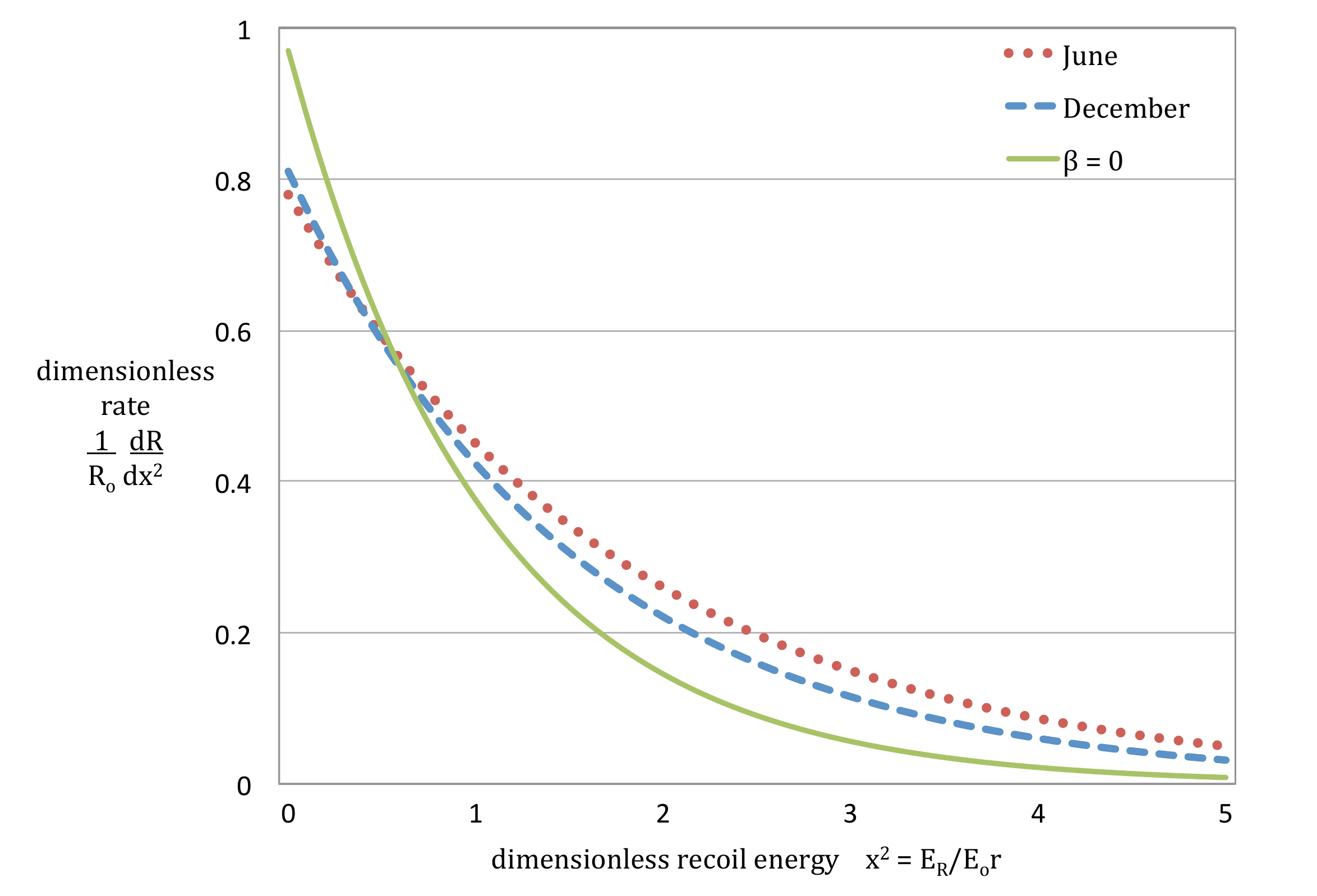}
\caption{Predicted annual fluctuation of dimensionless recoil energy
  spectrum arising from orbital motion of the earth combined with
  Galactic motion of the sun (from~\cite{Smith:1990}). The dashed line shows
  the spectrum expected for a stationary Earth.}
\label{fig:annmod}
\end{figure}

With availability of a sufficiently large sample of signal events, an
additional identifying signal would be the expected annual modulation
arising from the motion of the Earth around the Sun, as the sun moves
through the Galaxy. This can be conveniently parameterised as cyclic
variations in the values of $c_1$ and $c_2$ in Eq.\ref{eq:rate} of
Sec.\ref{subsec:recoil_spectra} (see~\cite{Smith:1990,Lewin:1996} for tables of values). The resulting
difference in differential energy spectrum between June and December
is shown, in dimensionless form, in Fig.\ref{fig:annmod}, which also
shows, for comparison, a dashed curve for the spectrum that would
result for an Earth stationary in the Galaxy.

The June and December curves cross at a recoil energy value $\sim
0.7~E_0\,r$, where $E_0$ and $r$ are defined in
Sec.\ref{subsec:recoil_spectra}, which gives a mass-dependent
crossover energy in the region $10-20$ keV for WIMP masses
$60-100$ GeV. Of interest in the case of Xe and Ar detectors is
the fact that, for that range of WIMP masses, the operating range of
Xe lies predominantly below the crossover energy, while the
operating range of Ar lies predominantly above the crossover
energy. Thus the Ar detector may observe signal events which show a
positive annual modulation, i.e. the June rate higher than the
December rate, while at the same time the xenon detector may observe
signal events showing a negative annual modulation -- i.e. the June
rate lower than December rate. This would provide an important
confirmation of the Galactic origin of the signal.
\begin{figure}[!htbp]
  \centering
\includegraphics[width=0.75 \columnwidth]{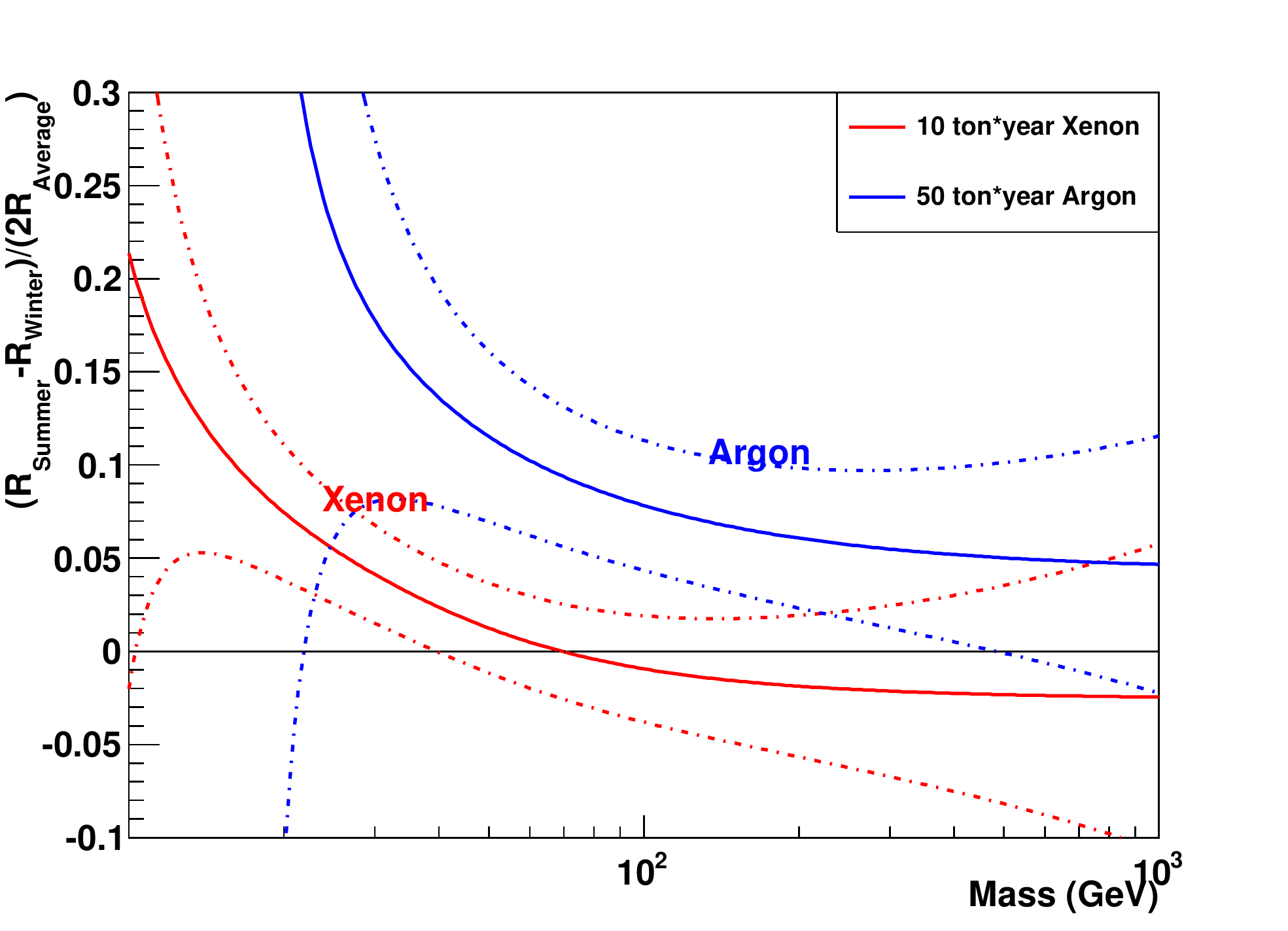}
\includegraphics[width=0.75 \columnwidth]{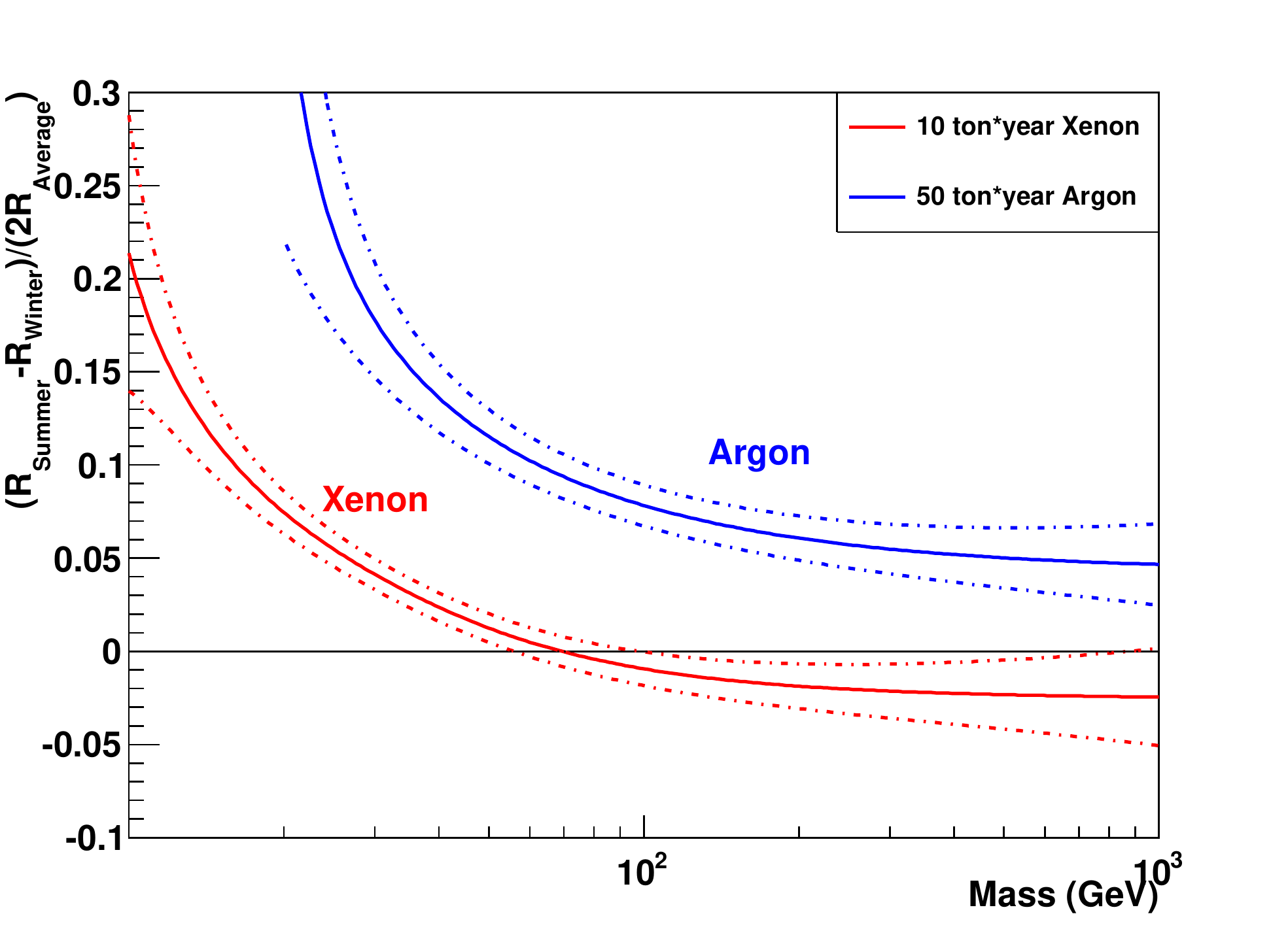}
\caption{Fractional annual modulation versus WIMP mass, shown
  separately for 10 ton-y Xe and 50 ton-y Ar. Full lines show
  predicted mean. Dashed lines show $1-\sigma $ error bands.\protect\\
(a) \textit{(upper plot ) }WIMP-nucleon cross section $10^{-45}~\n{cm}^2$\protect\\
(b) \textit{(lower plot ) }WIMP-nucleon cross section $10^{-44}~\n{cm}^2$\protect\\
The same plots will apply to 100 ton-y Xe and 500 ton-y Ar with cross sections a factor 10 smaller (i.e. giving the same event numbers)}
\label{fig:annmod_xear}
\end{figure}
The fractional modulation of the total number of events as a function
of WIMP mass is shown in Fig.\ref{fig:annmod_xear} (\textit{upper
  plot}) for a cross section $10^{-45}~\n{cm}^2$ and
Fig.\ref{fig:annmod_xear} (\textit{lower plot}) for a cross section
$10^{-44}~\n{cm}^2$, with running times 10 ton-y Xe and 50 ton-y Ar,
together with the $1\sigma$ error contours. Thus these examples
represent the G3 system running for one year or equivalently the same
number of events from the G4 system running for 1 year with cross
sections an order of magnitude smaller.

\begin{figure}[!htbp]
  \centering
\includegraphics[width=0.75 \columnwidth]{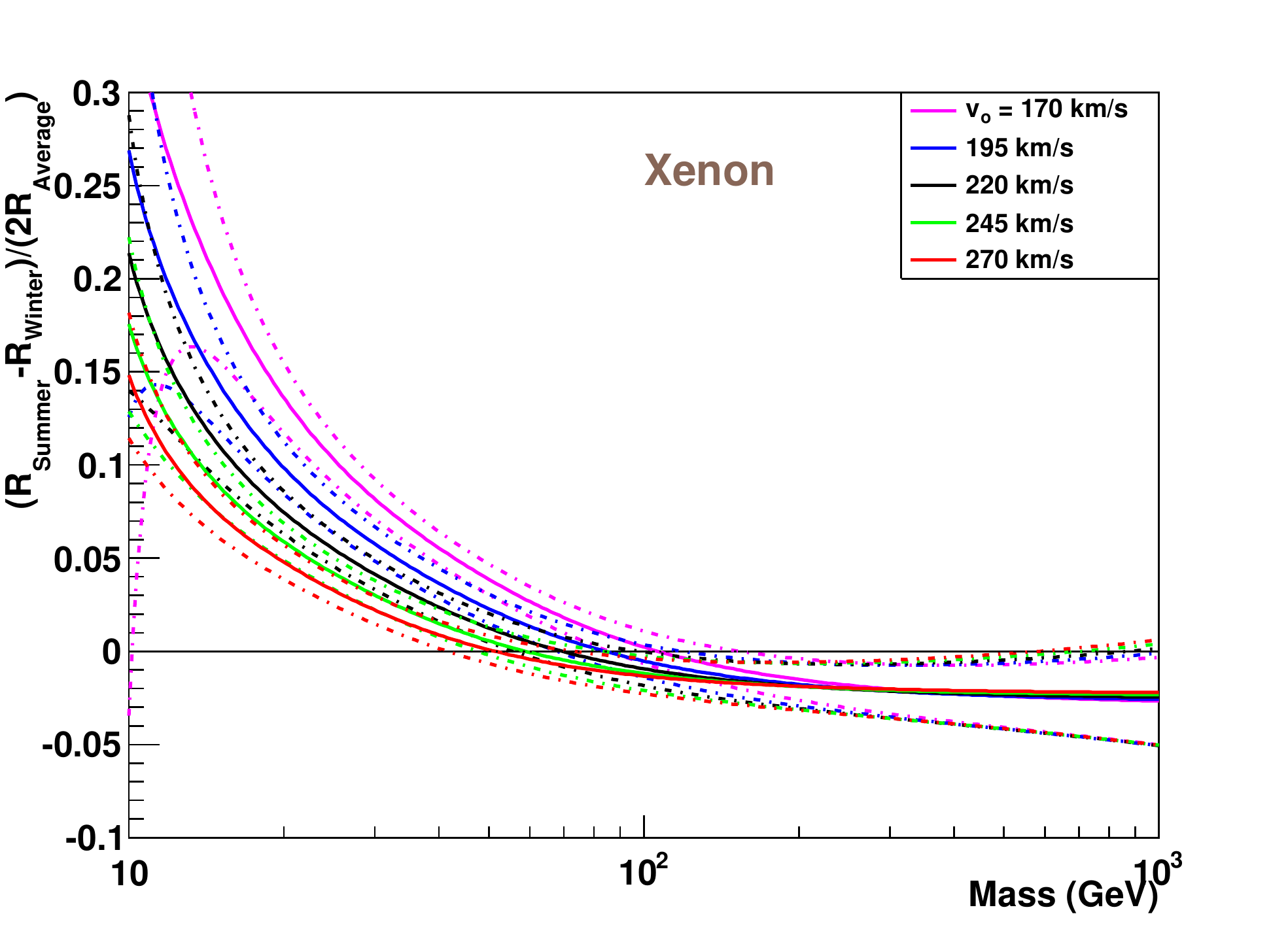}
\includegraphics[width=0.75 \columnwidth]{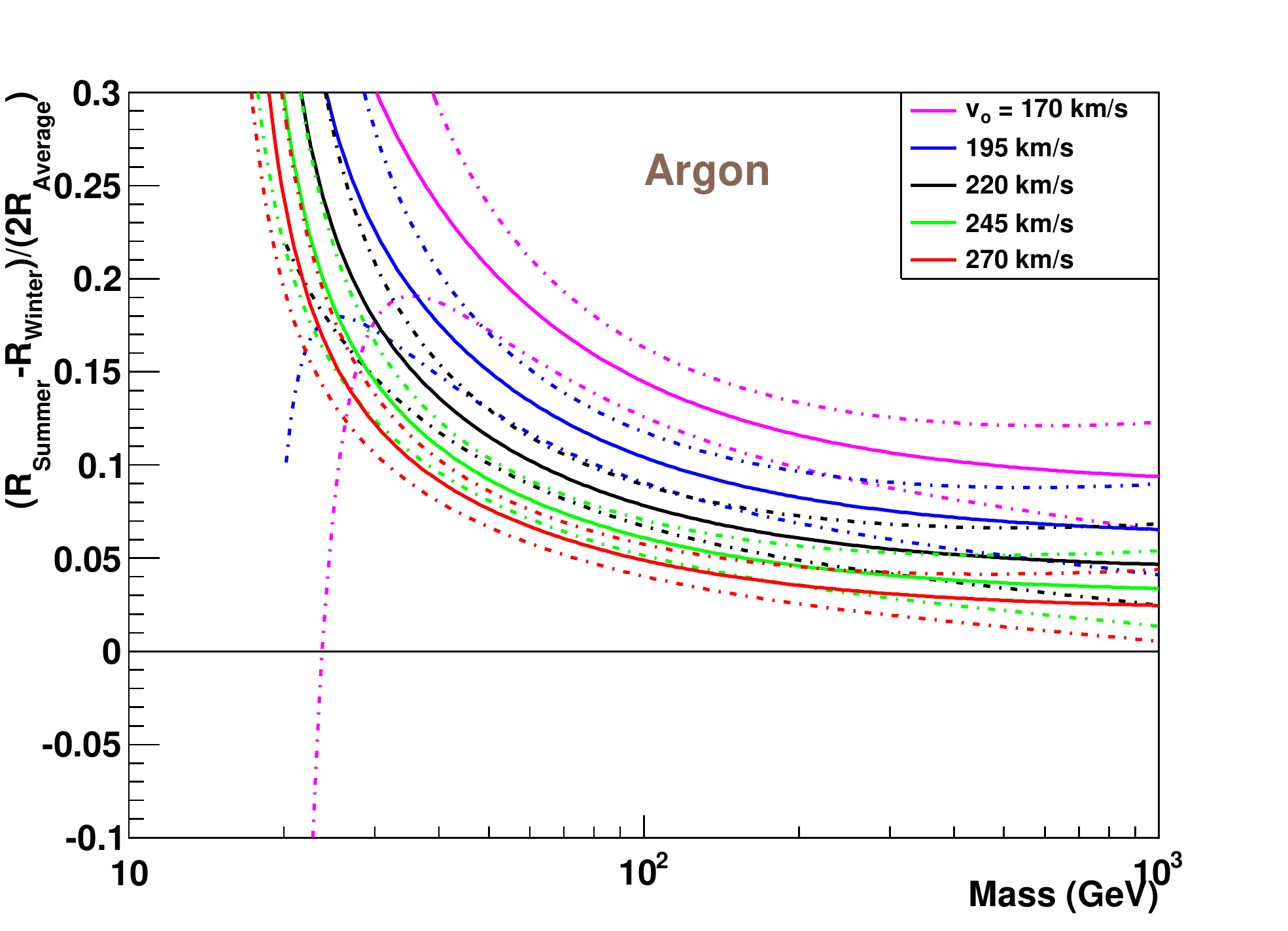}
\caption{Example of the variation of annual modulation amplitude with
  WIMP mean velocity varied by $\pm 50~\n{km/s}$ about the
  ``standard'' assumption of $220\, \n{km/s}$. Dashed lines show
  $1\sigma$ error bands for the illustrative case of 10 ton-y Xe, 50
  ton-y Ar, for a WIMP-nucleon cross-section $10^{-44}~\n{cm}^2$ (or
  equivalent numbers of events at $10^{-45}~\n{cm}^2$ for 100 ton-y Xe
  and 500 ton-y Ar).}
\label{fig:annmod_xear_vel}
\end{figure}
Following the difficulty of accounting for the persistent observed
annual modulation of low energy events in the DAMA/LIBRA NaI
arrays~\cite{Bernabei:2008,Bernabei_1:2008} there has been interest in
the effect of a possible uncertainty in the characteristic velocity
$v_0$ in the assumed WIMP velocity distribution~\cite{Lewin:1996}, which is
normally assumed to have the value $v_0 \sim 220~\n{km/s}$ (similar to
that of the sun's Galactic orbital velocity). We therefore illustrate
in Fig.\ref{fig:annmod_xear_vel} how the curves of annual modulation
versus WIMP mass vary when the value of $v_0$ is changed by $\pm
50~\n{km/s}$. It is seen that modulation percentages may reach $\sim
30\%$ at low WIMP masses if $v_0$ is as low as $170~\n{km/s}$. It is
also apparent from the $1\sigma $ error curves that the larger number
of events provided by the G3 or G4 systems would be required for
accurate investigation of annual modulation within a few-year time
period.

\subsection{Spin-dependent WIMP sensitivity}
\label{sec:spin_dependent}

Because of the $A^2$ multiplying factor for a coherent
spin-independent interaction (i.e. involving the whole nucleus), it is
usually believed that this will dominate over any spin-dependent
interaction, which would involve interaction with usually a single
unpaired nucleon in an odd-$A$ isotope. Nevertheless, spin-dependent
interactions might dominate in some theoretical scenarios where the
spin-independent term is suppressed. In addition, it has been
suggested that the DAMA annual
modulation~\cite{Bernabei:2008,Bernabei_1:2008} might result from a
low mass ($< 15$ GeV) WIMP with spin-dependent interaction. 
\begin{figure}[!htbp]
  \centering
\includegraphics[width=0.75 \columnwidth]{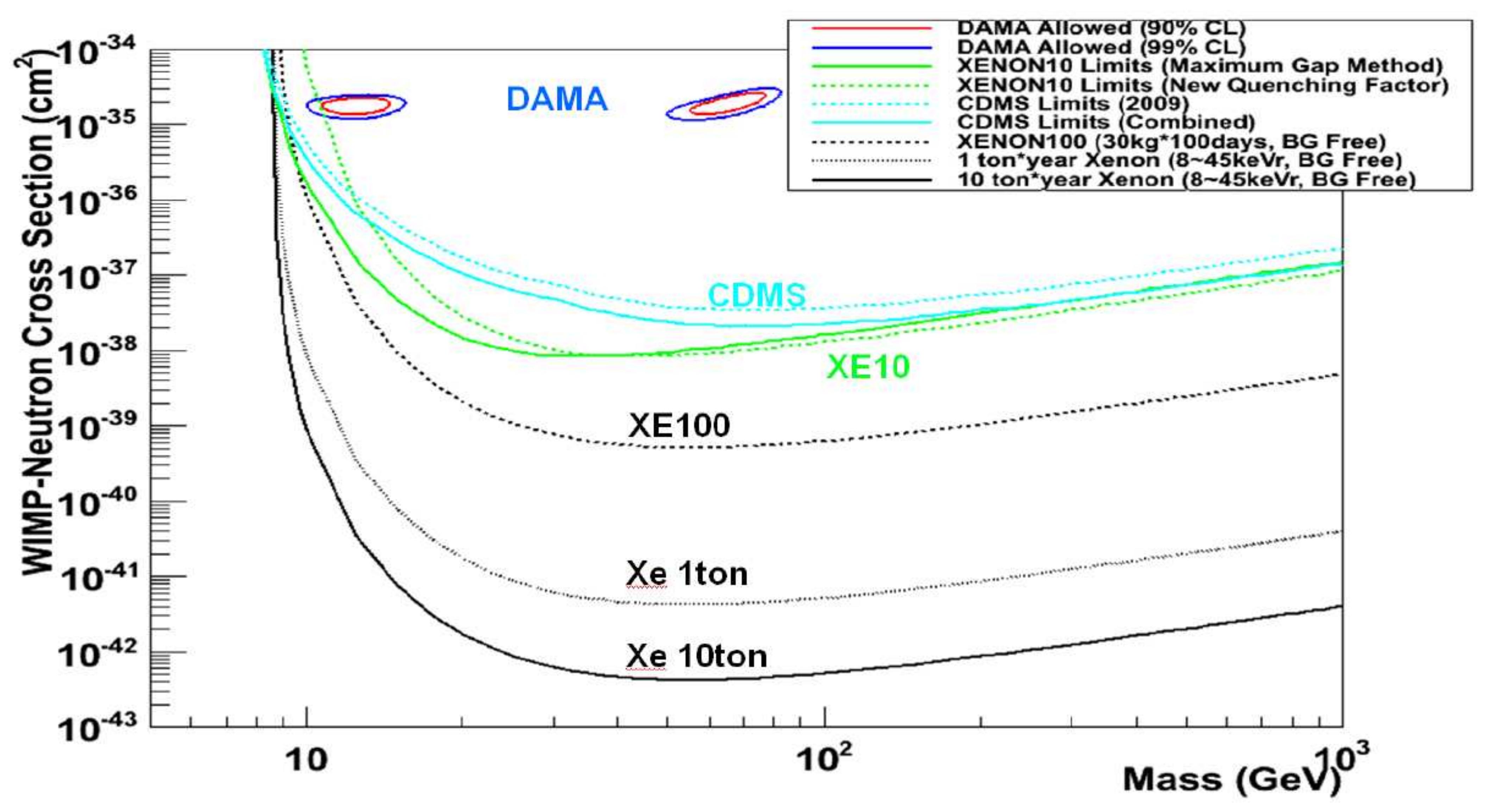}
\caption{90\% limits for WIMP-neutron spin-dependent interactions for
  1 ton-y and 10 ton-y Xe, also showing exclusion of DAMA
  limits by several previous experiments
  ~\cite{Bernabei:2008,Bernabei_1:2008, Angle:2009,Aprile_idm:2011,CDMS:2010}.}
\label{fig:sensitivity_spindep}
\end{figure}

Xe has two stable odd isotopes, $^{129}$Xe (spin-1/2), $^{131}$Xe
(spin-3/2), together constituting $47.6\%$ of natural xenon, and is
therefore able to observe signals from both spin-dependent and
spin-independent interactions. Ar has only even-$A$ spin-0 stable
isotopes, so that if WIMP-like signals were to be seen in liquid Xe
but not in liquid Ar, this would be suggestive of a pure
spin-dependent interaction (subsequently requiring confirmation by
running with isotopically-enriched xenon to produce an enhanced odd-$A$
signal). Calculation of spin-dependent cross-section limits from the
data is a more difficult process, requiring replacement of the
spin-independent form factor in Eq.\ref{eq:rate} by a spin-dependent
and nuclear structure-dependent form factor, and the replacement of
the $A^2$ factor in Eq.\ref{eq:xsection} by a non-trivial
combination of nuclear spin terms, which differ for neutrons and
protons. 

In the evaluation of spin-dependent limits for ZEPLIN
III~\cite{Lebedenko:2009}, the authors used the latest theoretical
studies on the required nuclear spin terms and form factors, and
concluded that the limit curve (versus WIMP mass) for spin-dependent
WIMP-neutron cross section lies close to a factor $\sim10^5$ higher
than the limit curve for spin-independent interactions
(Fig.\ref{fig:sensitivity}), while that for the WIMP-proton cross
section lies close to a factor $\sim 10^7$ higher. This factor $\sim
100$ difference is consistent with the fact that both $^{129}$Xe and
$^{131}$Xe nuclei contain unpaired neutrons, and hence are relatively
insensitive to WIMP-proton scattering. These two scaling factors are
almost independent of WIMP mass, so that the shapes of spin-dependent
and spin-independent limit curves are essentially identical. Thus, for
estimates of spin-dependent limits obtainable with the G2, G3, and G4
Xe detectors, it is necessary only to scale the Xe limit-curves in
Fig.\ref{fig:sensitivity} by the above factors $10^5$ for WIMP-neutron
and $10^7$ for WIMP-proton interactions. The odd-isotopes of Xe
provide the most sensitive experimental target element for detection
of spin-dependent WIMP-neutron interactions~\cite{Lebedenko:2009}. 

\subsection{Inelastic WIMP sensitivity}
\label{sec:sensitivity_idm}

The concept of a dark matter particle that could exist in two mass
levels, and thus recoil from a nucleus in an excited state, was
introduced as a possible explanation of the DAMA annual modulation
observations in a NaI
detector~\cite{Tucker-Smith:2001,Tucker-Smith:2005,Chang:2009}. The
recoil spectrum differs from that of a single state particle, being
suppressed at low energy, and peaking at a recoil energy of
$40-50$keV. For a hypothesised mass splitting $\delta$ in the
range $20-140$ keV the spectrum can be consistent with the DAMA
observations, while at the same time the annual modulation can be
larger because one is sampling a higher velocity component of the WIMP
velocity distribution~\cite{Chang:2009} where the percentage seasonal
fluctuations are larger~\cite{Smith:1990_1}. 
\begin{figure}[!htbp]
  \centering
 \includegraphics[width=0.75 \columnwidth]{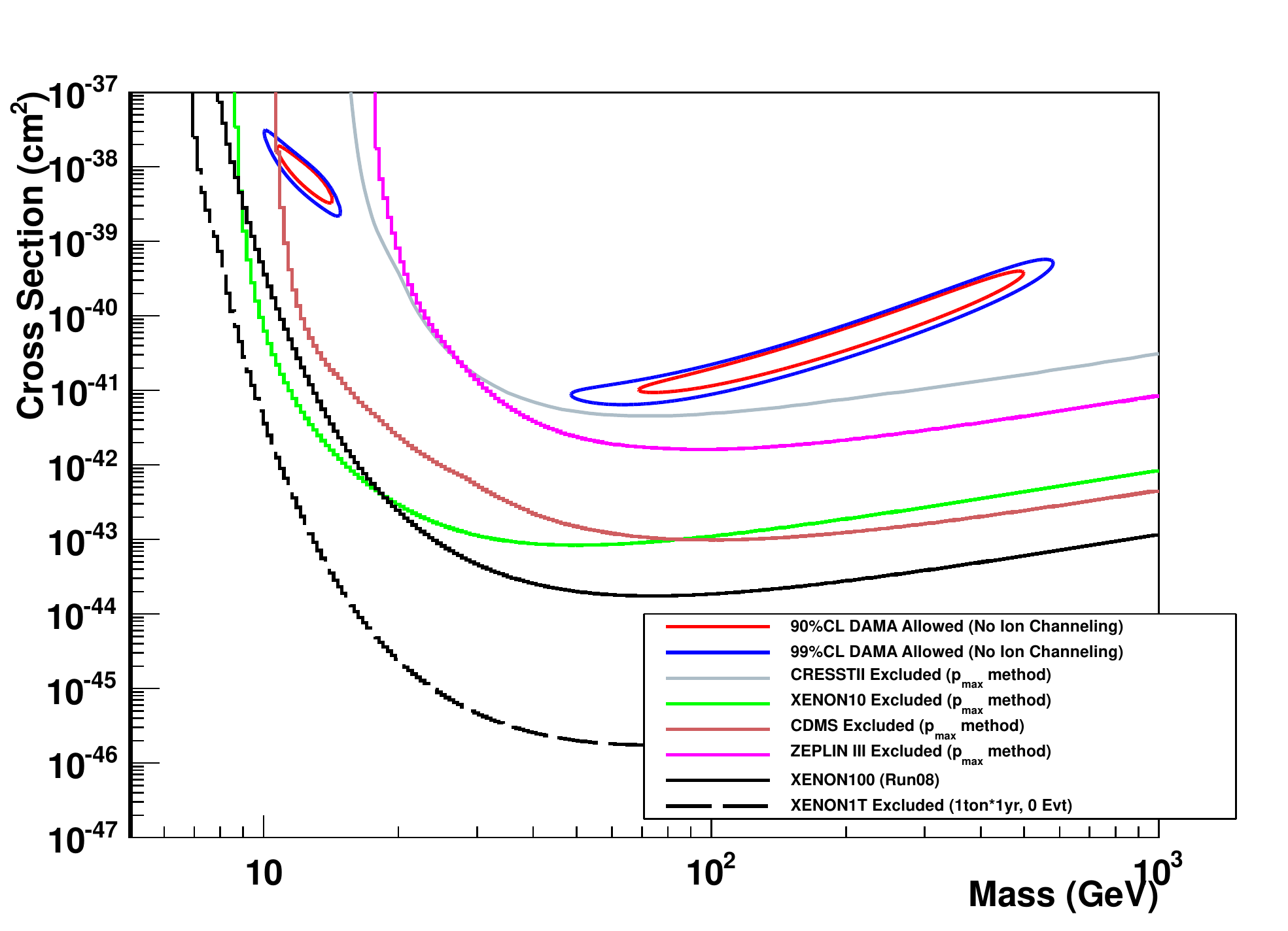}
  \includegraphics[width=0.75 \columnwidth]{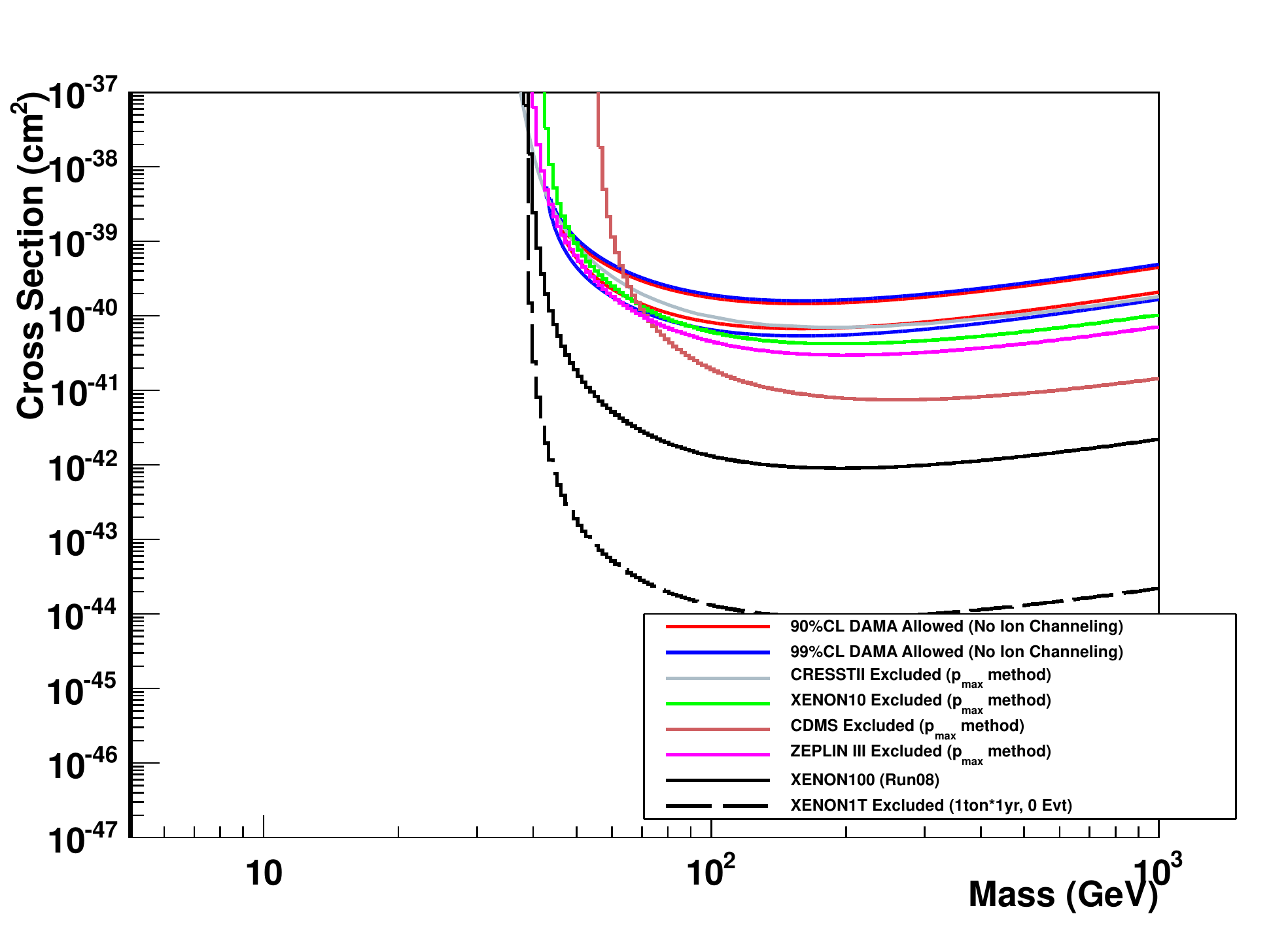}
  \caption{Inelastic WIMP-nucleon cross-section limits versus WIMP mass,
  showing DAMA~\cite{Tucker-Smith:2001,Tucker-Smith:2005,Chang:2009}
  allowed regions and exclusion boundaries for CRESSTII~\cite{Schmidt:2009}, CDMS~\cite{CDMS:2010},
  XENON10~\cite{Angle:2009}, XENON100~\cite{Aprile_idm:2011}, ZEPLIN
  III~\cite{Akimov:2010}, published limits, and projected 1 ton Xe limit, for Galactic escape velocity 550 km/s\protect\\
(\textit{upper plot}) mass-splitting $\delta = 20~\n{keV}$\protect\\
(\textit{lower plot}) mass-splitting $\delta = 100~\n{keV}$.}
\label{fig:sensitivity_idm}
\end{figure}

Areas of inelastic cross section versus WIMP mass compatible with the
DAMA annual modulation data for suitable choices of $\delta$ are shown
in Fig.\ref{fig:sensitivity_idm}, together with published exclusion
plots from XENON10~\cite{Angle:2009}, XENON100~\cite{Aprile_idm:2011},
CDMS~\cite{CDMS:2010}, and ZEPLIN III~\cite{Akimov:2010}. It can be seen that the inelastic
interpretation of the DAMA region appears now to be fully excluded. A
limit curve for a 1 ton Xe detector (G2) is shown which will be capable of excluding the
inelastic explanation of the DAMA modulation by a larger
margin. Fig.\ref{fig:sensitivity_idm} is drawn for a galactic escape
velocity of 550 km/s, and similar conclusions result from an
assumption of the escape velocity anywhere in the range $550\pm
70$ km/s.

\section{Backgrounds}
\label{sec:backgrounds}

\subsection{Background categories and detector background discrimination}
\label{subsec:categories}

The cross-section sensitivity and target mass estimates discussed in
Sec.\ref{sec:sensitivity} depend on the reduction of the sum
of all unrejected low energy (\textless 50 keV) backgrounds to below
$\sim 0.2$ events/y, in each of the detectors comprising the G2, G3
and G4 systems. Sources of background can be divided into three categories:
\begin{pbenumerate}
\item External backgrounds, i.e. gammas, neutrons from sources outside
  the detector. 
\item Target backgrounds, i.e. gammas, neutrons and betas from
  isotopic contamination of the liquid xenon and argon and by
  contamination with other radioactive materials.
\item Internal backgrounds, i.e. gammas, neutrons, betas and other
  interactions from internal radioactive contamination of the detector
  vessel, photodetectors, and structural materials.
\end{pbenumerate}

The detectors are two-phase Xe and Ar TPC configurations, which can
discriminate nuclear recoils from electron recoils (gamma and beta
interactions) by the ratio of primary (S1) and secondary (S2) scintillation
signals. There is some overlap between the two distributions, and for
an optimum ``cut'' between the two it is possible to remove all but
$0.1 -1\%$ of the gamma background in a fiducial target region. We
make the conservative assumption that this factor is $\sim 1\%$, so
that it is sufficient to reduce low energy electron recoil background
to $<20$ events/y, using the two-phase discrimination to reduce
this to $<0.2$ events/y. For neutron interactions in the target,
a basic multiple-scattering cut can be made, after which any remaining
neutron-nucleus collisions will be indistinguishable from WIMP-nucleus
collisions. For this reason the attenuation of neutron background to a
negligible level, by shielding and veto techniques, is the prime
requirement.

\subsection{External backgrounds}
\label{subsec:external}

Backgrounds from outside the detector arise from:
\begin{pbenumerate}
\item Neutrons produced by absorption and spallation of cosmic ray
  muons. These require operation in underground laboratories to reduce
  the muon flux by a factor $10^6 - 10^8$.
\item Neutrons from U/Th in surrounding materials, in particular the
  underground cavern rock, predominately from alpha-n reactions
  together with some U fission.
\item Gammas emitted by U/Th in surrounding materials, plus a smaller
  contribution from muon bremsstrahlung in the rock.
\end{pbenumerate}

Typical neutron production rates and spectra from sources 1 and 2
have been tabulated by Bungau et al.~\cite{Bungau:2005} for several site
depths. Neutrons from source 2 have energies $<10$ MeV,
while those from source 1 have energies up to the region $1-10$
GeV. The flux from source 1 is typically a factor 1000 less than
that from source 2, but is more penetrating and could result in a
higher target neutron background unless adequately
shielded. In~\cite{Bungau:2005} simulation results are also shown of
the attenuation of neutron flux by lead or iron outer shielding
followed by inner passive hydrocarbon shielding (or an active
scintillator veto) for detectors up to several hundred kilograms, and
which are applicable to detectors of higher mass and larger surface
area simply by increasing the thickness of the various shielding
layers. However, for multi-ton detectors the required larger mass and higher
cost of the metallic shielding, together with the availability of
larger underground experimental volumes, is such that it is now
thought to be simpler and more economical to use a several-meter
thickness of water shielding, possibly also with an inner active
scintillator veto. 

A tabulation of total muon, neutron, and gamma fluxes for some typical
underground laboratories is shown in \ref{app:externalbkg}, together with the
results of simulations of the attenuation of gamma and neutron fluxes
by water shielding. The conclusion is that the water shielding radial
thickness would need to be about 4 m for G2, increasing to 5.5 m for G3
and 7 m for G4, although self-shielding within the target volume can be
used to provide some reduction in these figures.

\subsection{Target material backgrounds}
\label{subsec:target}

Significant concerns arise from target contamination by any of the
following:
\begin{pbenumerate}
\item Contamination of liquid Xe with $^{85}$Kr, giving a
  population of low energy beta decay events and requiring Kr removal
  by distillation. 
\item Contamination of liquid Ar with $^{39}$Ar, giving both
  data pile-up and a population of low energy beta decays, reducible by
  the use of $^{39}$Ar-depleted Ar gas from an underground
  source, possible further centrifuging, by anti-coincident
  position resolution and pulse shape discrimination (see ~\ref{subapp:bkg_ar5t}).
\item Contamination of Xe or Ar with Rn, the latter giving Po or Bi
  decay products on the detector walls, some of which alpha-decay into
  the wall and Po or Bi recoils into the liquid, with recoil energies
  $< 100$ keVnr. These are known as ``surface events'' and can be
  rejected by a radial fiducial cut.
\item Contamination of Xe or Ar with U/Th, giving a population of low
  energy gammas, betas, or neutrons, largely but not always vetoed by
  nearly-coincident higher energy events in the U/Th chain, together
  with purification to the 0.1 ppt level.
\end{pbenumerate}

A more detailed and quantitative discussion of the above backgrounds
is given in \ref{app:targetbkg}.

\subsection{Self-shielding of internal backgrounds from detector
  materials and photodetectors.}
\label{subsec:material}

Shielding or rejection of low energy target events resulting from
gamma and neutron activity in nearby materials (in particular vacuum
vessels and photodetectors) can be achieved by a combination of the
following methods:
\begin{pbenumerate}
\item Rejection of $\sim 99\%$ of electron recoil events by the
  S2/S1 discrimination analysis in the case of both xenon and argon. 
\item An additional factor $>100$ rejection of electron recoils by pulse
  shape discrimination in the case of argon.
\item Rejection of neutrons multiply-scattered in the target,
  including tagging by gammas from neutron inelastic scattering.
\item Additional use of a 0.1\% Gd-loaded liquid scintillator (or
  Gd-interleaved plastic) around the detector, to register neutrons
  scattered out of the detector. A light attenuation length $>4$ m
  over a period of years can be achieved~\cite{Piepke:1999}.
\item Using an outer thickness of the target material as a region of
  passive/active shielding leaving an inner fiducial region of
  typically half the total mass. 
\end{pbenumerate}
\begin{table}[!htbp]
  \begin{center}
    \begin{tabular}{|c|l|c|c|c|c|c|c|c|c|c|}
      \hline
      & & \multicolumn{3}{c|}{\textbf{G2}} & \multicolumn{3}{c|}{\textbf{G3}} & \multicolumn{3}{c|}{\textbf{G4}} \\
      \hline
      \raisebox{-6.00ex}[0cm][0cm]{Xe} & fiducial mass (nominal,
      tons) & 1 & & & 10 & & & 100 & & \\
      \cline{2-11} 
      & 
      diameter \par (cm) & 100 & & & 200 & & & 400 & & \\
      \cline{2-11} 
      & 
      passive outer zone (cm) & 10 & & & 15 & & & 20 & & \\
      \cline{2-11} 
      & 
      background particle & & $\gamma/\n{e}$ & n & & $\gamma/\n{e}$
      & n & & $\gamma/\n{e}$ & n \\
      \cline{2-11} 
      & 
      fiducial events \par per year & & $\sim 0.1$ & $\sim 0.1$ & &
      $\sim 0.1$ & $\sim 0.1$ & & $\sim 0.2$ & $\sim 0.2$ \\
      \hline
      \raisebox{-6.00ex}[0cm][0cm]{Ar} & fiducial mass (nominal,
      tons) & 5 & & & 50 & & & 500 & & \\
      \cline{2-11} 
      & diameter \par (cm) & 200 & & & 400 & & & 800 & & \\
      \cline{2-11} & passive outer zone (cm) & 15 & & & 25 & & & 35
      & & \\
      \cline{2-11} 
      & background particle & & $\gamma/\n{e}$ & n & & $\gamma
      /\n{e}$ & n & & $\gamma /\n{e}$ & n \\
      \cline{2-11} 
      & fiducial events \par per year & & \textless 0.01 & $\sim
      0.1$ & & \textless 0.01 & $\sim 0.3$ & & \textless 0.1 & $\sim 0.8$ \\
      \hline
    \end{tabular}
    \caption{Summary of outer target zone passive thickness for G2,
      G3, G4, detectors, required to reduce gamma, electron \& neutron
      events to $<1$ event/year in the fiducial target region.}
    \label{tab:g2g3g4_bkg}
  \end{center}
\end{table}

To investigate the background levels achievable with these measures we
have carried out a detailed program of simulations, the results of
which are shown and discussed in \ref{app:detmaterial}. 

The overall conclusion is that the basic requirement of no more than
$0.2-0.5$ unrejected background events/year can be sustained for each
order of magnitude increase in target mass through G2, G3 and G4, by
using an increasing thickness of passive self-shielding, while
nevertheless keeping to the fiducial volumes indicated in
Tab.\ref{tab:g2g3g4_1} of Sec.\ref{sec:overview}.

A summary of passive shielding thicknesses for G2,G3 and G4 xenon and
argon detectors and corresponding residual gamma, beta, and neutron
background events/y is given in Tab.\ref{tab:g2g3g4_bkg}.

\section{Low energy neutrino astrophysics}
\label{sec:nulowenergy}

\subsection{Solar neutrino p-p spectrum}
\label{sec:ppsolar}

The p-p solar neutrino spectrum produces in a Xe or Ar target a
spectrum of electron recoils extending to 260 keV, and at low energies
gives a flat background of $10^{-5}$ electron recoil
events/keV/kg/d. The S2/S1 discrimination reduces this by a factor
$\sim 100$ to $10^{-7}$ electron recoil events/keV/kg/d (see
Fig.\ref{fig:gamma_rate_xe1t}, \ref{app:detmaterial}) at which it is below the background
level required for a dark matter sensitivity of even
$10^{-47}~\n{cm}^2$. However, above 20 keV (electron equivalent) the signal becomes
observable in the S1 channel as a measurable signal.

\begin{figure}[!htbp]
  \centering
\includegraphics[width=0.9 \columnwidth]{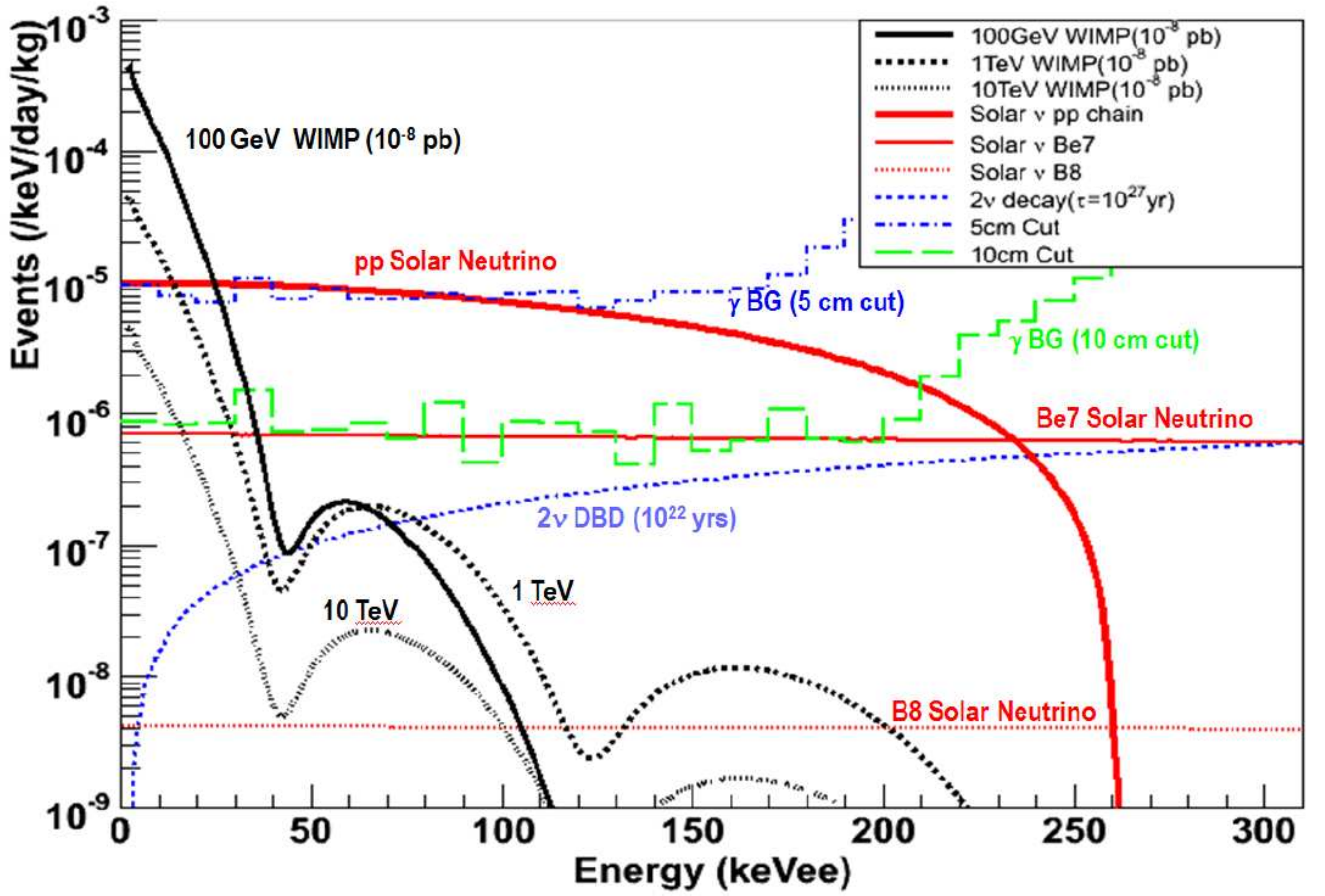}
\caption{Spectra for p-p \& $^{7}$Be solar neutrinos~\cite{Bahcall:1989}, depleted
  $^{136}$Xe two-neutrino double beta decay~\cite{Grotz:1990}, and WIMP spectra for
  $10^{-44}~\n{cm^2}$, together with gamma background self-shielding
  cuts. No S1/S1 discrimination is used for the solar neutrino
  electron spectra.}
\label{fig:ppsolar}
\end{figure}
\begin{figure}[!htbp]
  \centering
  \includegraphics[width=0.9\columnwidth]{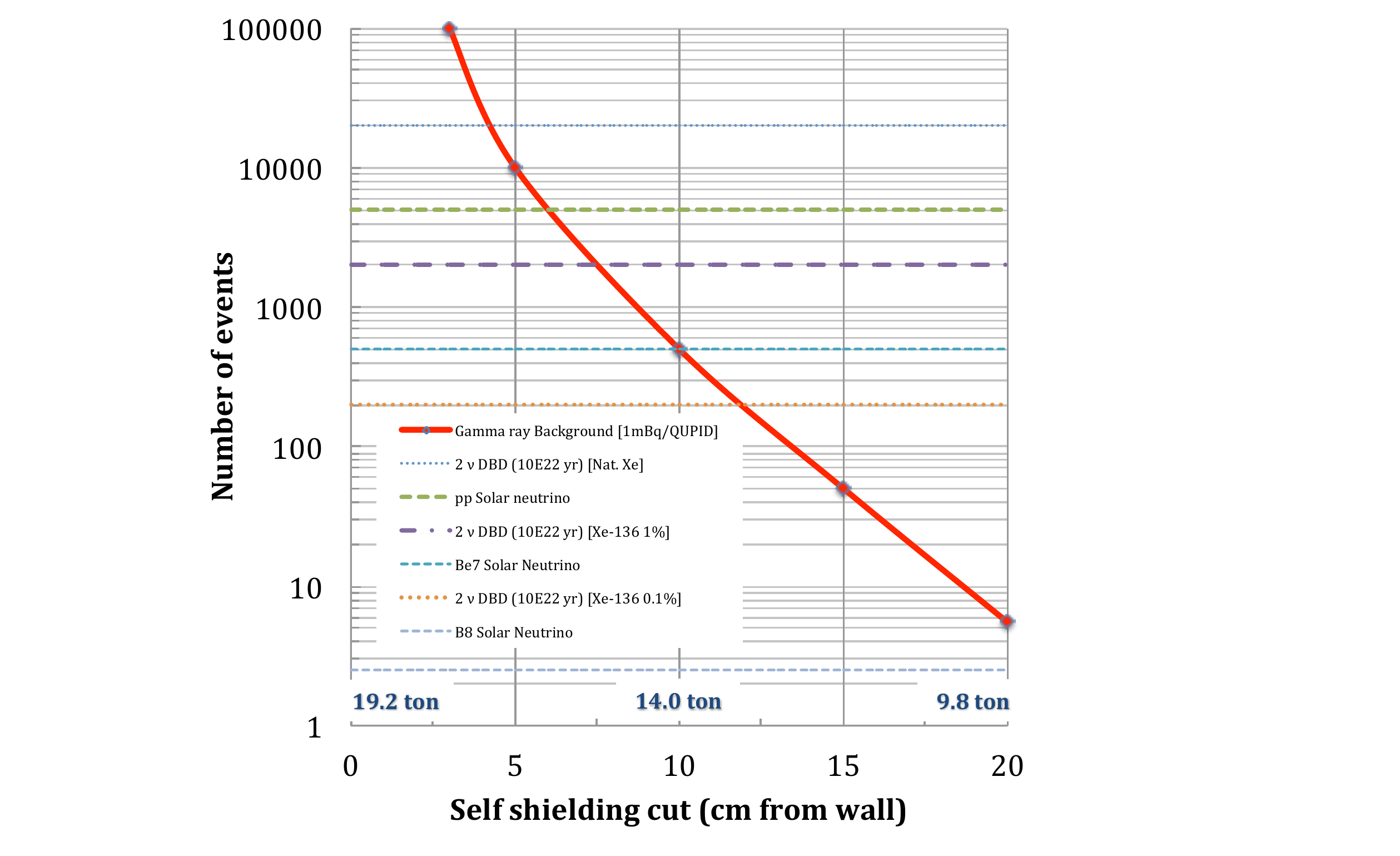}
  \caption{Number of solar neutrino events for 10 ton-y Xe and energy
    window 50-250 keV, showing reduction of electron recoil background
    to below solar neutrino signal by progressive increase of G3 Xe
    self-shielding cut, leaving 10-ton fiducial mass at 15 cm
    cut. 2$\nu \beta \beta $ levels are included for comparison,
    showing effect of $^{\mathrm{136}}$Xe depletion.}
  \label{fig:numsolarnu}
\end{figure}
However, as also shown in Fig.\ref{fig:gamma_rate_xe1t} in natural Xe, the
$2\nu \beta \beta$ decay of the $8.9\%$ $^{136}$Xe (using an
assumed half life $\sim 10^{22}$ years) exceeds the p-p
solar neutrino signal above about 50 keV. Thus for accurate
measurement of the solar neutrino spectrum it would be preferable to
deplete the $^{136}$Xe component by a factor 10 -100~\cite{Suzuki:2007,Suzuki:2000}. Such
depleted Xe may in fact become automatically available as a result of
the simultaneous demand for enrichment of $^{136}$Xe for
neutrinoless double beta decay experiments, as discussed in
Sec.\ref{sec:nulessdoublebeta}, but even with a natural xenon target
it would still be possible to extract a significant portion of the pp
solar neutrino spectrum by a two-population fit to the overlapping $2\nu \beta \beta$ and
solar distributions. Fig.\ref{fig:ppsolar} compares
absolute rates
and energy spectra of events from p-p and $^{7}$Be solar
neutrinos, two neutrino double beta decay (with Xe depleted to
0.1{\%}), and hypothetical dark matter fluxes for $10^{-44}~\n{cm^2}$
WIMP-proton cross section (proportionally lower for
$10^{-45}~\n{cm^2}$ and $10^{-46}~\n{cm^2}$). The S2/S1 cut is not applicable for this plot since
the required solar neutrino events are electron
recoils. Fig.\ref{fig:numsolarnu} shows the progressive reduction of
gamma background in a G3 detector to below the p-p (and $^{7}$Be) solar neutrino
rates by increasing the self-shielding cut, decreasing the 19-ton
total Xe volume to leave a low background 10-ton fiducial mass. A
similar gamma background reduction with self-shielding cut applies
also to the 100-ton G4 Xe detector, which would provide ten times as
many neutrino events per year but would require a correspondingly
larger quantity of depleted Xe. Using a $^{136}$Xe-depleted
target, the G3 Xe detector would provide sufficient events to measure
the p-p neutrino spectrum with a few \% precision. 

In principle, the p-p solar neutrino signal would be observable also in
the Ar detector (which has the same p-p neutrino event rate per ton)
but the $^{39}$Ar beta background at a level (assuming the
use of underground argon) of $\sim100$ events/keV/kg/d, and extending to
570 keV, would overwhelm the whole of the p-p neutrino spectrum
even with several orders of magnitude further depletion of the $^{39}$Ar. 

\subsection{Supernova neutrinos}
\label{sec:supernova}

Neutrino bursts from Galactic supernovae will be seen an order of
magnitude more frequently than those near enough to the Earth to be
visible optically, at a rate believed to be $\sim 3-5$ per
century~\cite{Fang:1990}. All three types of neutrino will be emitted, and estimates of
rates, fluxes, and spectra are summarised in \ref{app:nugalsupern}. Both
vacuum and MSW mixing will occur after production, making detailed
observation of these rare events using different types of
detector a potentially important contribution to neutrino physics.

The G2, G3 and G4 detector systems can most efficiently detect these
neutrinos through coherent neutral current elastic scattering from the
Xe and Ar nuclei~\cite{Horowitz:2003,Anderson:2011}, which will respond equally to
all flavours. The relative fluxes of the different neutrino flavors
will be affected by MSW mixing in the supernova environment, and by
vacuum mixing between the supernova and Earth, but the total flux of
all three types will remain constant. In addition, although the
non-zero neutrino masses will propagate as mass eigenstates (mixtures
of the three flavor eigenstates), the arrival times will differ by
less than 1 ms with neutrino masses all $<1$ eV, compared with
the $>1$ s overall time duration of the neutrino burst. The
typical distance of a Galactic supernova from the Earth is taken to be
10 kpc for the calculations in this section.

\begin{figure}[!htbp]
  \centering
  \includegraphics[width=0.75 \columnwidth]{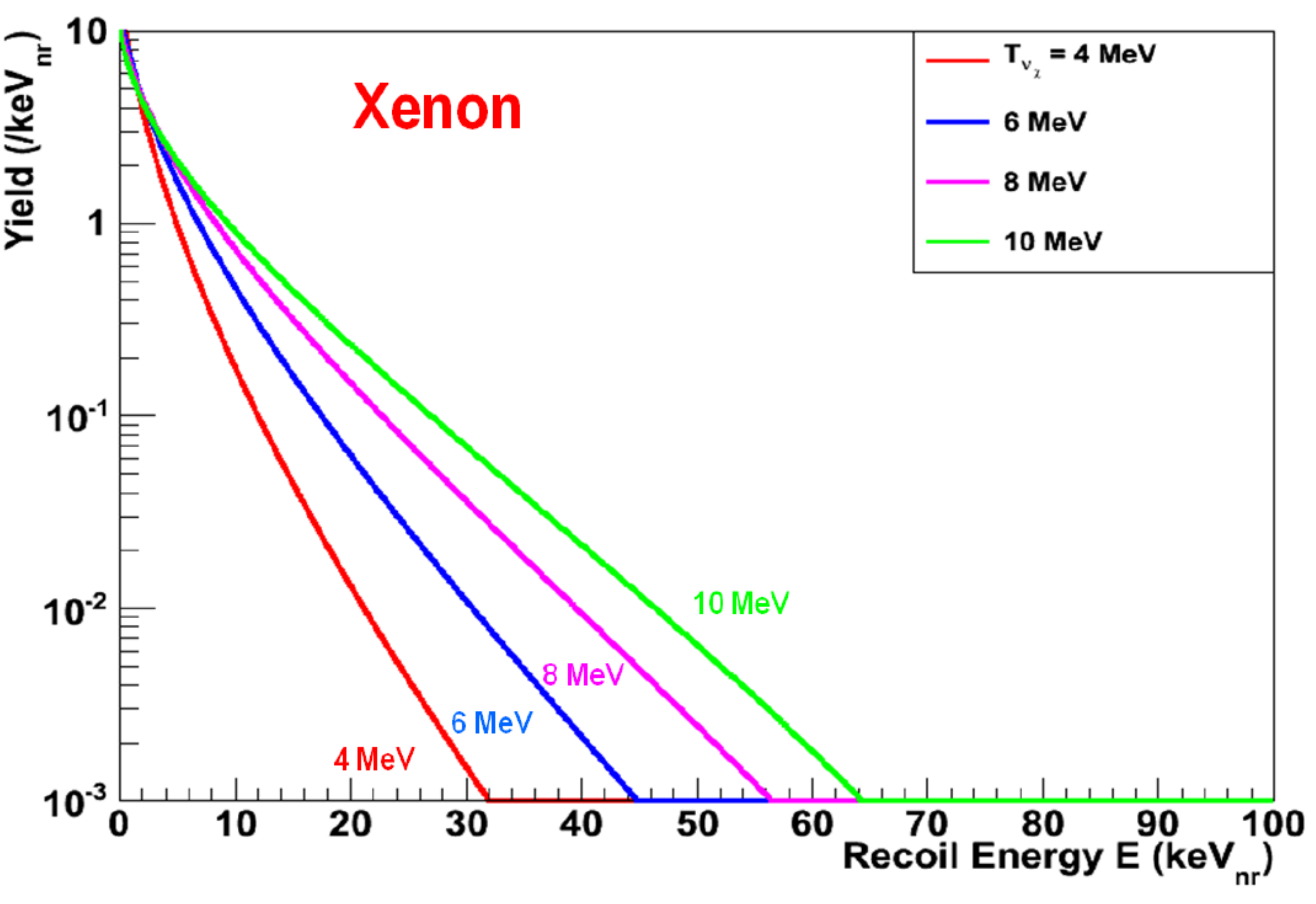}
  \includegraphics[width=0.75 \columnwidth]{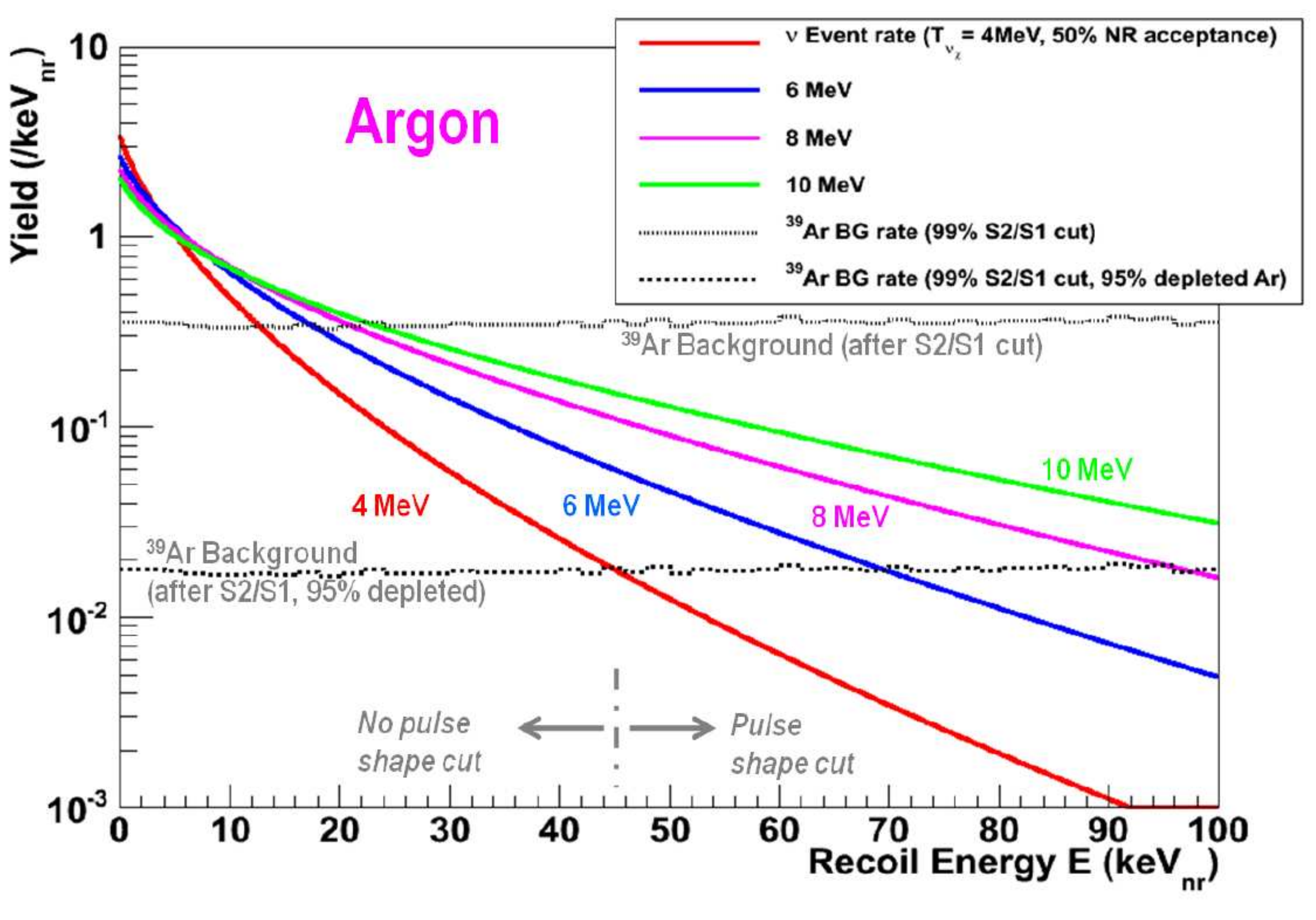}
  \caption{Energy spectrum of coherent nuclear scattering events from
    supernova neutrino burst at distance 10 kpc, for mean neutrino
    temperatures 4, 6, 8, 10 MeV.\protect\\
    (a) \textit{(upper)} 1-ton Xe detector.\protect\\
    (b) \textit{(lower)} 5-ton Ar detector, including 95\% depletion
    of $^{39}$Ar and pulse shape cut. \protect\\
    Note that for a 20 second neutrino burst, the levels of continuous
    background for dark matter or solar neutrinos (see Fig.~\ref{fig:gamma_rate_xe1t}) give
    20 second event yields below the vertical scale of these plots.}
  \label{fig:nusuper_xe_ar}
\end{figure}

\begin{figure}[!htbp]
  \centering
  \includegraphics[width=0.6 \columnwidth]{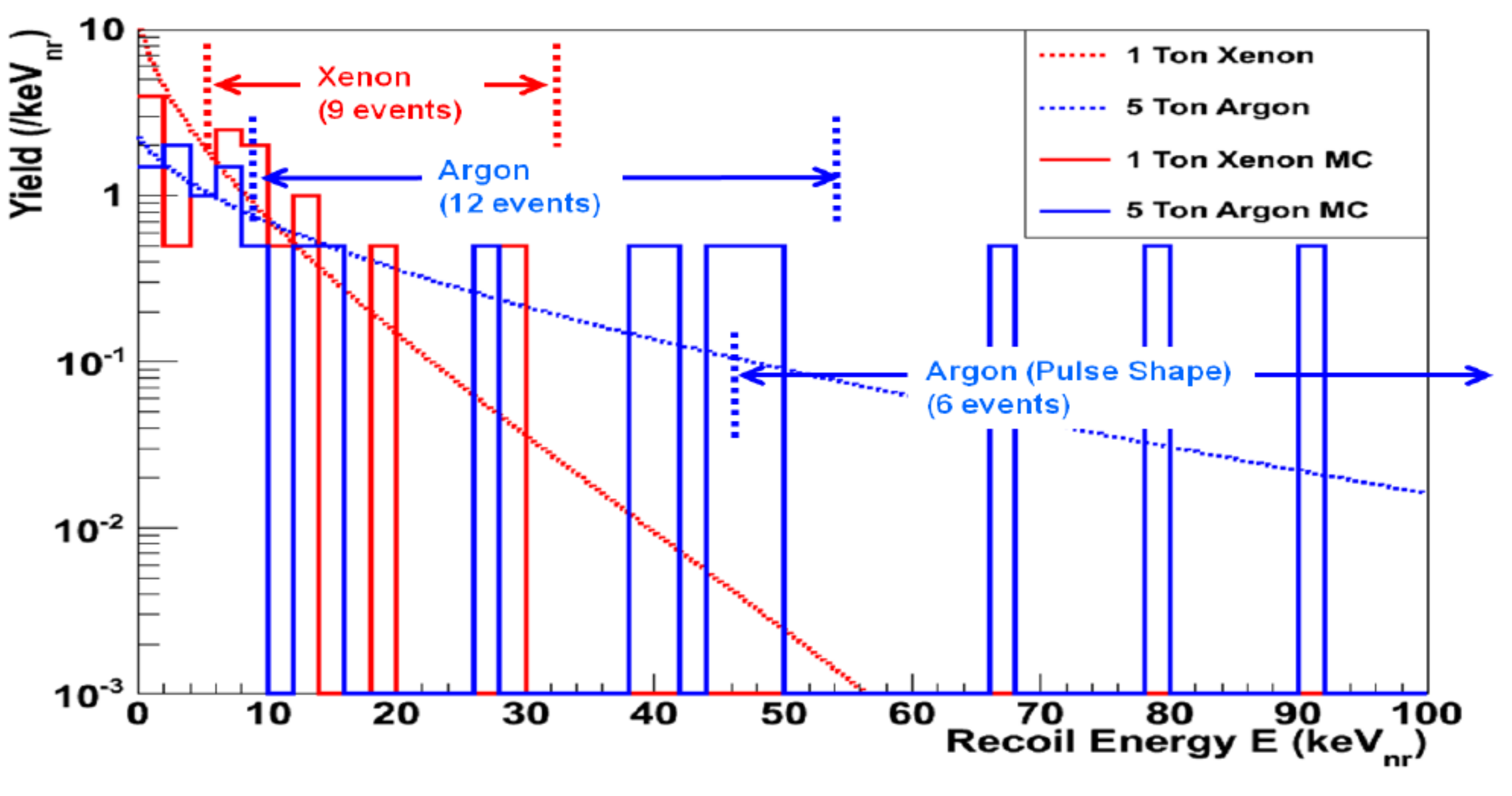}
  \includegraphics[width=0.6 \columnwidth]{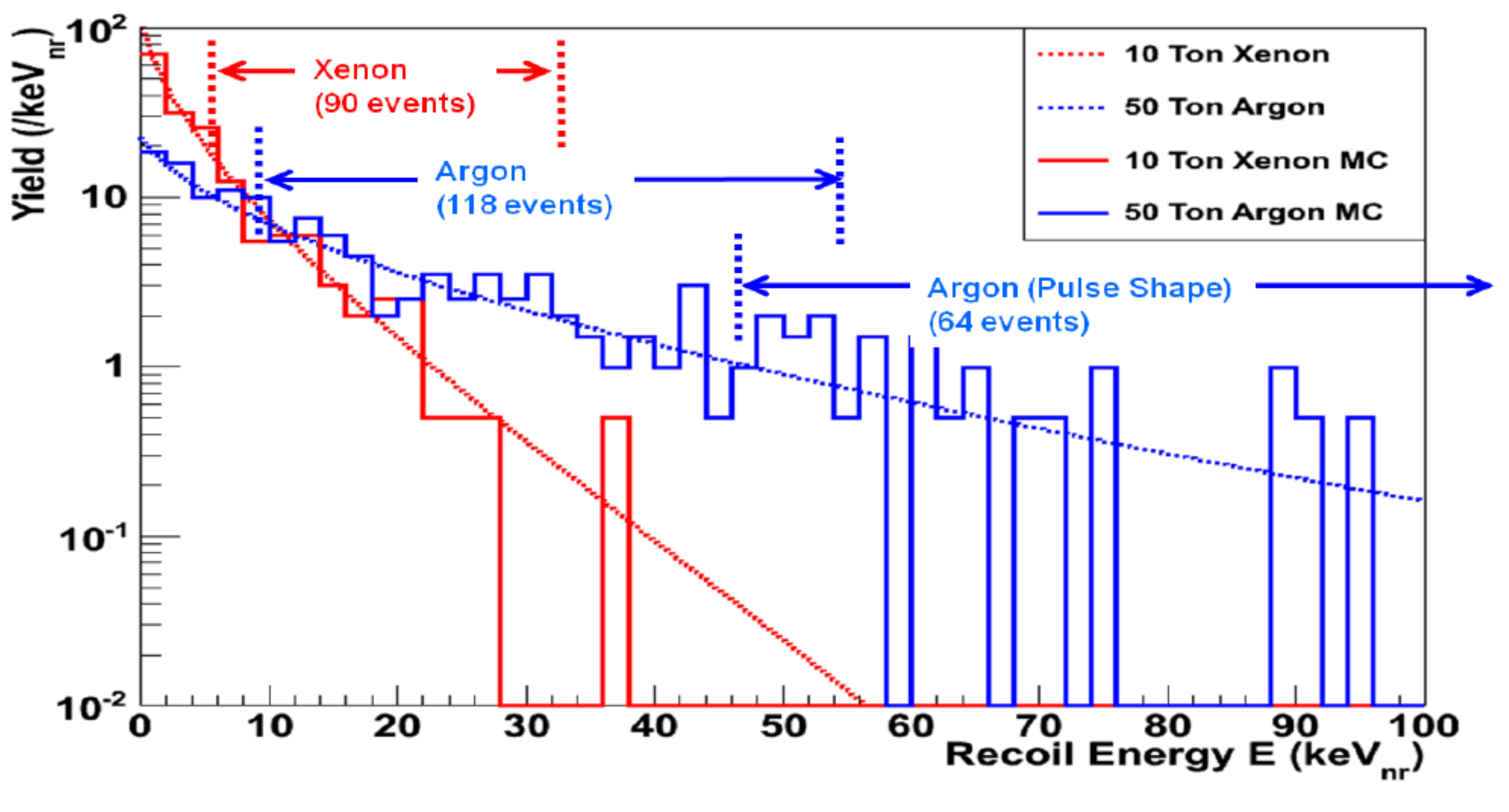}
  \includegraphics[width=0.6 \columnwidth]{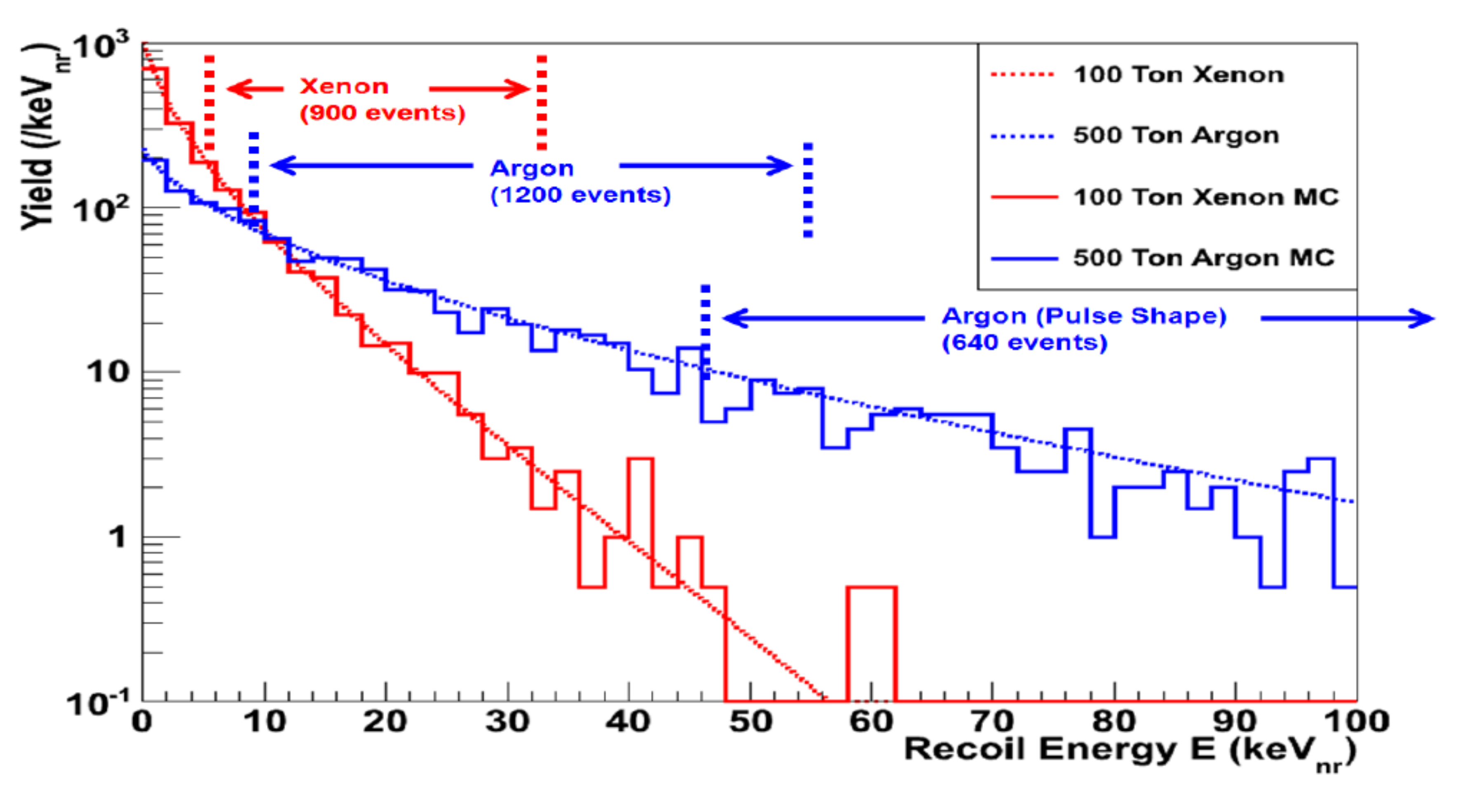}
  \caption{Simulated data for nuclear recoil spectra from supernova at 10 kpc\protect\\
    \textit{(upper plot)} G2: 1-ton Xe and 5-tons Ar\protect\\
    \textit{(middle plot)} G3: 10-ton Xe and 50-ton Ar\protect\\
    \textit{(lower plot)} G4: 100-ton Xe and 500-ton Ar.\protect\\
    Note that for a 20 second neutrino burst, the levels of continuous
    background for dark matter or solar neutrinos (see Fig.~\ref{fig:gamma_rate_xe1t}) give
    20 second event yields below the vertical scale of these plots.}
  \label{fig:spect_nusuper_g2g3g4}
\end{figure}

\begin{figure}[!htbp]
  \centering
  \includegraphics[width=1.1 \columnwidth]{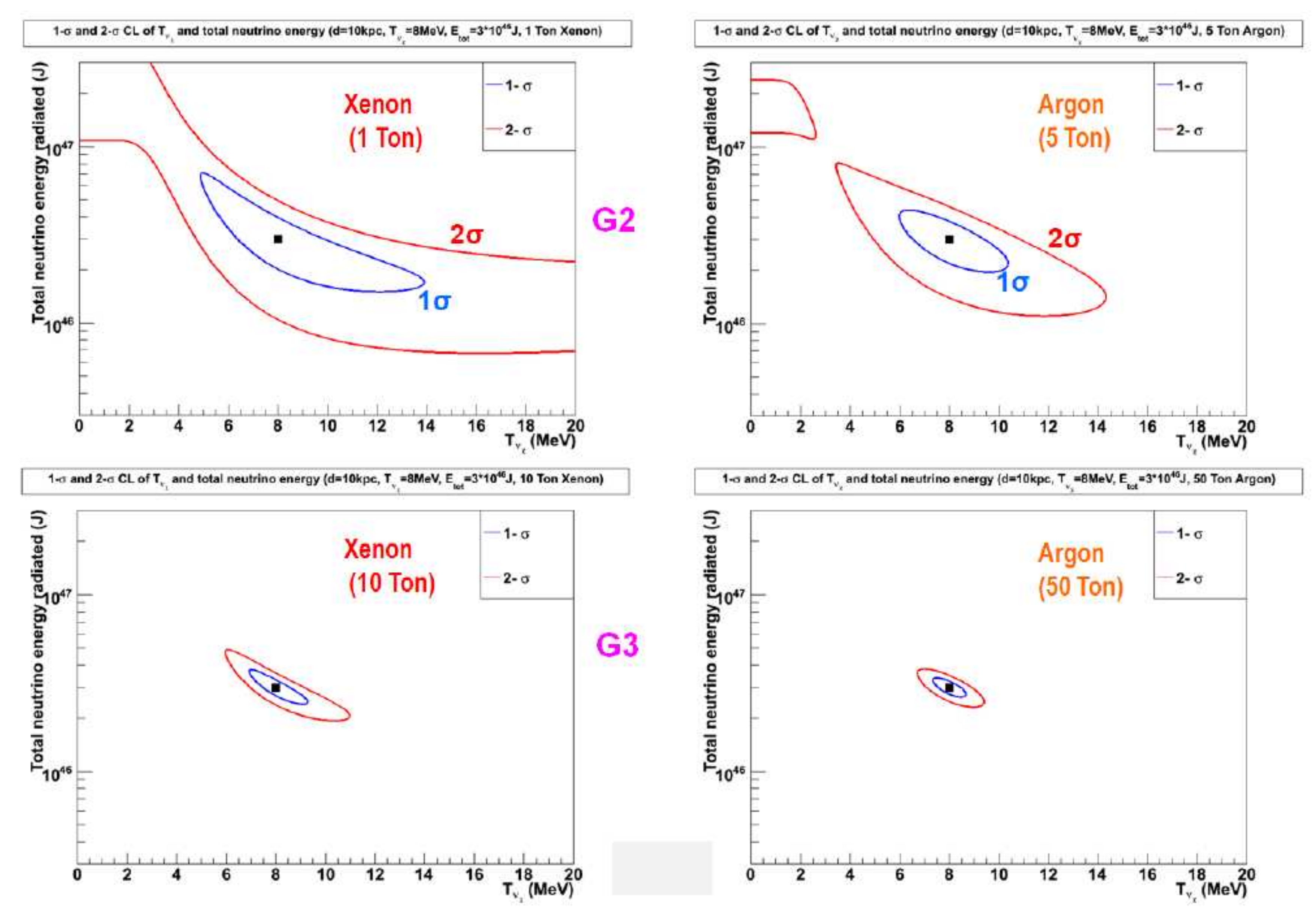}
\caption{Examples of combined precision of Supernova neutrino total
  energy and mean temperature, derivable from the spectrum of coherent
  nuclear recoil events shown in Fig\ref{fig:spect_nusuper_g2g3g4}.\protect\\
  \textit{(upper plots)} Predicted precision of the G2 system with fiducial 1-ton Xe, 5-ton Ar.\protect\\
  \textit{(lower plots)} Predicted precision of the G3 system, with
  fiducial 10-ton Xe, 50-ton Ar.}
\label{fig:nusuper_enetemp}
\end{figure}

Both Xe and Ar detectors can independently determine both the total
energy and the temperature of the neutrino burst. Fig.\ref{fig:nusuper_xe_ar}a
and Fig.\ref{fig:nusuper_xe_ar}b show the yield/ton versus energy of Xe and Ar
nuclear recoil events for a supernova at 10 kpc and several values of
mean neutrino temperature $T_{\nu } ( = <E>_{\nu}/3)$ averaged over all neutrino
types. Corresponding Monte Carlo simulations of G2, G3 and G4 data
sets are shown in Fig.\ref{fig:spect_nusuper_g2g3g4}, in turn leading to estimates of
total energy and neutrino temperature, and with the typical precision
shown in Fig.\ref{fig:nusuper_enetemp} for G2 and G3 detectors. Figs.Fig.\ref{fig:nusuper_xe_ar}
and \ref{fig:spect_nusuper_g2g3g4} are plotted against recoil energy,
so need to be considered in relation to the achievable future energy
threshold in Xe and Ar detectors, estimated in this paper to be $\sim
8$ keVnr for Xe and $\sim 20$ keVnr for Ar, but in principle
improvable in each case with sufficient gains in light
collection.

\section{Neutrinoless double beta decay}
\label{sec:nulessdoublebeta}

With the establishment, from neutrino mixing measurements, of a
non-zero neutrino mass scale in the region $0.01-0.1$ eV, the issue of
the Dirac or Majorana nature of neutrinos has become of increased
urgency for understanding the lepton family and the quark-lepton
relationship. For Majorana neutrinos, the equivalence of neutrino and
antineutrino would ensure that those nuclei that can undergo double
beta decay with emission of two neutrinos must also undergo
neutrinoless double beta decay. For the present study, we consider the
Xe detectors in the G2, G3 and G4, stages, focusing on the isotope
$^{136}$Xe which is present at 8.9\% in natural Xe, and
which could be enriched by an order of magnitude.

\begin{figure}[!htbp]
  \centering
  \includegraphics[width=0.9 \columnwidth]{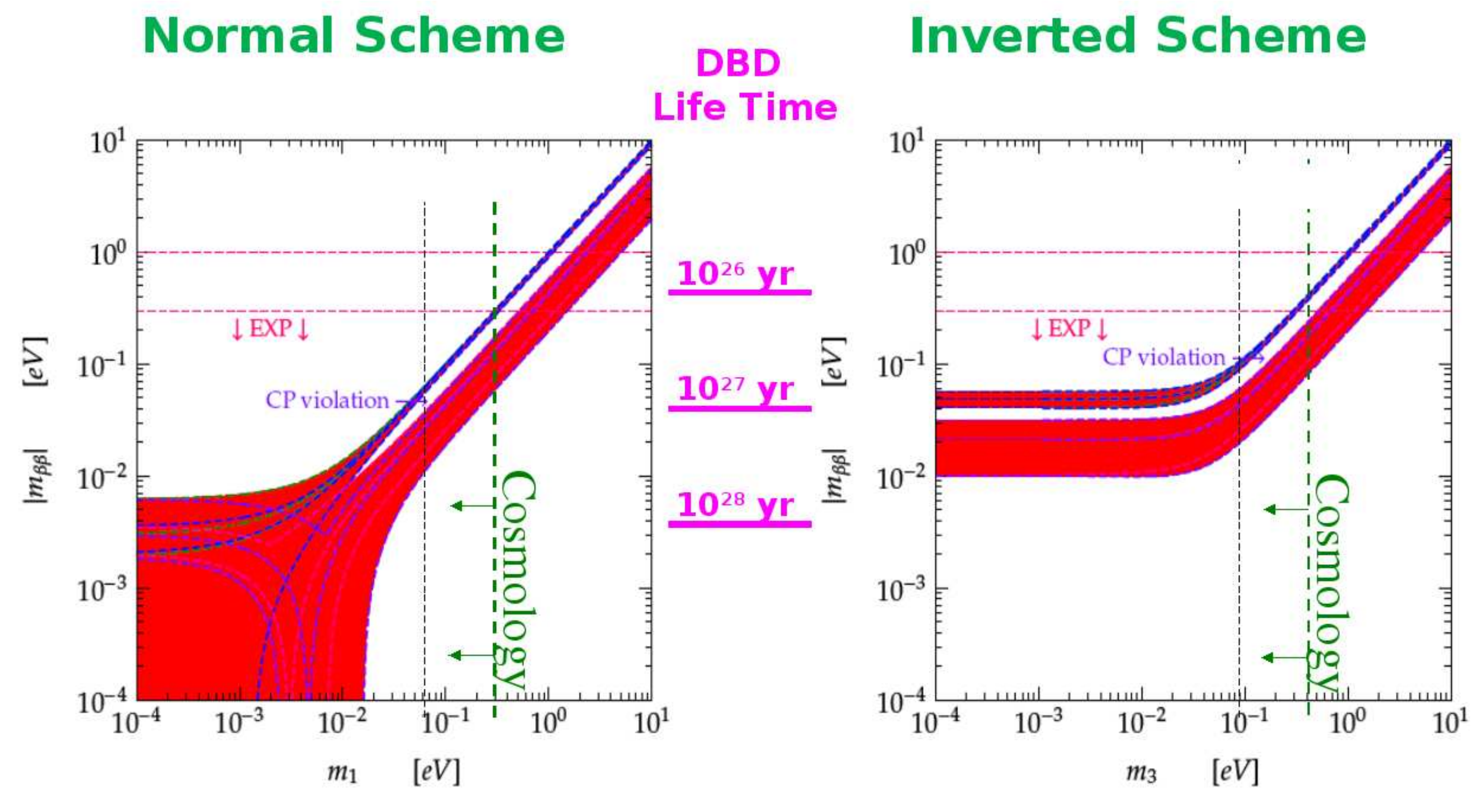}
  \caption{Plots of Majorana neutrino mass versus unknown lowest mass
    eigenstate~\cite{Horowitz:2003}, together with corresponding neutrinoless
    double beta decay half-lives.}
  \label{fig:maj_mass}
\end{figure}

The rate for this process is a function of nuclear matrix elements
multiplied by (the square of) a Majorana electron neutrino mass parameter
m$_{\beta \beta }$ which is expressible as a linear
combination of the three neutrino mass eigenstates, but uncertain in
the range $0.01 - 0.1$ eV owing to the unknown absolute value of the
lowest of the three mass eigenstates~\cite{Avignone:2005}. The resulting Majorana
masses are shown in Fig.\ref{fig:maj_mass} as a function of the unknown
$m_1$ for normal hierarchy, or the unknown $m_3$
for inverted mass hierarchy~\cite{Giunti:2002}. The same figure gives
corresponding half-lives for $^{136}$Xe, showing that
sensitivity to $10^{27} -10^{28}$ y is likely to be
necessary to provide a detection of $0\nu \beta \beta $ decay for the
first time, previous experiments having been sensitive to upper limits
of $10^{24} -10^{25}$ y.

\begin{figure}[!htbp]
  \includegraphics[width=0.9 \columnwidth]{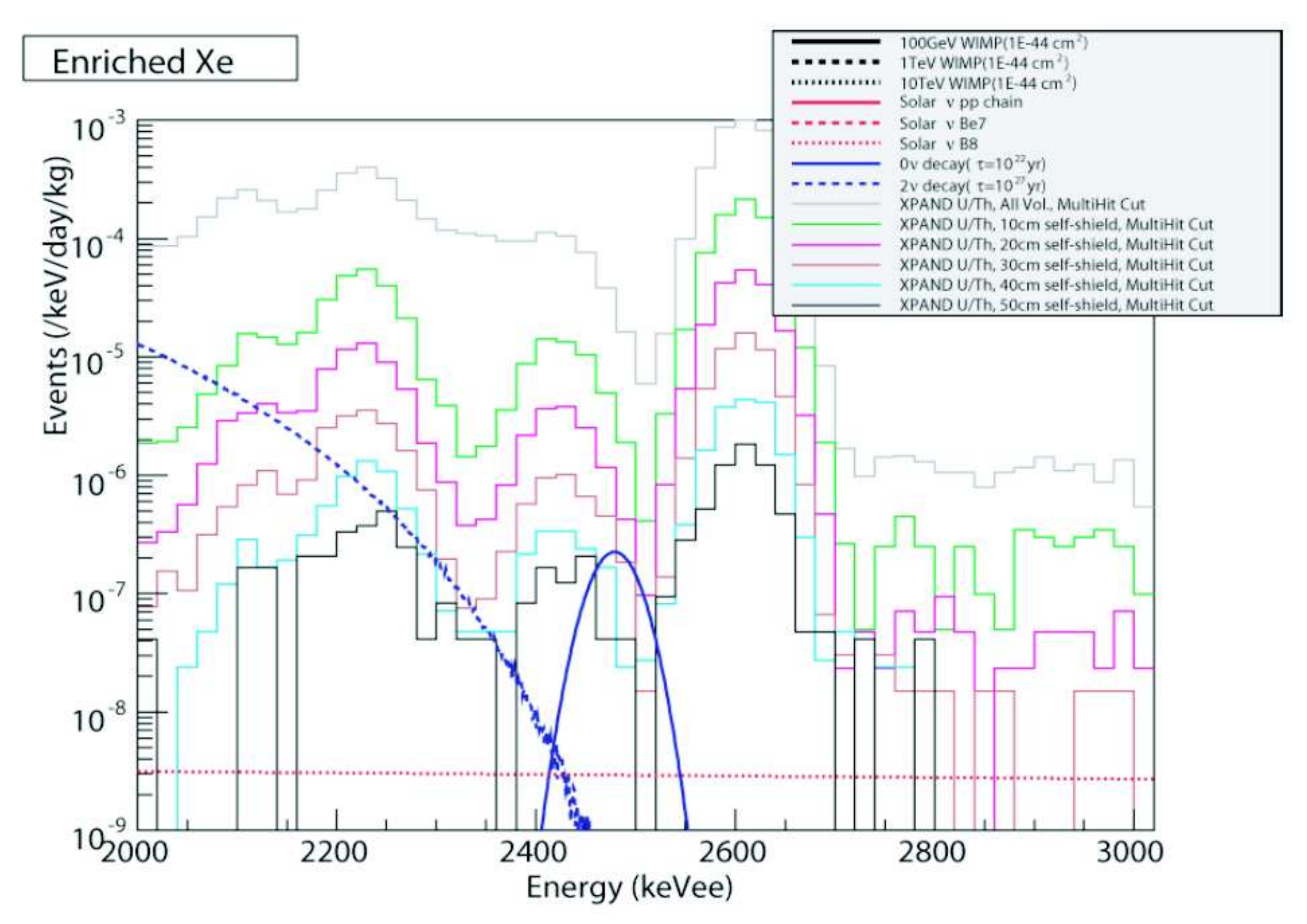}
\caption{Plot showing $0\nu \beta \beta $ signal (half-life $10^{27}$
  y) and $2\nu \beta \beta $ tail in a 2 m diam. detector with 90\%
  $^{136}$Xe, together with simulated gamma background after
  progressively increasing self-shielding cuts. The largest background
  is from the QUPID array, for which 1 mBq/QUPID is
  assumed.}
\label{fig:nuless_signal}
\end{figure}

The signature for $0\nu \beta \beta $ decay of $^{136}$Xe will be an
electron recoil signal at the specific energy 2479 keV, which is at
the end of the $2\nu \beta \beta $ decay spectrum, mentioned
previously as a background for pp solar neutrino detection. In this
case the tail of the $2\nu \beta \beta $ decay spectrum provides a
background to the $0\nu \beta \beta $ decay line, requiring the latter
to be observed with good energy resolution. Scaling from the
resolution achieved with $10-100$ kg Xe detectors, we estimate a FWHM
energy resolution of 50 keV at $\sim 2.5$ MeV, giving the separation
between the $0\nu \beta \beta $ signal and $2\nu \beta \beta $
spectrum tail shown on Fig.\ref{fig:nuless_signal}. The superimposed
gamma background versus self shielding, shown for the case of the 2 m
diameter G3 Xe detector after rejection of multiple scattering events,
arises mainly from the QUPID array, for which 1 mBq/QUPID is assumed.
Consideration is also needed for possible contamination of the Xe by
the $^{232}$Th chain, which includes a 100\% intensity $^{208}$Tl 2.61
MeV line, and by the $^{238}$U chain, which includes a 2\% intensity
$^{214}$Bi 2.44 MeV line~\cite{Martin-Albo:2010}. The peak/Compton
ratios for these are $\sim10^{-5}$ so the majority of background comes
from overlap with the Compton shoulder (at 2.38 MeV for $^{208}$Tl, and
2.32 MeV for $^{214}$Bi). Reducing this background to an acceptable
level thus depends firstly on ensuring U and Th contamination levels
are below 1 ppt, and secondly on achieving sufficiently good energy
resolution to fully separate the 2.48 MeV signal region from the above
Compton shoulders.

\begin{figure}[!htbp]
  \centering
  \includegraphics[width=0.9 \columnwidth]{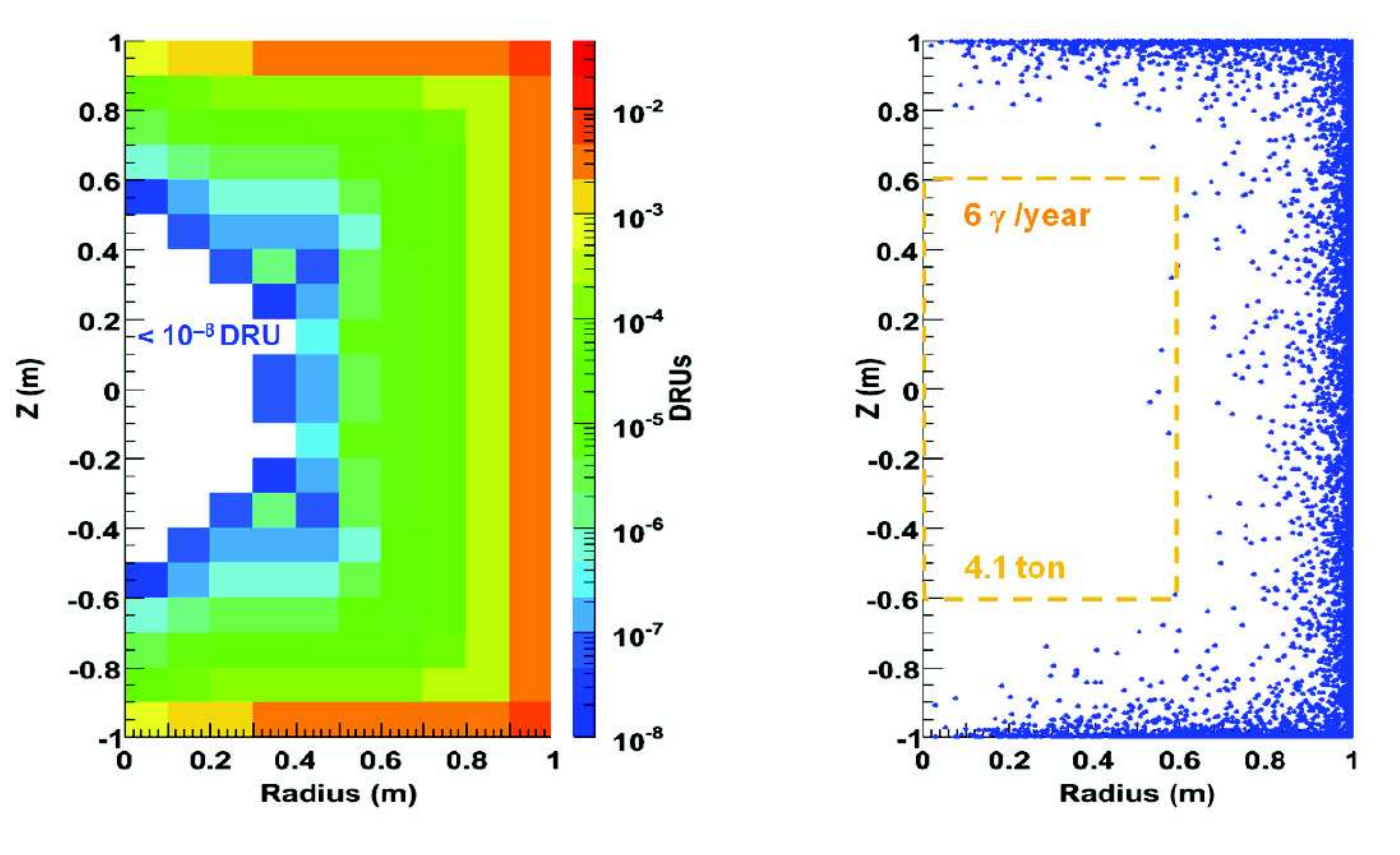}
  \caption{Spatial distribution of 1-y gamma background from QUPID array
    surrounding 2 m diameter Xe detector, after rejection of multiple
    scattering events and showing need for a $40-45$ cm radial cut for
    observation of a small (i.e. \textless 10 event) electron signal
    from $0\nu \beta \beta $.}
  \label{fig:qupid_gamma_spatial}
\end{figure}

\begin{figure}[!htbp]
  \centering
  \includegraphics[width=0.9 \columnwidth]{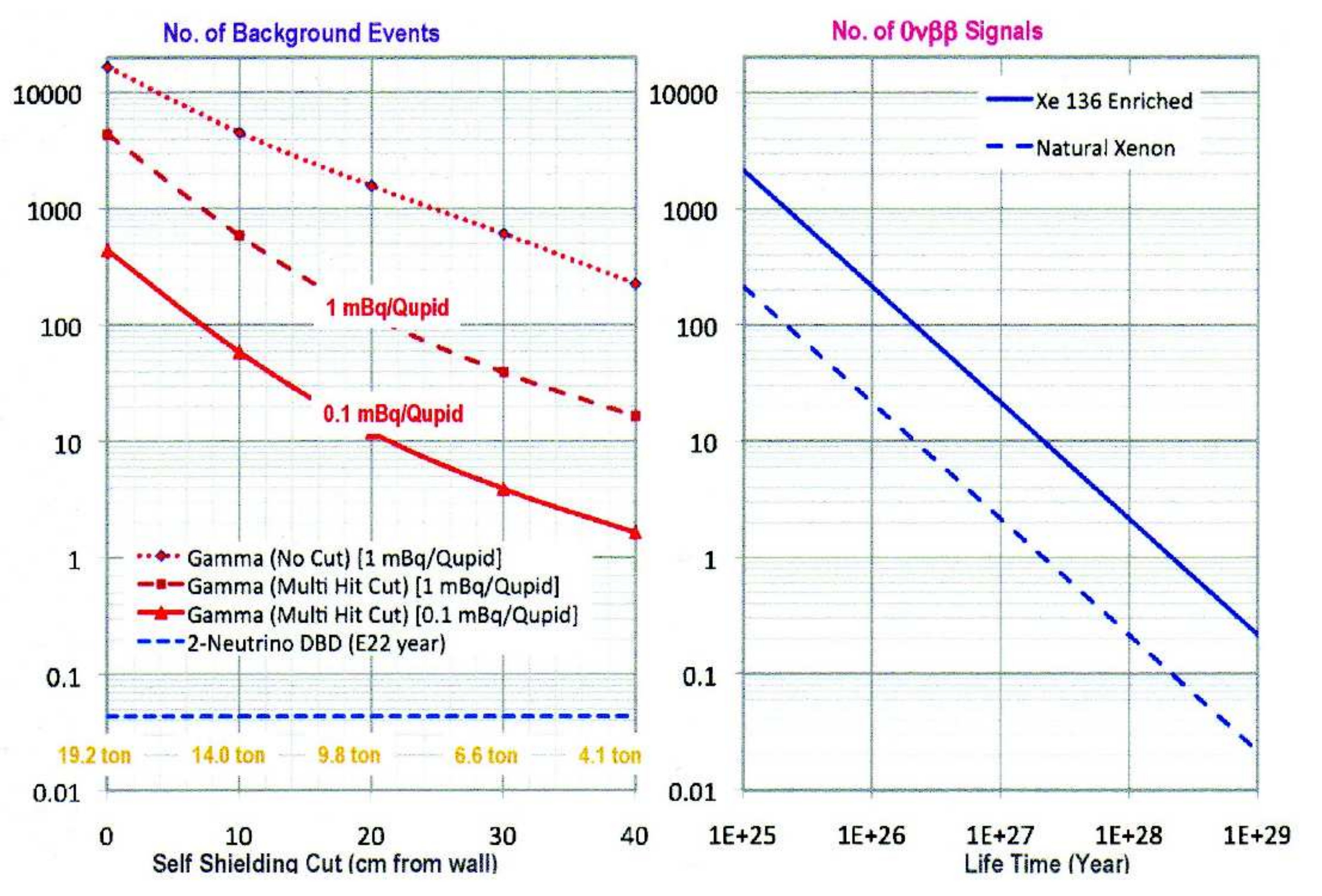}
  \caption{(a) \textit{(left)} Gamma background events remaining in Xe
    fiducial region versus self-shielding cut, for 10 ton-y operation
    and photodetector activity of 1 or 0.1 mBq/QUPID\protect\\
    (b) \textit{(right)} Number of $0\nu \beta \beta $ events for 10
  ton-y operation in enriched and unenriched $^{136}$Xe
  versus $0\nu \beta \beta$ half-life.}
\label{fig:gamma_selhshield}
\end{figure}

Fig.\ref{fig:qupid_gamma_spatial} shows the spatial distribution of gamma background
events in the $0\nu \beta \beta $ energy range $2450-2500~\n{keV}$
(including energy resolution) requiring a larger self-shielding cut
($\sim 40~\n{cm}$) than in the case of the dark matter background discussed
in Sec.\ref{sec:backgrounds}, because the S2/S1 cut is not relevant to
the electron recoil signal. This would reduce the G3 fiducial target mass
to $\sim$ 8 tons, but a foreseeable reduction of QUPID activity from
1 mBq to 0.1 mBq/QUPID would reduce the required cut to 15cm and restore
the fiducial mass to $\sim$ 10 tons. Fig.\ref{fig:gamma_selhshield}a
summarizes background number per year versus self-shielding cut, and
Fig.\ref{fig:gamma_selhshield}b plots $0\nu \beta \beta $ signal in events per 10
ton-y operation versus $0\nu \beta \beta $ lifetime, for both natural
Xe, and Xe enriched by a factor 10 to $\sim 90\%$
$^{136}$Xe. The calculation for a unit of 10 ton-y enables
the plots to be scaled for G2, G3 and G4 detectors with corresponding
fiducial masses and running times.

Tab.\ref{tab:nuless_summary} summarizes the estimated $0\nu \beta \beta $
lifetime sensitivity for most of the currently proposed world
experiments aimed at reaching $>10^{26}$ years
(reviewed in~\cite{Avignone:2005,Elliot:2002,Avignone:2008,Barabash:2007}).
For comparison, the approximate sensitivity levels that would be
achievable by the G2, G3 and G4 detectors in this paper are also
shown, estimated from Fig.\ref{fig:gamma_selhshield}(a) and
\ref{fig:gamma_selhshield}(b) assuming sufficient
events to give a non-zero 90\% Poisson (Feldman-Cousins) lower signal
limit, together with a conservative 50\%
energy resolution and detection
efficiency. The table also shows the
ultimate sensitivity obtainable using
enriched $^{136}$Xe in the 10 ton G3
detector. In the latter case, it would be preferable to restrict the scarcer
$^{136}$Xe specifically to the fiducial region, using the
concentric XAX configuration proposed previously~\cite{Arisaka:2009}. In all
cases the detector design and component activity need to be chosen to
keep the background in the $0\nu \beta \beta $ peak within the $1-10$
events/y level, in turn arising from the need to constrain this below a
potential signal $\sim 5-10$ events/y. In each project this is
believed by the proposing groups to be an attainable objective, but
subject to further R\&D. 

As concluded in~\cite{Arisaka:2009}, if the estimated energy
resolution and backgrounds for a 10-ton Xe detector can be matched in
practice, it will be possible to achieve a half-life sensitivity
\textgreater 10$^{\mathrm{28}}$ y without the needing to tag the Ba
nucleus as proposed by the EXO project~\cite{Umehara:2010}. It is at present
premature to include in Tab.\ref{tab:nuless_summary} specific figures for an
enriched version of G4, but it is evident that if it proves realistic
to produce $60-80$ tons enriched $^{136}$Xe, then this could
achieve a sensitivity $>10^{28}$ y in only 1 year
of operation. 
\begin{table}[!htbp]
  % \centering
  % \begin{tabular*}{1.1\textwidth}{ @{\extracolsep{\fill}} |c|c|c|c|c|c|c| }
  \begin{tabular}{|c|c|c|c|c|c|c|}
      \hline
      Future Project & ref & Isotope & Fid. isotopic & Exp. bkg &
      $0\nu \beta \beta~\tau_{1/2}$ (y) & $0\nu \beta \beta~\tau_{1/2}$ (y)\\
      & & & mass (tons) & in peak (evts/y) & 1y run & 5y run \\
      \hline
      MAJORANA & \cite{Avignone:2008} & $^{76}$Ge en. & 0.5 &
      $2 - 5$ & $6\times10^{26}$ & $2\times10^{27}$ \\
      \hline
      COBRA & \cite{Avignone:2008} & $^{113}$Cd en. & 0.4 &
      $10-20$ & $1 \times 10^{26}$ & $3 \times 10^{26}$ \\
      \hline
      CUORE & \cite{Avignone:2008} & $^{130}$Te nat. & 0.2 & $4 -
      20$ & $2 \times 10^{26}$ & $6 \times 10^{26}$ \\
      \hline
      NEXT & \cite{Gomez:2009,Yahlali:2010} & $^{136}$Xe en. &
      0.1 & $2 - 5$ & $6 \times 10^{25}$ & $2 \times 10^{26}$ \\
      \hline
      GERDA & \cite{Avignone:2008} & $^{76}$Ge en. & 0.05 & $2
      - 10$ & $4 \times 10^{25}$ & $1 \times 10^{26}$ \\
      \hline
      SuperNEMO & \cite{Avignone:2008,Simard:2010} & $^{82}$Se
      en. & 0.1 & $2 - 10$ & $1 \times 10^{26}$ & $3 \times 10^{26}$ \\
      \hline
      MOON III & \cite{Ejiri:2009} & $^{100}$Mo, $^{82}$Se, en. &
      0.5 & $10-20$ & $1 \times 10^{26}$ & $3 \times 10^{26}$ \\
      \hline
      CANDLES III & \cite{Avignone:2008,Umehara:2010} & $^{48}$Ca
      nat. & 0.1 & $5 - 10$ & $1 \times 10^{26}$ & $3 \times 10^{26}$ \\
      \hline
      & & & & & & \\
      \hline
      EXO200 & \cite{Avignone:2008,Gornea:2010} & $^{136}$Xe
      en. & 0.2 & $ 4 - 8 $ & $3 \times 10^{25}$ & $1 \times 10^{26}$ \\
      \hline
      EXO 1T & \cite{Gornea:2010} & $^{136}$Xe en. & 1 & $ 4 -
      8$ & $4 \times 10^{26}$ & $1 \times 10^{27}$ \\
      \hline
      EXO 10T + Ba tag & \cite{Gornea:2010} & $^{136}$Xe en. &
      8 & $1 - 2$ & $6 \times 10^{27}$ & $2 \times 10^{28}$ \\
      \hline
      & & & & & & \\
      \hline
      Xe G2 1T & this work & $^{136}$Xe nat. & 0.04 & $3 - 6$ &
      $2 \times 10^{25}$ & $6 \times 10^{25}$ \\
      \hline
      Xe G2 1T & this work & $^{136}$Xe en. & 0.32 & $3 - 6$
      & $1 \times 10^{26}$ & $4 \times 10^{26}$ \\
      \hline
      Xe G3 10T & this work & $^{136}$Xe nat. & 0.4 & $4 - 8$
      & $2 \times 10^{26}$ & $6 \times 10^{26}$ \\
      \hline
      XAX 10T & \cite{Arisaka:2009} & $^{136}$Xe en. & 8 & $4 -
      8$ & $4 \times 10^{27}$ & $1 \times 10^{28}$ \\
      \hline
      Xe G4 100T & this work & $^{136}$Xe nat. & 9 & $2 - 4$ &
      $7 \times 10^{27}$ & $2 \times 10^{28}$ \\
      \hline
      Xe G4 100T & this work & $^{136}$Xe en. & see text & & & see
      text \\
      \hline
    \end{tabular}
    \caption{Summary of projected $0\nu \beta \beta $ performance of
      G2,G3,G4, Xe detectors, and other planned world projects to reach
      half-lives $10^{26} - 10^{28}~\n{y}$ (all background estimates are
      approximate and subject to further R\&D). In the table ``en.'' stands
      for enriched, while ``nat.'' for natural.}
    \label{tab:nuless_summary}
\end{table}

\section{Xe/Ar mixtures}
\label{sec:xear}

In addition to the above possibilities for separate liquid Xe and Ar
detectors, it is of interest to consider possibilities for combining
the two elements in one detector, for example using a low
concentration of Xe as a dopant for Ar, or single detectors containing
higher percentage mixture of the two.

A number of papers have reported measurements of scintillation light output and wavelength,
electron mobility, and pulse shape, in liquid Ar doped with Xe
at the $10-100$ ppm level~\cite{Suzuki:1993,Borghesani:1993,Kim:1994,Conti:1996,Piefer:2008}. It was also found that
measurements could be extended to much larger Xe concentrations, up
to $5-20\%$ (Mol fraction). By visual tests in transparent vessels, it was found that there was
settling out of frozen Xe at 20\% mol fraction, but that the mixture remained apparently stable up
to 5\%, and possibly 10\%~\cite{Suzuki:1993}. Mol fractions $5-10\%$ correspond to mass fractions $14-26\%$. This
unexpectedly high solubility was suggested to be the result of the formation of clusters and
compounds. More recently Peifer et al.~\cite{Suzuki:1993} found possible evidence of inhomogeneous mixing, or
stratification, which could impair the performance of detectors utilising such mixtures, and thus
needs further investigation.

Some key features of the published results are
\begin{pbenumerate}
\item an admixture of 100 ppm Xe was found to produce complete wavelength shifting of
  the scintillation light from 128 nm to 178 nm~\cite{Conti:1996}.
\item Adding Xe to Ar resulted in a progressive increase of scintillation light output, by up to
  a factor 2, for Xe concentrations up to 1\%, in zero electric field, this increase being
  suppressed by electric fields $\sim 1-8$ kV/cm~\cite{Suzuki:1993}.
\item The addition of only 100 ppm Xe produces considerable changes in scintillation pulse
  shape, including a small improvement in neutron/gamma pulse shape discrimination
  for only 300 ppm Xe~\cite{Piefer:2008} suggesting further investigation of pulse shape discrimination
  at larger Xe fractions.
\end{pbenumerate}

These observations suggest several possible applications to the Xe and Ar detectors:
\begin{pbitemize}
\item[a)] ability to use Xe-wavelength photodetectors for both Xe and Ar detectors;
\item[b)] prospects for improved energy threshold and resolution in Ar detectors, from the
  increased scintillation light per keV.
\item[c)] possibility of detectors combining Xe and Ar targets in the same vessel, if separate
  identification of Xe and Ar recoils proves feasible, for example by pulse shape. In
  particular, a 7\% molar ($\sim20\%$ by weight) Xe concentration would provide the 5:1 Ar/Xe
  ratio required in the G2 and G3 two-vessel detector systems, but in
  the single Ar vessel;
\item[d)] an alternative possibility of determining the Xe/Ar signal ratio (and verifying an $A^2$ signal
  dependence) by comparison of signals before and after the addition
  of Xe to an Ar detector;
\item[e)] A G3-level $0\nu\beta\beta$ detector, using the (4 m diameter) G3 Ar detector but with a target
  consisting of a 20\% Xe/Ar mass ratio (13 tons $^{136}$Xe, 65 tons Ar). This could provide a
  $0\nu\beta\beta$ experiment with a majority of self-shielding from the Ar, to give a lower
  background and higher resolution than the use of the enriched $^{136}$Xe in the G3 Xe detector
  (and without requiring the two concentric vessels of the XAX configuration). This offers
  the additional possibility that some of the β track pairs (range
  $\sim 8~\n{mm}$ for the factor $\sim2$
  lower mixture density) might be identifiable by timing in the vertical direction. This is
  part way towards the high pressure (5 --10 bar) Xe gas TPC detectors proposed by Nygren
  et al.~\cite{Nygren:2007,Chinowski:2007,Novella:2005} with a factor
  10 lower density than that of the Xe/Ar mixture and allowing
  more complete reconstruction of the two $\beta$ tracks.
\end{pbitemize}

\section{Conclusions}
\label{sec:conclusion}

We have studied the capabilities of a several stage program of construction and operation of
multi-ton liquid Xe and liquid Ar two-phase TPC detectors. The principal objective is to
measure the energy spectrum of nuclear recoils from weakly interacting massive particles that
may constitute the Galactic dark matter, with sufficiently low background to reach WIMPnucleon
cross-sections $\sim 10^{-46}~\n{cm^2}$ and lower, with sufficient events to allow determination of the
shape of the recoil energy spectrum and from this to estimate the WIMP mass. As previously
proposed~\cite{Smith:2005} it is important to measure this signal in two detectors with different target
elements, to confirm the expected $A^2$-dependence of the signal from a coherent spin-independent
cross-section, and to provide agreement between two independent determinations of the incident
particle mass.

The first stage of this program, referred to as G2, would consist of a liquid Xe detector of 1-ton
fiducial mass, paired with a liquid Ar detector with a 5-ton fiducial mass. The second stage,
referred to as G3, would use the G2 argon vessel as a 10-ton fiducial Xe detector, then paired
with a 50-ton fiducial Ar detector. A further order of magnitude scale-up, referred to as G4,
could be achieved by using the G3 argon vessel for a 100 ton Xe
detector, and adding if required, a 500 ton argon detector. It has been shown in this paper that sufficiently low gamma and
neutron backgrounds can be achieved by combining the following
methods:
\begin{pbitemize}
\item[a)] by the now proven two-phase discrimination of nuclear recoils from gamma/beta events;
\item[b)] by using in each case a total target mass approximately twice the fiducial mass and using the
  outer half of the mass as self-shielding;
\item[c)] by the development, now in progress, of a new high quantum efficiency QUPID
  photodetector, with a radioactivity level lower by $1-2$ orders than the current best
  photomultipliers. This would allow for the first time the use of
  full $4\pi$ photodetector arrays
  with improved light collection, energy resolution, and event
  position resolution.
\end{pbitemize}

Other internal backgrounds, from radioactive impurities in structural materials, have been shown
to be reducible to negligible levels. In addition to achieving the best sensitivity for measurement
of signals from spin-independent interactions, the data will at the
same time provide the best limits on spin-dependent or inelastic
WIMP scattering, again with independent data from Xe and
Ar detectors.

The two detectors of the G3 system are further capable of observing the annual WIMP-spectrum
modulation arising from the combined Earth-Sun motion in the Galaxy. The Xe and Ar detectors
will provide two independent measurements of this, confirming the Galactic origin of the signal.

The sensitivity of liquid noble gas detectors to signals below 100 keV provides the opportunity to
measure signals from two sources of astrophysical neutrinos: solar
and supernova. The excellent position resolution of the two-phase
design and light collection system allows the gamma  
background to be reduced sufficiently to allow pp solar neutrinos to be clearly seen as an
neutrino-electron scattering spectrum up to 250 keV and, with a Xe target depleted in $^{136}$Xe,
sufficient events could be obtained to measure the spectrum to a precision of a few percent. The
total neutrino burst from a Galactic supernova could be observed as a coherent neutral current
nuclear recoil spectrum, in both the Ar and Xe detectors, with sufficient accuracy to measure both
the total flux and mean energy (temperature) of the neutrino
source. At a distance $\sim 10$ kpc, the
precision would be $\sim 10\%$ with the G3 detection system.

We have discussed the further gains to be achieved with the G4 100-ton
(fiducial) Xe scale-up, with or without a corresponding 500-ton
(fiducial) Ar detector. These include order of magnitude gains in dark
matter event numbers and mass determination, more precise annual
modulation signal, and increased numbers of neutrino events from a
supernova burst, giving sensitivity to more remote Galactic
supernovae.

An important application of the G3 and/or G4 systems would be a high
level of sensitivity to neutrinoless double beta decay from
$^{136}$Xe. This would be achievable with the 10-ton G3 Xe 
detector, using either natural or enriched $^{136}$Xe, reaching a
lifetime sensitivity $10^{27} - 10^{28}$ y which, for Majorana
neutrinos, lies within the $0.1-0.01$ eV majorana mass range expected
from existing neutrino mixing data, and hence an expectation of a positive $0 \nu
\beta \beta $ signal for the first time. We
consider also the option of using Xe/Ar mixtures for this and/or dark
matter detection, for potentially higher resolution and improved self
shielding.

This study concludes that the ultra-low backgrounds needed for these
searches are achievable with existing techniques, and that
combinations of multi-ton Xe and Ar detectors provide the most
sensitive method of identifying dark matter signals, providing at the
same time new high sensitivity measurements at the frontier of neutrino physics.

\section*{Acknowledgements}
\label{sec:acknowledgements}

We are grateful for helpful discussions with E. Aprile, F. Calaprice,
C. Galbiati, B. Sadoulet, R. Gaitskell, M. Tripathi and A. Hime. We also acknowledge contributions from E. Brown. 

QUPID is the outcome of collaboration with Hamamatsu Photonics Co., in particular A. Fukasawa, S. Muramatsu,  M. Suyama, J. Takeuchi and T. Hakamata.

This work was supported in part by US DOE grant DE-FG-03-91ER40662, and by NSF grants PHY-0130065/PHY-0653459/PHY-0810283/PHY-0919363/PHY-0904224. We gratefully thank H. Nicholson and M. Salamon from the DOE, and J. Whitmore from the NSF. Additional support was provided by the INPAC Fund from the UC Office of the President, the UCLA Dean, and UCLA Physics Chair funds, and we are thankful to R. Peccei, J. Rudnick, and F. Coroniti, J. Rosenzweig for financial support and encouragement.

\setcounter{equation}{0}
\setcounter{figure}{0}
\setcounter{table}{0}
\appendix
\section{Underground background fluxes and water shielding
  simulation}
\label{app:externalbkg}

A comparison of the orders of magnitude of muon, neutron and gamma
fluxes for several underground sites is shown in Tab.~\ref{tab:undergr_bkg}.
\begin{table}[!htbp]
  \begin{center}
    % {\scriptsize 
      \begin{tabular}{|c|c|c|c|}
        \hline
       %  &  LNGS Hall B/C \par 3600 mwe & LSM \par 4800 mwe & DUSEL \par 4800 mwe \\
        &  LNGS Hall B/C & LSM & DUSEL \\
        &  3600 mwe & 4800 mwe & 4800 mwe \\
        \hline
        $\mu$ flux & $3\times 10^{-8}$ & $3-4\times 10^{-9}$ &
        $3-4\times 10^{-9}$ \\
        ($0.1~\n{MeV} - 10~\n{GeV}$) cm$^{-2}$s$^{-1}$ & (meas.) &
        (meas.) & (meas.) \\
        \hline
        n flux from $\mu$ in rock & $3\times 10^{-9}$ & $3\times
        10^{-10}$ & $3\times 10^{-10}$ \\
        ($1~\n{MeV} - 1~\n{GeV}$) cm$^{-2}$s$^{-1}$ & (sim.) &
        (sim.) & (sim.) \\
        \hline
        n flux from rock activity & $4\times 10^{-6}$
        (sim.) & $4\times 10^{-6}$ & $6-10\times 10^{-6}$ \\
        ($1~\n{MeV} - 6~\n{MeV}$) cm$^{-2}$s$^{-1}$ & (50\% from
        concrete) & (meas.) & (scaled from $\gamma$ flux) \\
        \hline
        $\gamma$ flux from rock activity & 0.2 (sim.)
        & $0.2-0.3$ & $0.3-0.5$ \\
        ($0-3~\n{MeV}$) cm$^{-2}$s$^{-1}$& (80\% from concrete) &
        (activity similar to LNGS) & (sim.)\\
        \hline
      % \end{tabular}}
      \end{tabular}
    \caption{Comparison of muon, neutron, and gamma fluxes for some underground sites}
    \label{tab:undergr_bkg}
  \end{center}
\end{table}

To provide a suitable external background for the detectors discussed
in this paper, we need to reduce the external neutron and gamma fluxes
to below the level $10^{-10}~~\n{events/cm^2/s}$. Although this could
be achieved by sufficient thicknesses of conventional outer metallic
and inner hydrocarbon shielding, which provide the most compact
shielding solution, the potential availability of larger underground
experimental volumes has increased preference for an all-water shield,
$3-4~\n{m}$ thick, with the innermost $50-60~\n{cm}$ instrumented as an active veto
or replaced by liquid scintillator. This offers the additional
possibility of instrumenting the scintillator with photomultipliers
placed outside the water volume, and hence with their activity
shielded by the water. The latter can at the same time function as a
muon veto by detecting Cerenkov signals in the water. 

\begin{figure}[!htbp]
  \centering
  \includegraphics[width=1. \columnwidth]{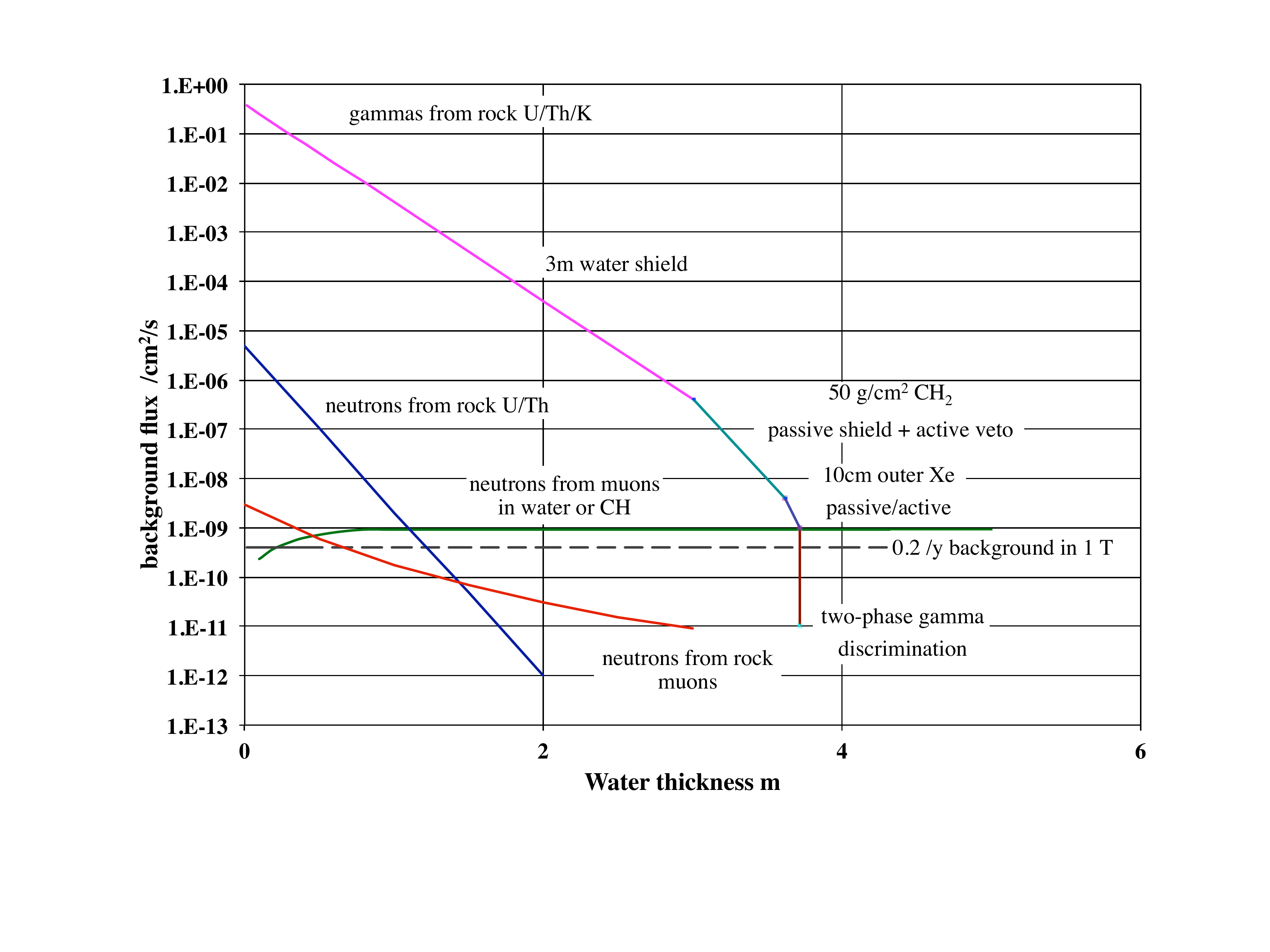}
  \caption{Chart showing progressive attenuation of external gamma and
    neutron fluxes through 3 m water shielding, an active and/or
    passive hydrocarbon veto, and a 10 cm passive outer layer of the
    target. An additional factor $\sim 100$ two-phase detector
    discrimination for gammas is also shown. The dashed line shows the
    level required for sensitivity to a signal at WIMP-nucleon
    cross-section $10^{-46}~\n{cm^2}$ with a fiducial one ton Xe
    target. These simulations used RAL-developed shielding
    code FAUST (1992, unpublished) using cross-section data files
    identical to MCNP and GEANT4, cross checked against MCNP
    in~\cite{Smith:2001}, and subsequently against GEANT4 (ZEPLIN III internal
    notes, unpublished)}
    \label{fig:fluxes_atten}
\end{figure}

We simulate the performance of this by tracking first a typical cavern
gamma flux from rock U/Th/K activity (using the example of the LNGS
activity given by Wulandari et al.~\cite{Wulandari:2004})
through successive 1 m thicknesses of water, then through
$50~\n{g/cm^{2}}$ liquid scintillator, and through an outer 10 cm
passive layer of liquid Xe to an internal fiducial target 80 cm
diameter, finally assuming a 2-phase detector discrimination factor of
100. The results of these steps are summarized in the right hand path
of Fig.~\ref{fig:fluxes_atten}, which traces the decrease in total gamma flux
through the shielding. The simulations also keeps track of the gamma
energy spectrum after each stage, allowing the final number of low
energy events in the fiducial Xe region to be counted. The horizontal
dashed line at $2 \times 10^{-10}~\n{events/cm^2/s}$ is the
level corresponding to an unrejected background of 0.2 events/y
( $<40$ keV) in an inner 1-ton (fiducial) Xe target.

The corresponding attenuation of the principal neutron sources is
shown by the three curves on the left hand side of the plot, again
each needing to fall below the horizontal dashed line to achieve a
$0.1-0.2$ event/y background. The one source that remains above this
level is that of neutrons from muons interacting in the metre of water
closest to the detector. However, this can be rejected by using the
water itself as a Cerenkov veto, or by means of additional
scintillating muon veto panels above the water shield. For these
calculations, the muon flux at 3600 m.w.e. depth (the LNGS site depth)
has been used as an example, together with the typical gamma flux from
LNGS rock activity. 

The shielding requirement for other target masses and materials, can
be estimated by scaling of fiducial surface area, and noting the
approximate linearity of the water attenuation, as follows:
\begin{enumerate}
\item For G2 detector combination $\sim 1$-ton Xe \& $\sim 5$-ton Ar.\\
  Fig.~\ref{fig:fluxes_atten} applies to a 1 m diameter, 1 m high, Xe
  detector volume with 0.8 m diameter fiducial region (density 2.9
  g/cc). Thus for an Ar detector of 5 times the fiducial mass (and
  density 1.45 g/cc) the fiducial diameter is a factor 2 larger and
  the fiducial surface area a factor 4 larger. At the same time the
  required background/ton is reduced by a factor 5 (to achieve the
  same signal sensitivity in Ar). Hence the dashed line in
  Fig~\ref{fig:fluxes_atten} is lower by a factor 20 for Ar, which nevertheless
  remains within the attenuation curves and hence requires no
  additional shielding.
\item For G3 detector combination $\sim 10$-ton Xe \& $\sim 50$-ton Ar.\\
  To achieve all of the physics objectives of a factor 10 increase in
  Xe and Ar masses, one will require a factor 10 gain in
  background/ton for both detectors. At the same time, the fiducial
  surface areas would increase by a factor $\sim 4$. Thus the combined
  drop in the dashed curve of Fig~\ref{fig:fluxes_atten} becomes a factor 40 for
  10 T Xe, and (taking account also of the size factor in (a)) a
  factor $\sim$ 800 for a 50-ton Ar detector. The former remains within
  the attenuation curves of Fig.\ref{fig:fluxes_atten}, but for the 50-ton Ar this
  would require extending the water shielding by an additional 1.5 m
  radially to lower by a factor 1000 the end point of the
  gamma-shielding curve. It is also apparent that the slower
  attenuation of neutron events from high-energy muons in rock (red
  curve) may be marginal for a 50-ton Ar detector, and would benefit
  from a deeper site (see Tab.~\ref{tab:undergr_bkg}) to gain a factor 10 in
  muon flux.
\item For G4 detector combination $\sim 100$-ton Xe \& $\sim 500$-ton Ar.\\
  For the G4 Xe detector, a further factor 40 gain over G3 is already
  covered by the increased water radius specified in (b) for the G3 50
  ton Ar detector. A further 1.5 m radial increase is needed to gain an
  additional reduction factor 800 for the 500 ton G4 Ar detector, now
  increasing the overall water radius to 7 m, and the overall G4 water
  tank diameter to over 22 m. However, this would be significantly
  reduced by the increasing thickness of the target self-shielding layer
  outside the fiducial volumes, in progressing from G2 to G3, and from
  G3 to G4, to reduce backgrounds from detector vessel and
  photodetectors (see Sec.\ref{sec:backgrounds}, Tab.\ref{tab:g2g3g4_bkg}).
\end{enumerate}

It is important to keep in mind that the above water shielding
discussion applies to continuous ($4\pi$) coverage of the shield
around the detector. Electrical, cryogenic, and mechanical
feed-throughs will provide channels along which external gamma and
neutron fluxes can reach the target. Since the external fluxes can be
6-8 orders greater than the shielded flux, even small areas of such
channels could substantially increase the target background. The
solution to this is to ensure that all such channels have two or three
90 degree bends, the scattering at each decreasing the flux (reaching
the detector) by typically a factor $10^3$. This will be
straightforward for cryogenic pipes and electrical connections, but
will need novel design for the metallic structures that support the
detector within the water shield. 

\setcounter{equation}{0}
\setcounter{figure}{0}
\setcounter{table}{0}
\section{Radioactive contamination of target elements}
\label{app:targetbkg}

\subsection{$^{85}$Kr background in natural Xe}
\label{subapp:85kr}

Natural Kr is present as an impurity in commercial Xe, at a level
typically 50 ppb, and in turn contains $\sim10^{-11}$
cosmogenic $^{85}$Kr, which emits a continuous beta decay
spectrum with maximum energy 690 keV and a 10.7 y half-life. This would
contribute a low energy beta background $\sim 1$ events/keV/kg/d to be
added to the gamma population and, at this contamination level,
limiting the dark matter sensitivity to $10^{-44}~\n{cm^2}$. The need to remove this, to reach
lower cross sections, has led to the development of reflux
distillation columns that will reduce the level of Kr in Xe by a
factor $10^3$ per pass~\cite{Abe:2006}. Thus, using several
passes, it will be possible to reduce the $^{85}$Kr content
to a negligible level.

\subsection{$^{39}$Ar background in natural Ar}
\label{subapp:39Ar}

In the case of the Ar target, natural (atmospheric) $^{40}$Ar contains
a fraction $\sim 10^{-15}$ of cosmogenically produced $^{39}$Ar, which
beta-decays with a half-life 269 y and a continuous spectrum with
maximum energy 570 keV. This gives a total background rate $\sim 1.2$
events/kg/s, a low energy differential rate $\sim 3 \times 10^2$
events/kg/d/keV, and thus a total event rate $\sim$ 6000 events/kg/d
in the 5 - 25 keVee electron equivalent dark matter energy range. This is too high a starting point
for the two-phase discrimination technique alone, and can also produce
``pile-up'' in the data acquisition system. However, using argon depleted
in $^{39}$Ar from an underground source, this can be immediately
reduced by a factor $>$ 100
~\cite{Galbiati:2008,Acosta:2008,Calaprice:2012} and then by a further
factor $\sim 10^8$ using pulse shape discrimination.  By means of
these factors it becomes possible to reduce the $^{39}$Ar background
to $<$ 1 event/5 tons/year for a nuclear recoil energy threshold of 20
keVnr.  This is calculated in detail in ~\ref{subapp:bkg_ar5t}
below. We note in addition that if the 5 ton and 50 ton argon
detectors can be instrumented with a  4$\pi$ photodetector array, this
provides a position resolution $\sim$ 1\% of the detector volume,
permitting rejection of events that do not correspond to a position
coincidence of S1 and S2 in the same 1\% of detector volume, also
reducing $^{39}$Ar ``pile-up'' by a factor 100.

\subsection{Rn-related background}
\label{subapp:rn}

Radon in the air ($^{222}$Rn) has a lifetime $\sim 4$ days,
continuously replenished from uranium in the ground and neighboring
materials. Hence it does not survive in stored xenon or argon gas, but
has been seen to cause background problems in the Xe
detectors~\cite{Alner:2007,Lebedenko:2009,Angle:2008}, by contaminating the target or gas systems
during assembly and/or transfer and depositing its decay products (for
example $^{210}$Po, lifetime 138 days) on the detector walls
where it further decays by alpha emission. The alphas themselves are
in the MeV range, and hence do not result in low energy signals if
emitted into the liquid. But if emitted into the wall, the Po nucleus
recoils into the liquid, with (by momentum conservation) \textless 100
keV recoil energy and thus can produce a signal comparable to that of
a recoil Xe or Ar nucleus. These ``wall-events'' can be eliminated from
the data set by using the S2 position sensitivity to make a radial
cut, but a $10~\n{cm} - 20~\n{cm}$ cut is already envisaged as an essential part
of the target shielding and included in the estimates in
Sec.\ref{subsec:material} and \ref{app:detmaterial}. Thus for
large volume detectors the fiducial target is already protected from
this source of background.

\subsection{U \& Th contamination of liquid Xe or Ar}
\label{subapp:uth}

Contamination of the target material with radioactive isotopes might
in principle occur during production and purification, or during
recirculation through getters if used in the detector cryogenic
system, so we need to estimate some permissible limits on the
tolerable U or Th content in the target material. The presence of U/Th
chains would give rise to both gamma emission, alpha emission, and
beta decay, the latter being the sum of \textgreater 100 different
beta spectra in the U/Th chains. Summing these contributions, we
estimate that to reduce the absolute level of electron recoil events
(after the S2/S1 cut) to \textless 1 events/ton/y below 20 keV would require
a stringent U/Th concentration level no more than $\sim
10^{-3} - 10^{-4}$ ppt. However, the
majority of these events will be automatically vetoed by
near-simultaneous higher energy events in the chain (gammas,
electrons, alphas). Within a time window range of, say, $\pm 1$ ms to
$\pm 1000$ ms (dependent on overall trigger rate), we estimate that
99.9\% of U/Th events \textless 20 keV will be in coincidence with a
higher energy decay in the chain, thus automatically reducing this
background by a factor 1000. On this basis it would appear that the
permissible upper limit on the contamination of Xe or Ar by U/Th is
eased to about $0.1 - 1$ ppt. If the levels of U/Th in liquid Xe or Ar
are found to be higher than this, some additional purification would
be necessary.

\setcounter{equation}{0}
\setcounter{figure}{0}
\setcounter{table}{0}
\section{Self-shielding of radioactivity in detector components}
\label{app:detmaterial}

For simulations of detector background from radioactive impurities in
the detector vessels and photodetectors, the assumed limits on U/Th/K
contamination, based on recent sample selection and testing, are
summarized in Tab.\ref{tab:bkg_detmaterial}.
\begin{table}[!htbp]
  \begin{center}
    \begin{tabular}{|c|c|c|c|c|}
      \hline
      \hline
      \textbf{material} & \textbf{unit} & \textbf{U (mBq/unit)} & \textbf{Th (mBq/unit)} & \textbf{K (mBq/unit)} \\
      \hline
      Copper & kg & $\le 0.07$ & $\le 0.03$ & $\le 0.5$ \\
      \hline
      Titanium & kg & $\le 0.25$ & $\le 0.20$ & $\le 1.3$ \\
      \hline
      PTFE & kg & $\le 0.31$ & $\le 0.16$ & $\le 2.2$ \\
      \hline
      QUPID (3'') & single QUPID & $\le 0.5$ & $\le 0.4$ & $\le 2.4$ \\
      % QUPID (3'') & single QUPID & $< 17$ & $ 0.4$ & $ 5.5 $ \\
      \hline
      QUPID (6'') & single QUPID & $\le 4.0$ & $\le 3.2$ & $\le 20$ \\
      \hline
      & & & & \\
      \hline
      \hline
      Conversion & 1 mBq (0.001 parent decays/s) & 0.08 ppb & 0.25 ppb & .03 ppm \\
      \hline
    \end{tabular}
    \caption{Assumed radioactivity levels in detector materials and components}
    \label{tab:bkg_detmaterial}
  \end{center}
\end{table}

We need estimates of both gamma and neutron backgrounds from these
sources. Gammas are emitted directly by U/Th chains, and by K, with a
known energy spectrum extending up to $\sim 3$ MeV (tabulated, for
example, in~\cite{Smith:1990}). Neutrons are produced by the MeV-range alpha
emission from U/Th, interacting with the host nuclei to produce an
MeV-range neutron spectrum with a total material-dependent rate
calculated by Heaton et al~\cite{Heaton:1990}, also incorporated in the \texttt{SOURCES}
program~\cite{Wilson:1999}. The latter also provides a prediction the emitted neutron
energy spectrum, but which appears not to be consistent with
underground measurements~\cite{Bungau:2005,Wulandari:2004,Chazel:1998}. We therefore consider it
safer to use, for background simulations, a more conservative (higher
mean energy) generic neutron spectrum shape as proposed in~\cite{Bungau:2005}. We
consider in turn gamma and neutron backgrounds for each of the four
detectors illustrated in Fig.\ref{fig:g2inner} (G2) and Fig.\ref{fig:g3inner}
(G3).

\subsection{G2 system: backgrounds for 1 ton Xe detector}
\label{subapp:bkg_xe1t}

\begin{figure}[!htbp]
  \centering
  \includegraphics[width=0.75 \columnwidth]{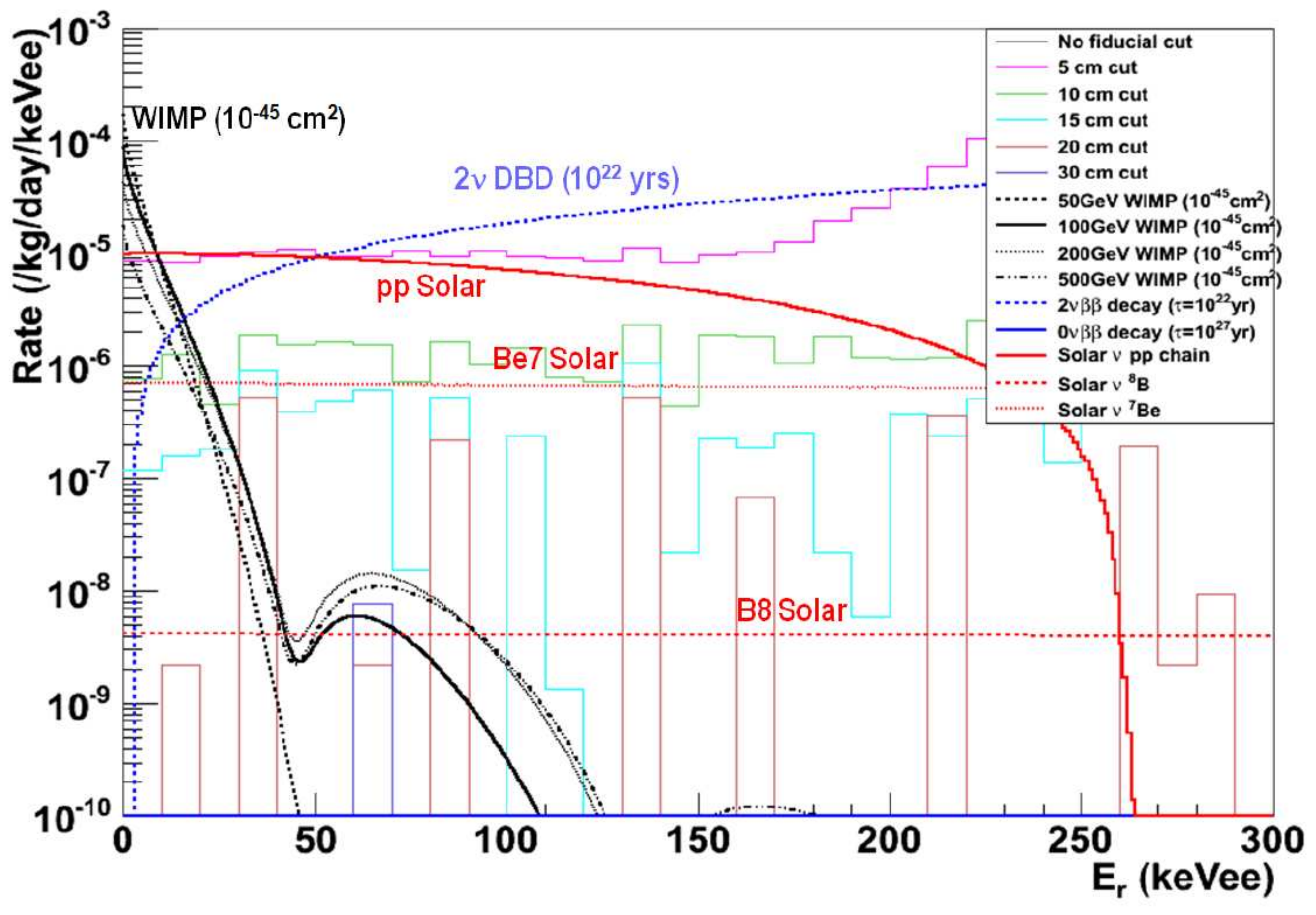}
  \includegraphics[width=0.75 \columnwidth]{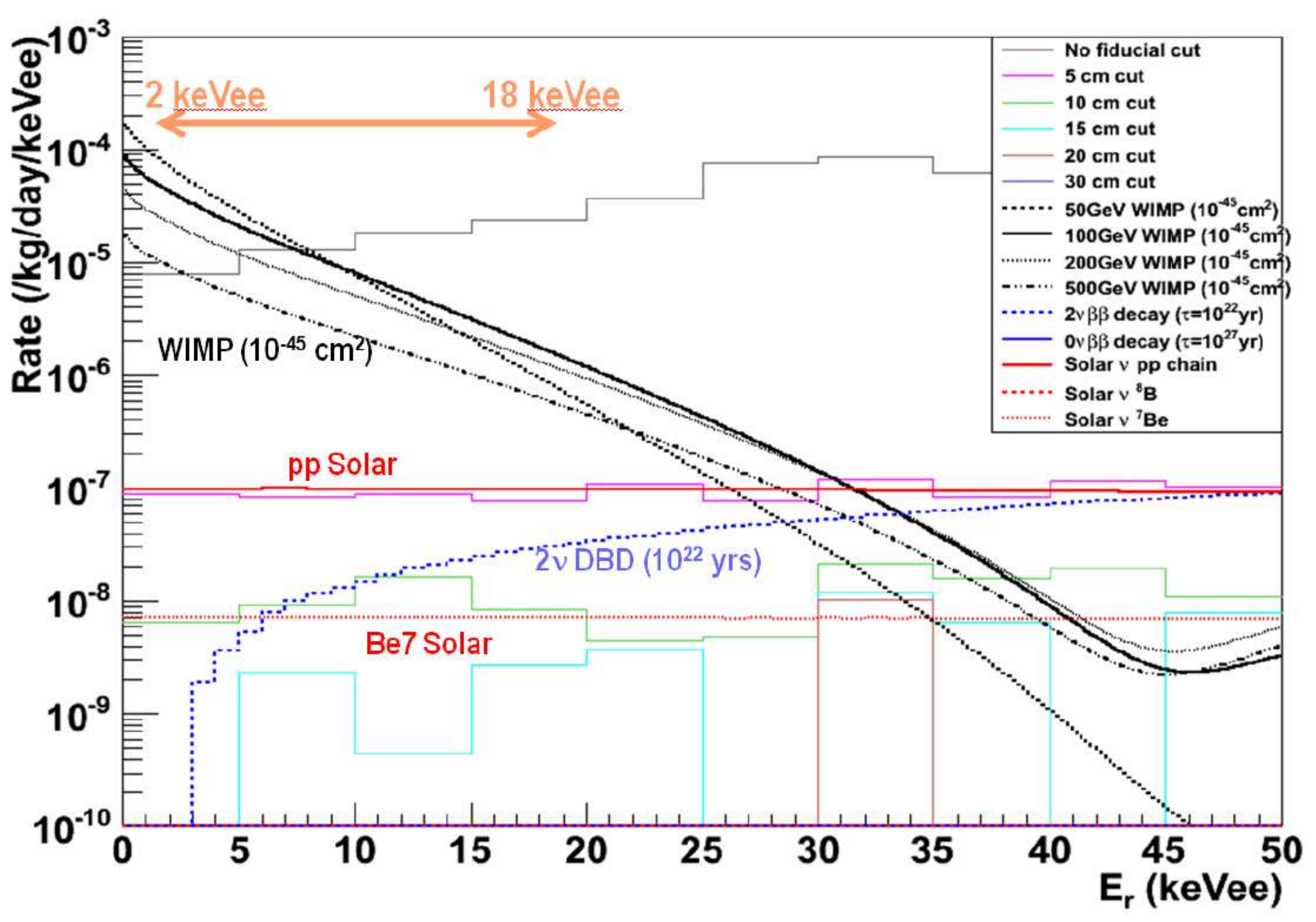}
  \caption{\textit{(upper plot)} Gamma rates from cryostat and QUPID
    radioactivity for 1 m diameter Xe detector (2.2 tons total Xe
    mass) with 0, 5, 10, 15, 20, 30 cm outer Xe shield, with
    multiple scattering cut, and no factor S2/S1 cut. Superimposed
    spectra show nuclear recoils from WIMPs at $10^{-45}~\n{cm^2}$ cross
    section, electron scattering from pp and $^7$Be solar neutrinos, and
    a $2\nu \beta \beta $ spectrum.\protect\\
    \textit{(lower plot)} Same as upper plot but expanded low energy scale
  and S2/S1 cut.}
\label{fig:gamma_rate_xe1t} 
\end{figure}

For the 1 ton Xe detector, Fig.\ref{fig:gamma_rate_xe1t}(\textit{upper}) shows the
results of a \texttt{GEANT4} simulation~\cite{Agostinelli:2003} of the energy spectrum of single scatter gamma
events in the fiducial region, assuming 99\% rejection by the
two-phase S2/S1 discrimination and various Xe thicknesses from zero to
30cm used as an outer passive shield. Position sensitivity also allows
events in the latter region to be identified and used as an active Xe
veto. The results are shown as a rate in unrejected
events/kg/day/keVee (electron equivalent) sometimes abbreviated to dru
(differential rate unit~\cite{Lewin:1996}). Superimposed on these results are
spectra for WIMPs of various masses, a WIMP-nucleon cross section of
$10^{-45}~\n{cm^2}$, also, for later discussion, pp and
$^7$Be solar neutrinos, and a two-neutrino double beta decay spectrum
(emitting 2.6 MeV electron recoil energy). Fig.\ref{fig:gamma_rate_xe1t}(\textit{lower}) shows the
same results for the expanded energy region \textless 50 keVee showing
also the $\sim 2-20$ keVee range most relevant to WIMP detection with
the liquid Xe TPC. For a 1-ton target, the results show that a 10 cm
outer passive Xe layer is sufficient to reduce the unrejected gamma
background to 0.1 events/y.

\begin{figure}[!htbp]
  \centering
  \includegraphics[width=0.9 \columnwidth]{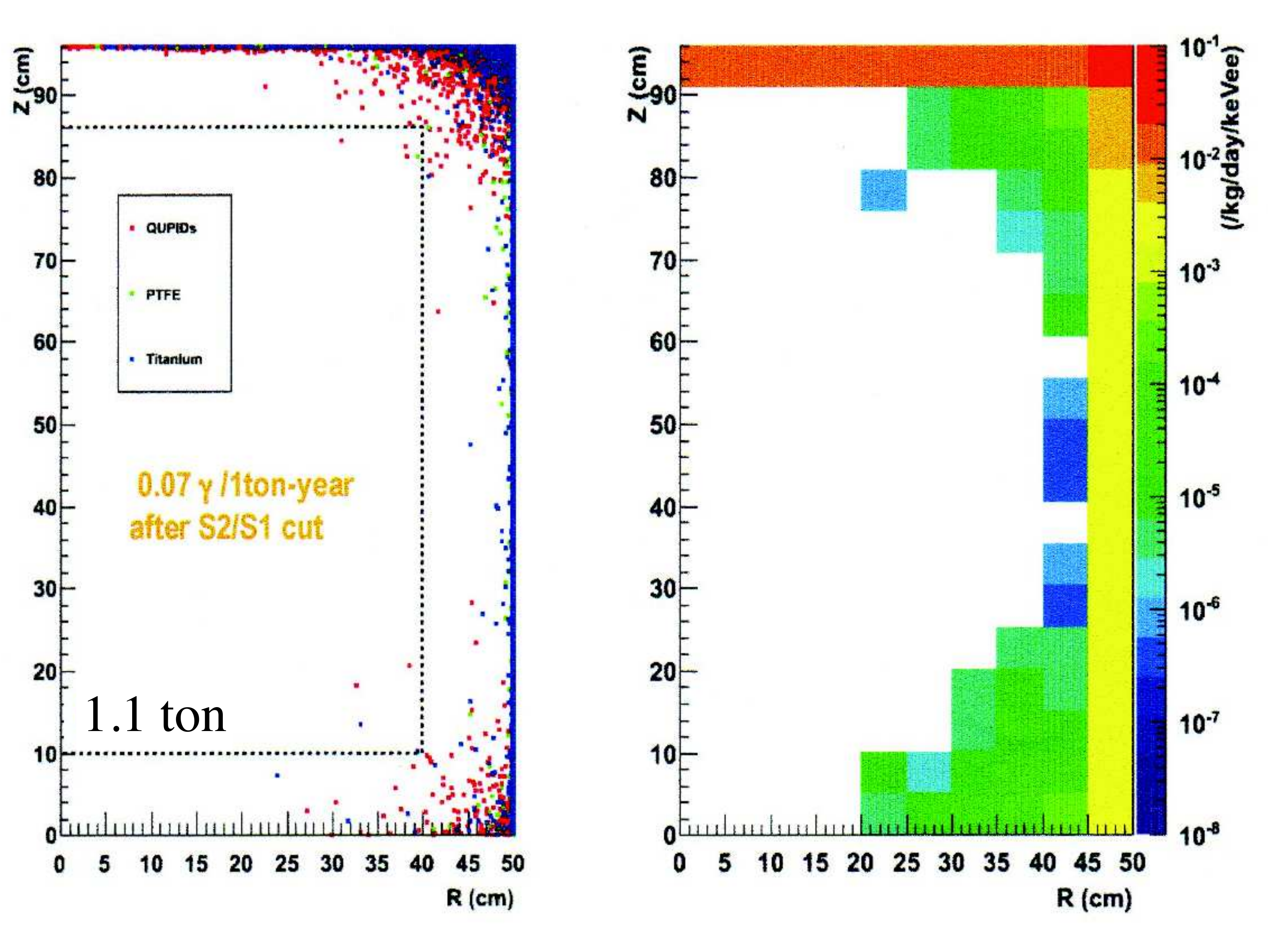}
  \caption{Simulated spatial distribution of gamma background in 2.2 ton
    (total) Xe detector after 1 year, with multiple scattering cut but
    without S2/S1 discrimination.}
  \label{fig:gamma_spatial_xe1t}
\end{figure}

Fig.\ref{fig:gamma_spatial_xe1t} shows the spatial distribution of single scatter
gammas for a simulated 1 year running, without an S2/S1 cut. Removing
an outer 10 cm leaves only 7 events/y in the fiducial region, which then
reduces to \textless 0.1 events/y after the two-phase S2/S1 cut. The
electron recoil spectrum from pp-chain solar neutrinos, shown in
Fig.\ref{fig:gamma_rate_xe1t}(\textit{upper}), is about a factor 10 higher and thus
represents an additional known background (constant with energy below
50 keVee and hence fully subtractable) $\sim 0.7$ event/y, or lower if
foreseeable improvements in the S2/S1 cut are made.

An outer cut of 10 cm leaves a fiducial mass 1.1 ton and removes all
but 7 events which, after a factor 100 S2/S1 discrimination, leaves
0.07 events/ton/y unrejected. 

\begin{figure}[!htbp]
  \centering
  \includegraphics[width=0.75 \columnwidth]{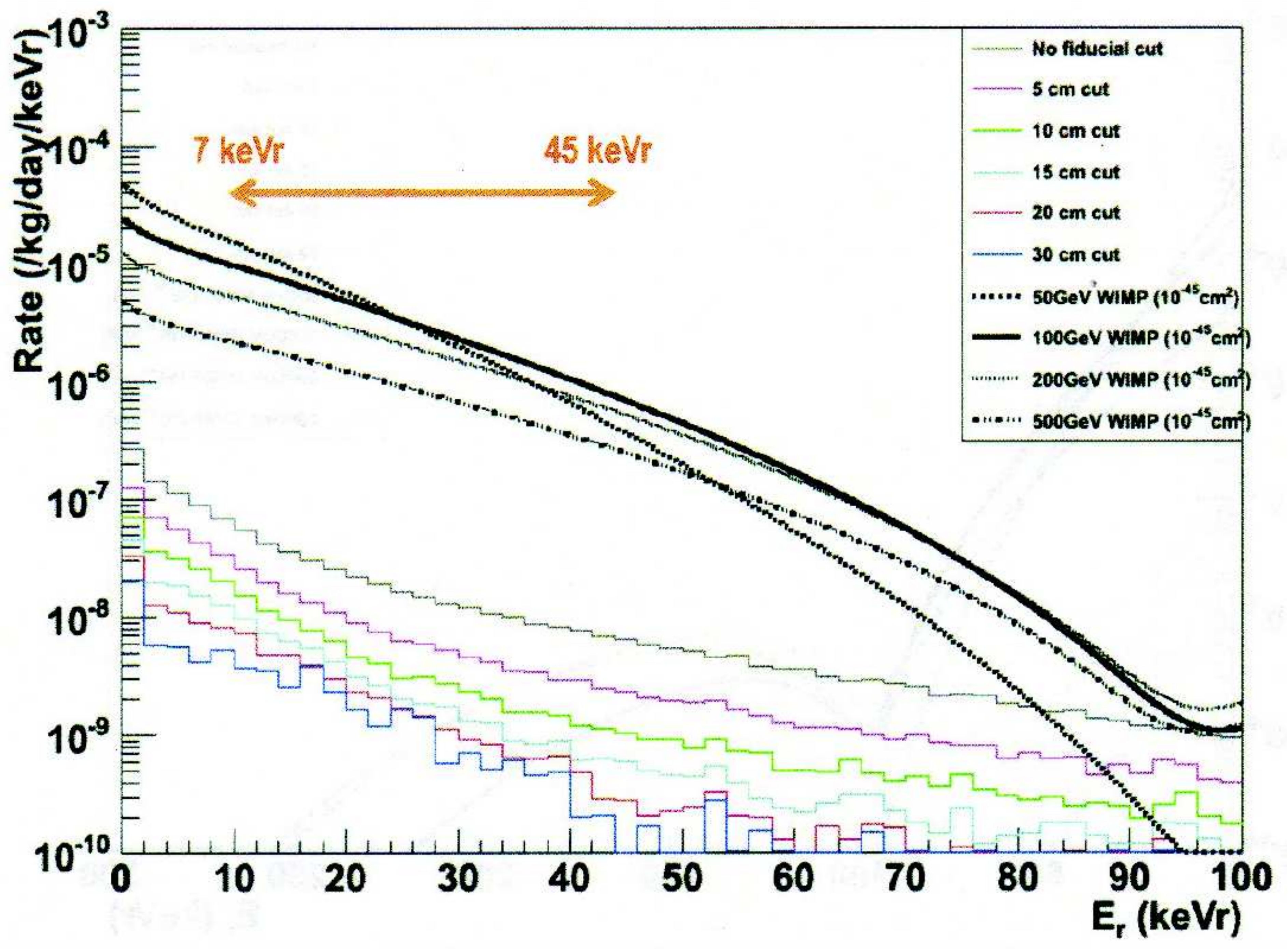}
  \includegraphics[width=0.75 \columnwidth]{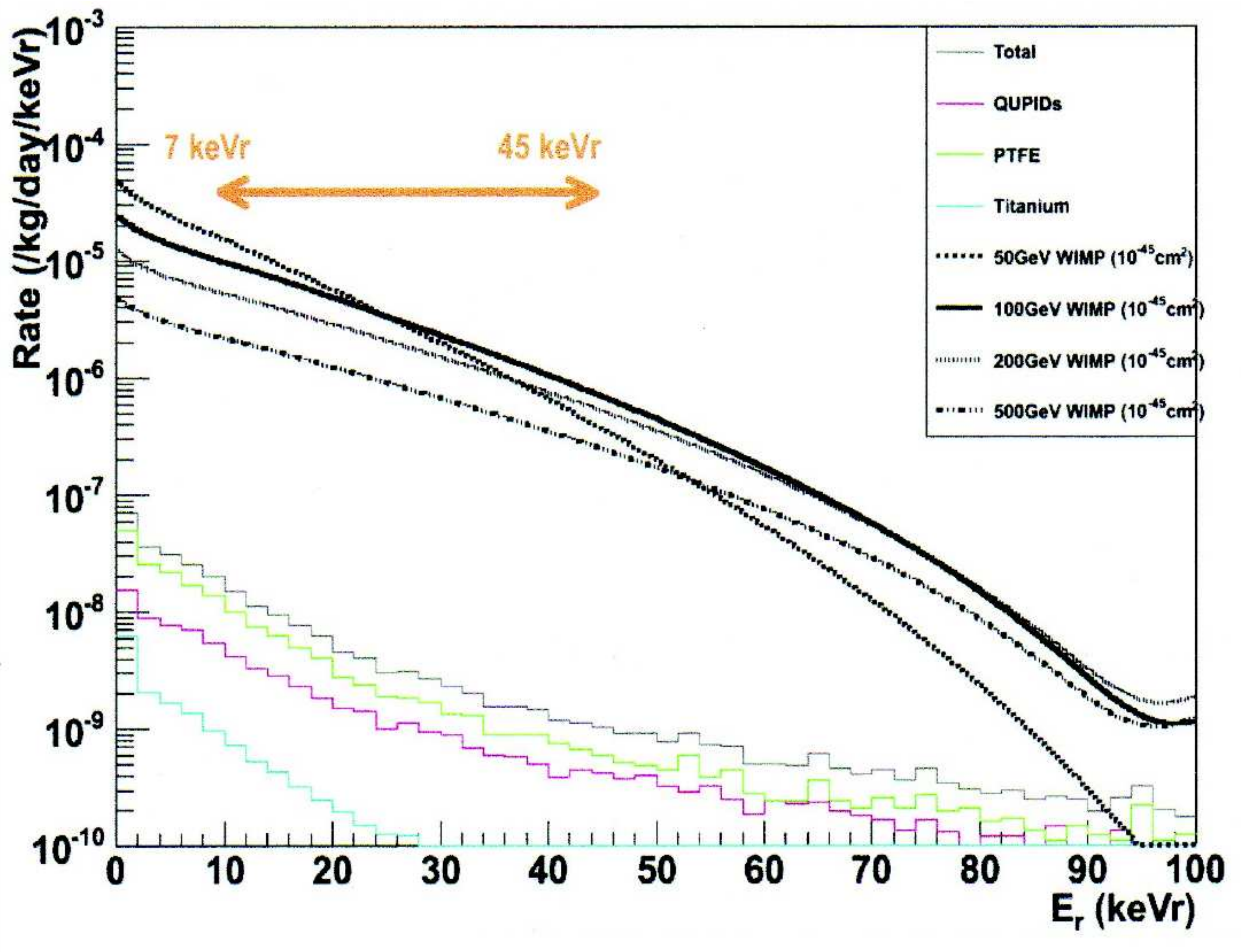}
\caption{Neutron single-scatter spectra in 1.1 ton fiducial Xe. Energy axis (keVr) is nuclear recoil energy\protect\\
\textit{(upper plots)} Showing several thicknesses of passive Xe.\protect\\
\textit{(lower plots)} Fixed 10 cm thickness passive Xe, but the
showing individual contributions from QUPIDs and different nearby
detector materials.
Dark matter spectra for WIMP-nucleon $\sigma = 10^{-45}~\n{cm^2}$ shown for
several WIMP masses.}
\label{fig:neutron_single_xe1t} 
\end{figure}

\begin{figure}[!htbp]
  \centering
  \includegraphics[width=0.9 \columnwidth]{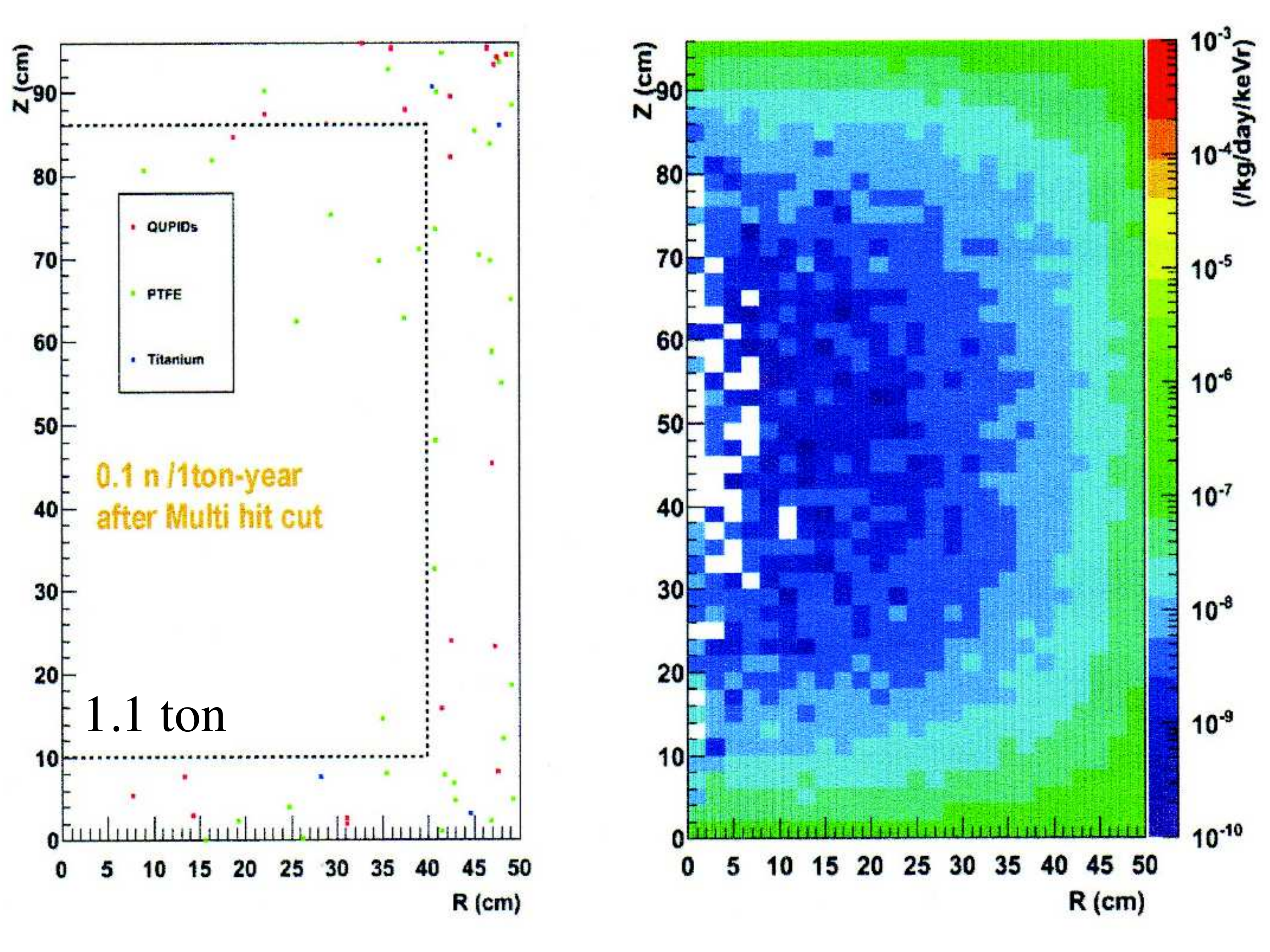}
\caption{Simulated spatial distribution of single scatter neutron
  events in 2.3 ton (total) Xe detector for 100 years. An outer cut of
  10 cm leaves 10 events in a fiducial mass of 1 ton giving 0.1
  events/ton/y unrejected.}
\label{fig:neutron_spatial_xe1t}
\end{figure}

Results of simulations of the corresponding single scatter neutron
backgrounds in 1 ton Xe are shown in Fig.\ref{fig:neutron_single_xe1t}
and Fig.\ref{fig:neutron_spatial_xe1t}. The production of neutrons from U and Th is
material-dependent but typically a factor $10^5$ lower in
absolute rate than the production of gammas. However, neutrons are
less attenuated than gammas by the outer 10 cm passive Xe and not
discriminated at all by the S2/S1 cut. Fig.\ref{fig:neutron_single_xe1t}(\textit{upper}) shows the
total single scatter neutron background spectrum in the Xe fiducial
region after outer cuts of various thicknesses. Fig.\ref{fig:neutron_single_xe1t}(\textit{lower}) uses
a fixed 10 cm outer cut, and plots the individual contributions from
QUPIDs and the different nearby detector materials. This shows the
largest contributions come from the QUPIDs and the PTFE (if used on
the side walls as a reflector in place of a full array of
QUPIDs). Thus in the absence of the PTFE, the QUPIDs are the dominant
source of background, a conclusion utilised below to simplify the
estimates for larger detectors.

Fig.\ref{fig:neutron_spatial_xe1t} is the neutron counterpart to Fig~\ref{fig:gamma_spatial_xe1t}, and
shows the individual single scatter events in the 1-ton fiducial
region for the equivalent of 100-year data, and for the specific case of
a 10 cm passive Xe thickness. In this simulation the fiducial region
contains 10 neutron events for 100 years, thus achieving the desired
background of $\sim 0.1$ events/ton/y.

\subsection{G2 system: backgrounds for 5-ton Ar detector}
\label{subapp:bkg_ar5t}

As discussed in  Sec.\ref{sec:overview}, the G2 system consists of a
$\sim$1-ton fiducial Xe detector partnered by a $\sim$5-ton fiducial
Ar detector. The liquid argon detector will also be constructed in a
two-phase TPC configuration, but operated at a higher nuclear recoil
threshold than liquid xenon, typically $\sim$20 keVnr compared with
$\sim$8 keVnr for Xe.  This is to achieve separation from the
$^{39}$Ar beta decay background discussed in~\ref{app:targetbkg}, and
calculated numerically in the next paragraph. The upper recoil energy range remains higher than that for Xe owing to the lower form factor correction as shown in Fig.\ref{fig:spectrum} (subject to the Galactic escape velocity cut-off in Tab.\ref{tab:recoil_ar}).

The nuclear recoil energy threshold for argon arises from an interplay
between the scintillation light output and the pulse shape
discrimination between electron and nuclear recoils.  But for the
presence of $^{39}$Ar it would be trivial to obtain a nuclear recoil
threshold even as low as a few keV.   With a measured light output of
40 photons/keV, 75\% photodetector coverage, and a $>$ 36\% quantum
efficiency for wavelength-shifted argon QUPIDs ~\cite{TPBAr:UCLA} the scintillation output is $>$ 10.8
photoelectrons/keVee, and 9 pe/keVee even with conventional PMTs.
Thus, with a trigger $\sim$5 pe this high pe/keVee output could give
a nuclear recoil threshold as low as 2 - 3 keVnr, were it not for the
presence of a $^{39}$Ar flat background at low energy. We now show that the
latter can be reduced to $<$ 1/y in detectors containing 5 - 50 tons
argon, for an energy threshold of 20 keVnr.

Using a quenching factor of 25\%~\cite{Boulay:2006} the required
electron recoil threshold is 5 keVee.  We assume an analysis energy
span of 5 - 25 keVee, corresponding to 54 - 270 photoelectrons in the
case of the QE already achieved for argon QUPID photodetectors. From
~\ref{subapp:39Ar} the $^{39}$Ar background in this energy span is 6000
events/kg/d = 10$^{10}$ events/5 ton/y, reduced to $10^8$ events/5 ton/y by
the use of argon from an underground source. Simulations of pulse
shape discrimination based on measured pulse time constants have been
presented as a function of photoelectron number by
A. Hime~\cite{Hime:2012}. These give an average psd of $6\times
10^{-9}$ over the (flat) energy range 5 - 25 keVee (54 - 270
photoelectrons), which reduces the background to 0.6 events/5 ton/y.

Hence, even without any additional low energy two-phase
discrimination, the $^{39}$Ar background in a 5 ton detector is
reduced by psd to $<$ 1/year with a 20 keVnr threshold. For the case of
detectors larger than 5 tons, we can utilize the additional
discrimination available from the two-phase S2/S1 ratio. From the
measurements of~\cite{Benetti:2008} the measured ratio of the quantity
S2/S1, for electron and nuclear recoils, averages 8 in the range 5 -
25 keVee, giving a further discrimination factor conservatively 50 -
100, which further reduces the background in both 5 and 50 ton
detectors below 0.1 event/year. The electric field requirement for
this modest additional discrimination can be adjusted to minimize
reduction in light output by suppression of
recombination~\cite{Kubota:1978}.  There is also likely to be an
additional gain of an order of magnitude or more from underground
argon, since the latest measurements have not yet seen any $^{39}$Ar
in underground samples.

From the preceding estimates of electron recoil background, we can assume that in the case of two-phase argon detectors, the residual neutron background rate will dominate in the fiducial target zone, and this will now be discussed. 

\begin{figure}[!htbp]
  \centering
\includegraphics[width=0.75 \columnwidth]{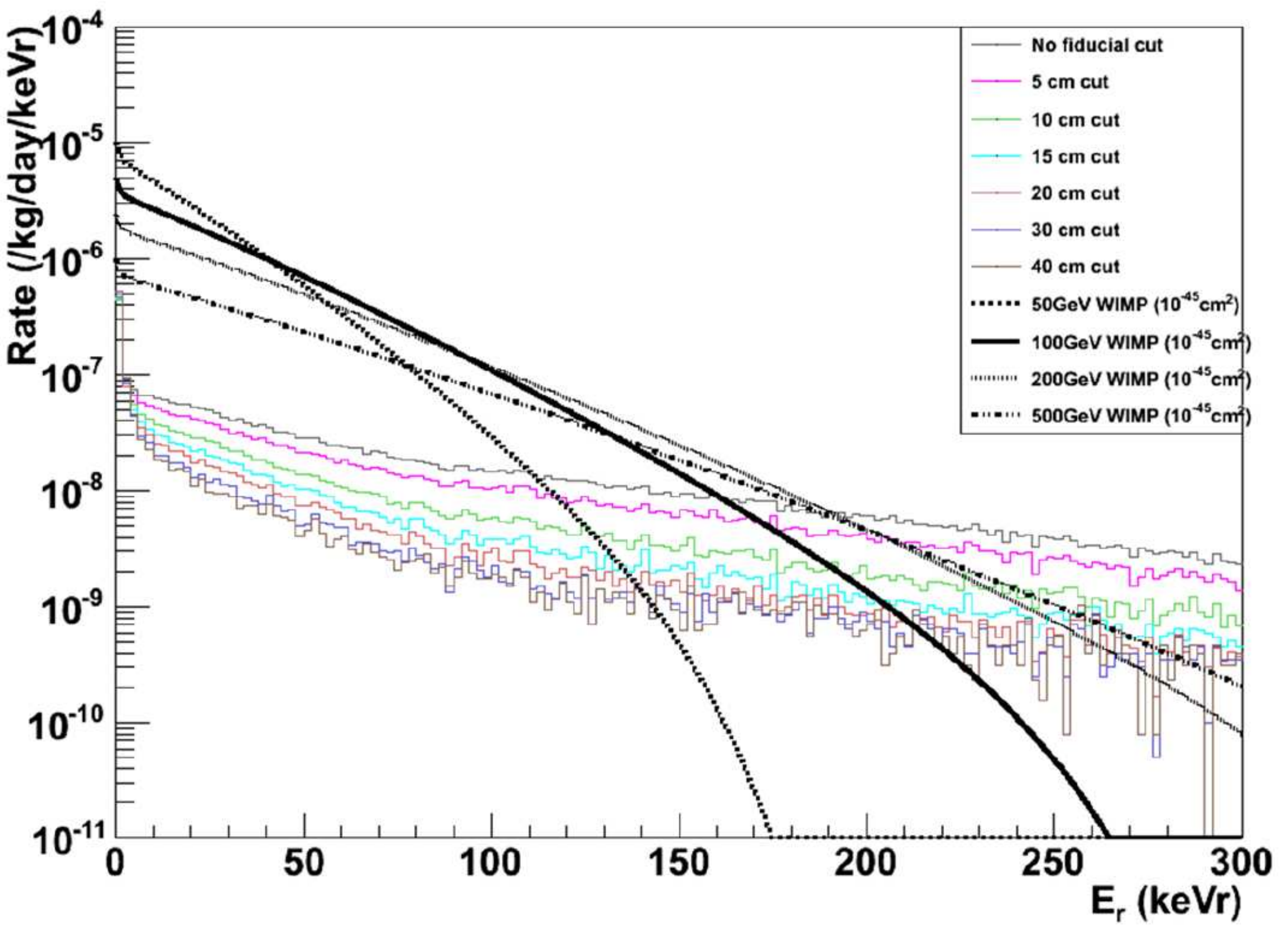}
\includegraphics[width=0.75 \columnwidth]{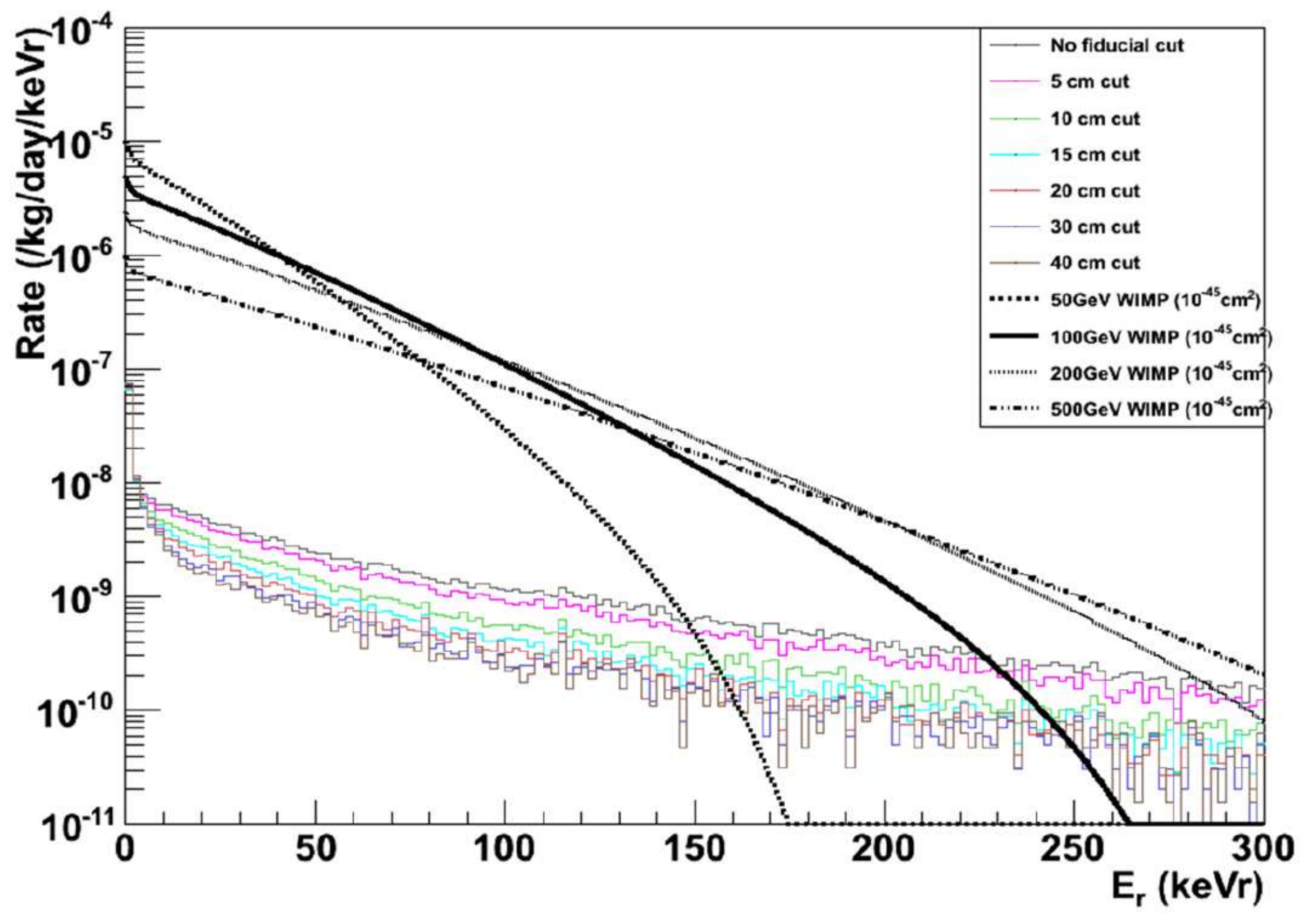}
\caption{Neutron single scatter spectra in 5-ton (fiducial) liquid Ar,
  assuming dominant contribution from residual U/Th in QUPIDs, and
  showing mass-dependent DM spectra for WIMP-nucleon $\sigma =
  10^{-45}~\n{cm^2}$.\protect\\
  \textit{(upper plots) } No outer liquid scintillator veto\protect\\
  \textit{(lower plots) } 0.5\% Gd-loaded liquid scintillator with
  veto threshold 300 keV.}
\label{fig:neutron_single_ar5t}
\end{figure}

As mentioned previously, Fig.\ref{fig:neutron_single_xe1t}(\textit{lower}) shows that
the dominant contributions to fiducial neutron background arise from
the QUPIDs and the PTFE, the latter providing an optional reflector on
the detector sides in place of a full array of QUPIDs. However, the G2
Ar design under consideration here would utilize a full array of
QUPIDs, and no PTFE reflecting walls, so that only the neutron
background from U/Th in the QUPIDs need be included in the
simulations. Figs~\ref{fig:neutron_single_ar5t}(\textit{upper})
and~\ref{fig:neutron_single_ar5t}(\textit{lower}) show the resulting
spectrum of nuclear recoils, with a multiple scattering cut, for
various thicknesses of outer liquid Ar cut, and with or without the
use of signals from a 0.5\% Gd-loaded liquid scintillator veto
outside the detector. Without the Gd, neutrons are slowed and absorbed
by H releasing a 2.2 MeV gamma. With 0.5\% Gd loading, $>95\%$ of
neutrons are absorbed on Gd nuclei releasing 4 gammas totalling 8 MeV. 
The veto rejects two types of event:
\begin{enumerate}
\item neutrons emitted from the QUPIDs into the outer shield,
  then scattered into the target;
\item neutrons emitted first into the target, then scattered
  out into the scintillator. 
\end{enumerate}
\begin{figure}[!htbp]
  \centering
  \includegraphics[width=0.75 \columnwidth]{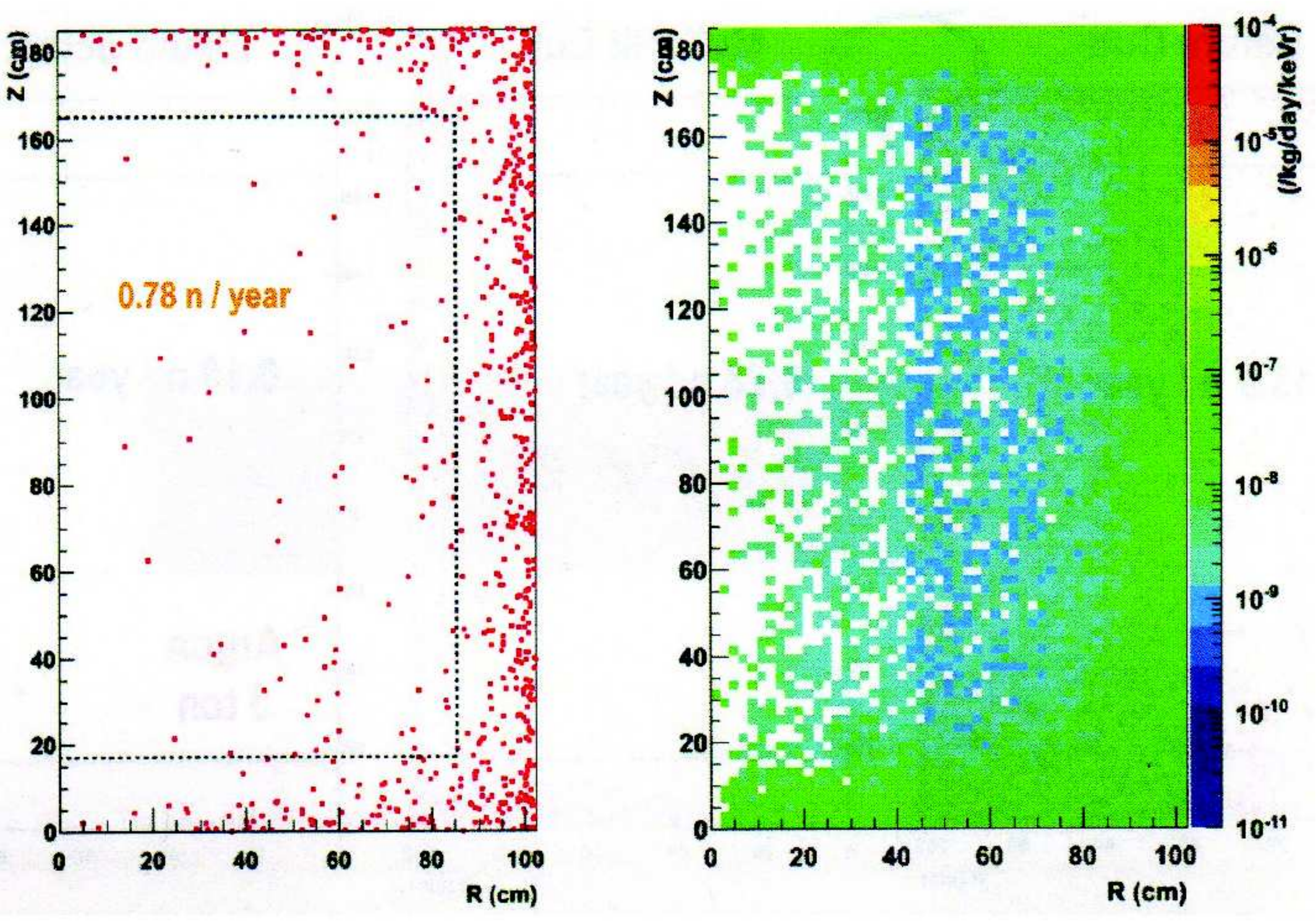}
  \includegraphics[width=0.75 \columnwidth]{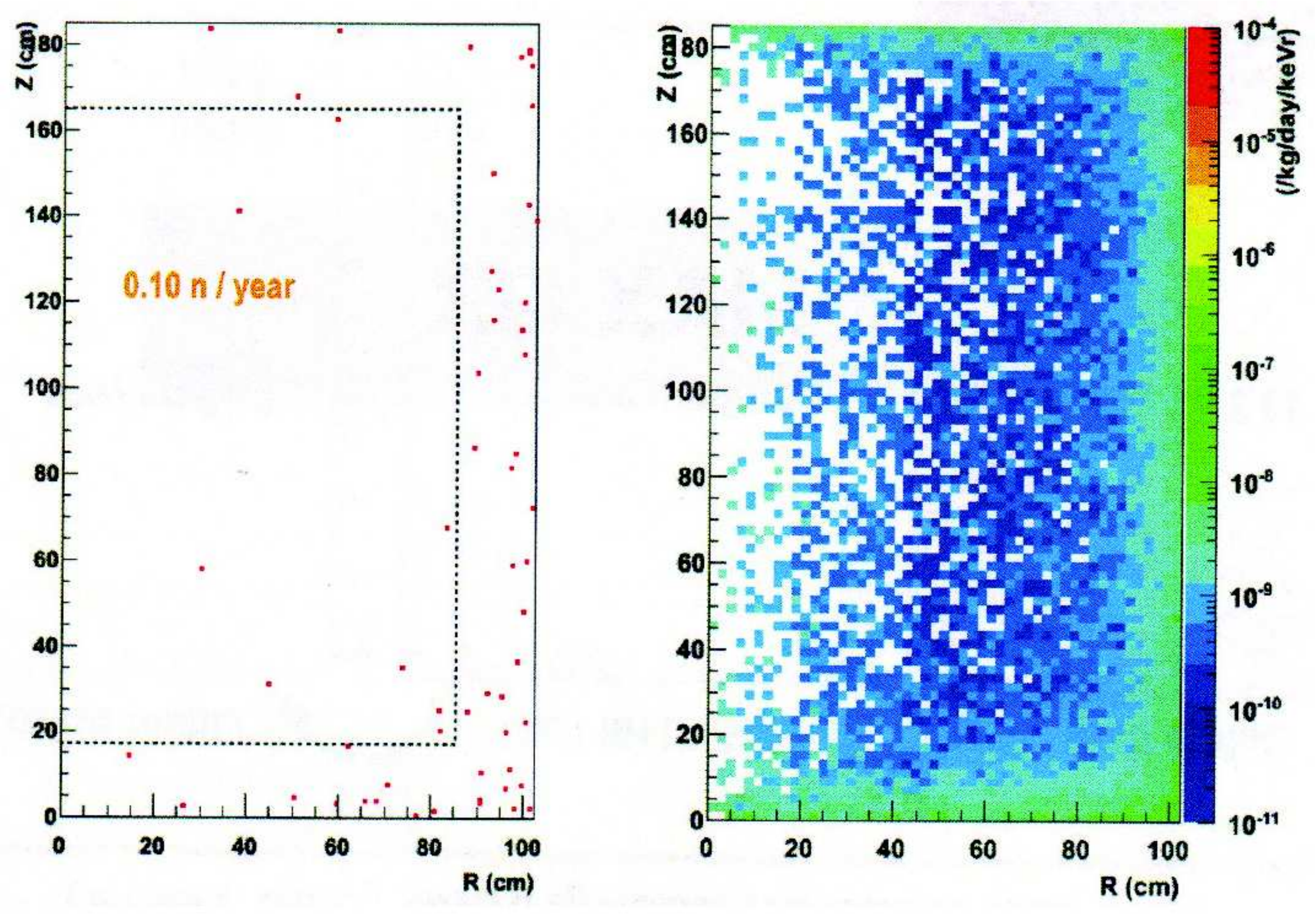}
  \caption{Simulated spatial distribution of single scatter neutron
    events in 100 years from U/Th in full QUPID array around 9-ton
    (total) Ar detector for 100 years,\protect\\
    \textit{(upper) } with no outer liquid scintillator veto. An outer
    cut of 15 cm leaves 78 events in a fiducial mass of 5 ton giving
    $\sim 0.8$ events/y (in the dark matter recoil energy range).\protect\\
    \textit{(lower) } using 0.5\% Gd-loaded liquid scintillator veto with 300 keV signal threshold. 
An outer cut of 15 cm leaves 10 events in a fiducial mass of 5 tons,
giving $\sim$ 0.1 events/y (in the dark matter nuclear recoil energy
range).}
\label{fig:neutron_spatial_ar5t} 
\end{figure}

\begin{figure}[!htbp]
  \centering
  \includegraphics[width=0.9 \columnwidth]{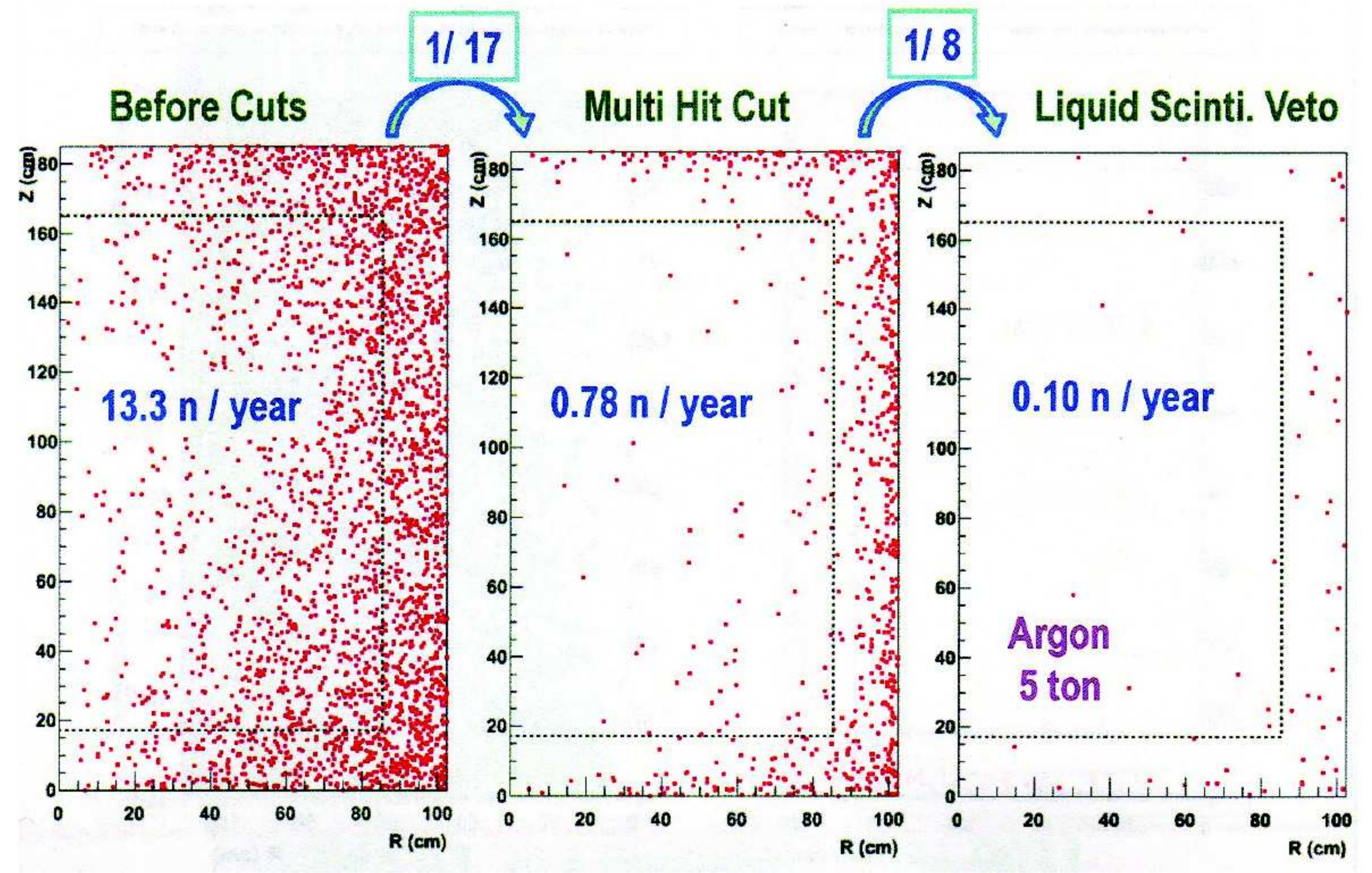}
  \caption{Summary of neutron background reduction in 5 ton Ar
    fiducial region. The multiple scattering cut and the liquid
    scintillator veto each produce a factor $\sim 10$ reduction to
    leave the desired level of $\sim 0.1$ neutrons/y (in the dark
    matter recoil energy range).}
  \label{fig:neutron_summary_ar5t}
\end{figure}

Fig.\ref{fig:neutron_spatial_ar5t} is the Ar counterpart to
Fig.\ref{fig:neutron_spatial_xe1t} for Xe, and shows the spatial
distribution of individual single scatter events in the 5 ton fiducial
region for 100 years data, and for the specific case of a 15 cm outer
passive Ar cut, and in this case shown with or without rejection of
events by a liquid scintillator veto surrounding the detector. With
both the multiple scattering cut and veto cut, about 10 events (in the
dark matter recoil range) remain in the fiducial region in 100 years
equivalent running time, thus achieving the background objective of
$\sim 0.1$ events/y. Fig.\ref{fig:neutron_summary_ar5t} provides a further illustration of the
stages of neutron background reduction, showing that the multiple
scattering cut and veto cut each produce about an order of magnitude
reduction in the events remaining in the fiducial region. Thus from
the simulations for the Xe and Ar detectors constituting the G2
system, we conclude that a neutron background $\sim 0.1$ events/y can be
achieved in both detectors, but the additional use of the liquid
scintillator veto appears essential in the larger argon detector,
unless a sensitivity limited by $\sim 1$ neutrons/y background is
judged acceptable. 

\subsection{G3 system: backgrounds for 10-ton Xe detector}
\label{subapp:bkg_xe10t}

\begin{figure}[!htbp]
  \centering
  \includegraphics[width=0.9 \columnwidth]{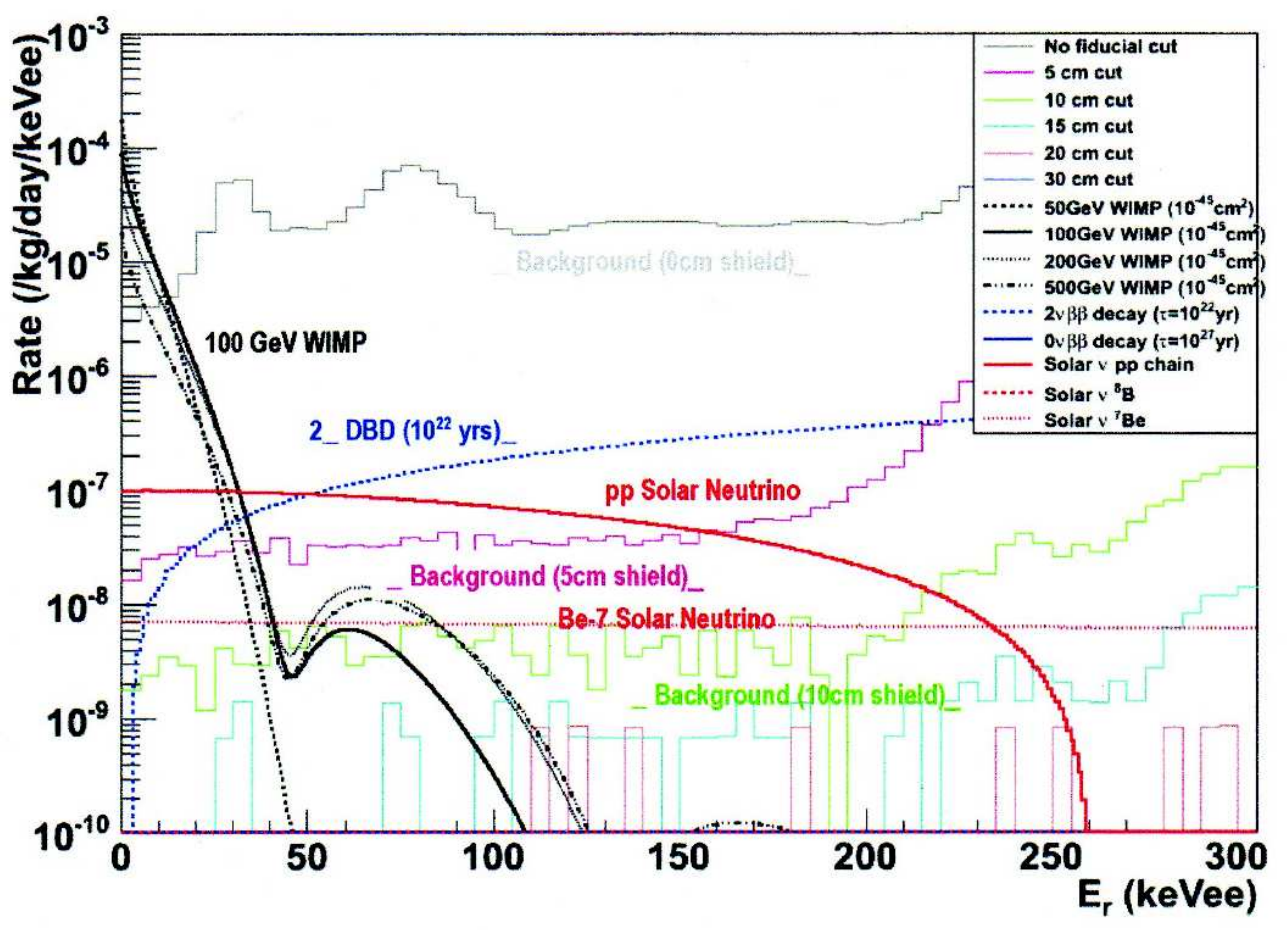}
\caption{Gamma rates from cryostat and QUPID radioactivity for 2 m
  diameter Xe detector (18 tons total Xe) using multiple scattering
  cut, S2/S1 cut, and 0, 5, 10, 15, 20, 30 cm outer Xe
  shield. Superimposed spectra show nuclear recoils from WIMPs at
  $10^{-45}~\n{cm^2}$ cross section, electron scattering from pp and
$^7$Be solar neutrinos, and a $2\nu \beta \beta $ spectrum.}
\label{fig:gamma_rate_xe10t} 
\end{figure}

Using the same assumptions for materials radioactivity levels as for
the 1-ton Xe detector, Fig.\ref{fig:gamma_rate_xe10t} shows \texttt{GEANT4} simulation
results for the spectrum of electron recoils in the fiducial region,
after a multiple scattering cut and the two-phase S2/S1 cut, shown for
different outer radial cuts ranging from 0 to 30 cm.

An outer cut of $\sim 15$ cm reduces the 18 tons total mass to a
10-ton fiducial mass, in which, after multiple scattering cut and
S2/S1 cut of a factor 100, the residual gamma rate is reduced to
\textless 0.1 events/y. As in the G2 1-ton Xe detector, there is, below 20
keVee, a spatially uniform and energy independent background of
electron recoils from pp solar neutrinos, at the level $\sim 8$ events/y,
which is acceptable (compared with the $\sim 1000$ events for dark
matter at $10^{-45}~\n{cm^2}$ or $\sim 100$ events at
$10^{-46}~\n{cm^2}$) since it has a subtractable spectrum of known
shape and magnitude.

\begin{figure}[!htbp]
  \centering
  \includegraphics[width=0.75 \columnwidth]{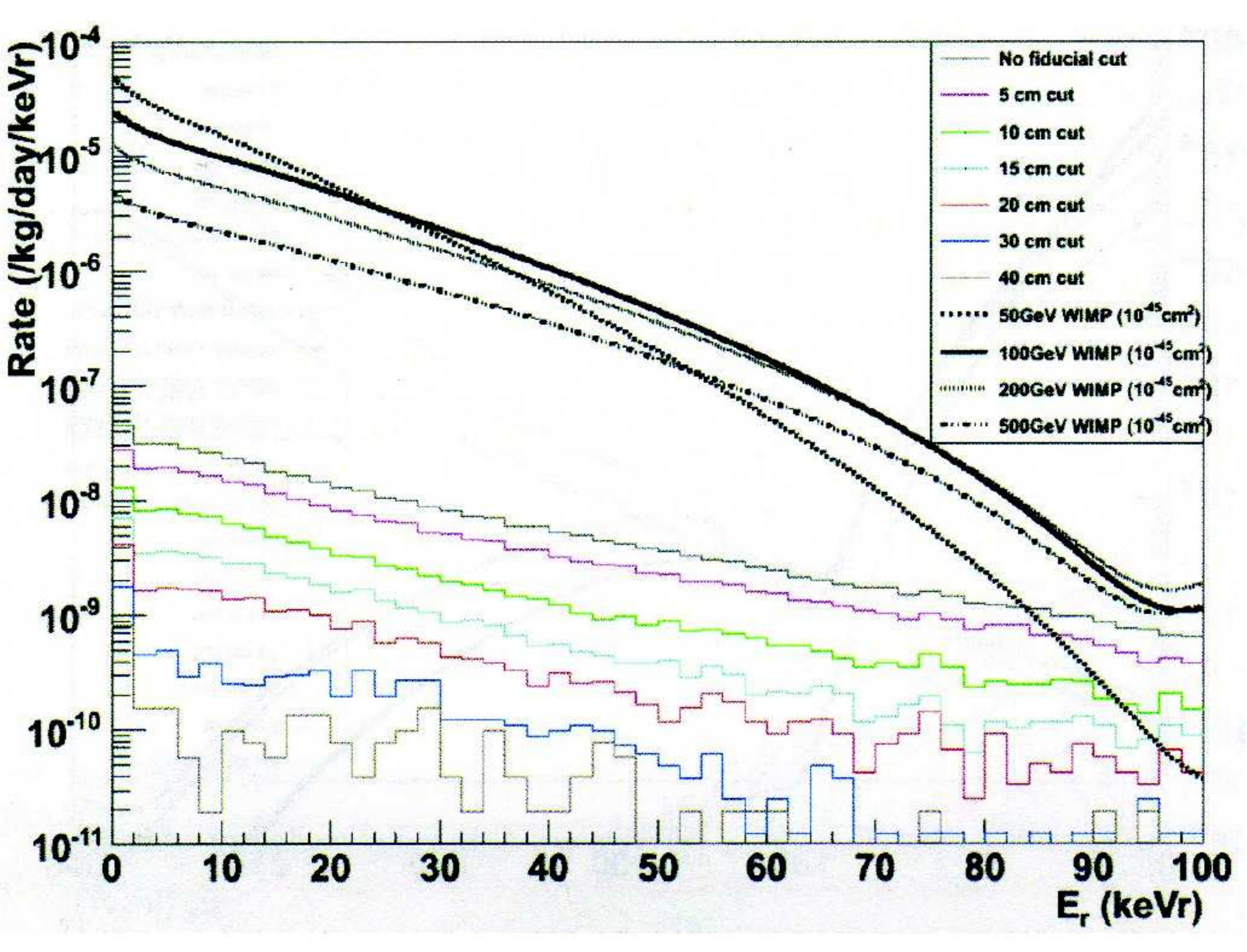}
  \includegraphics[width=0.75 \columnwidth]{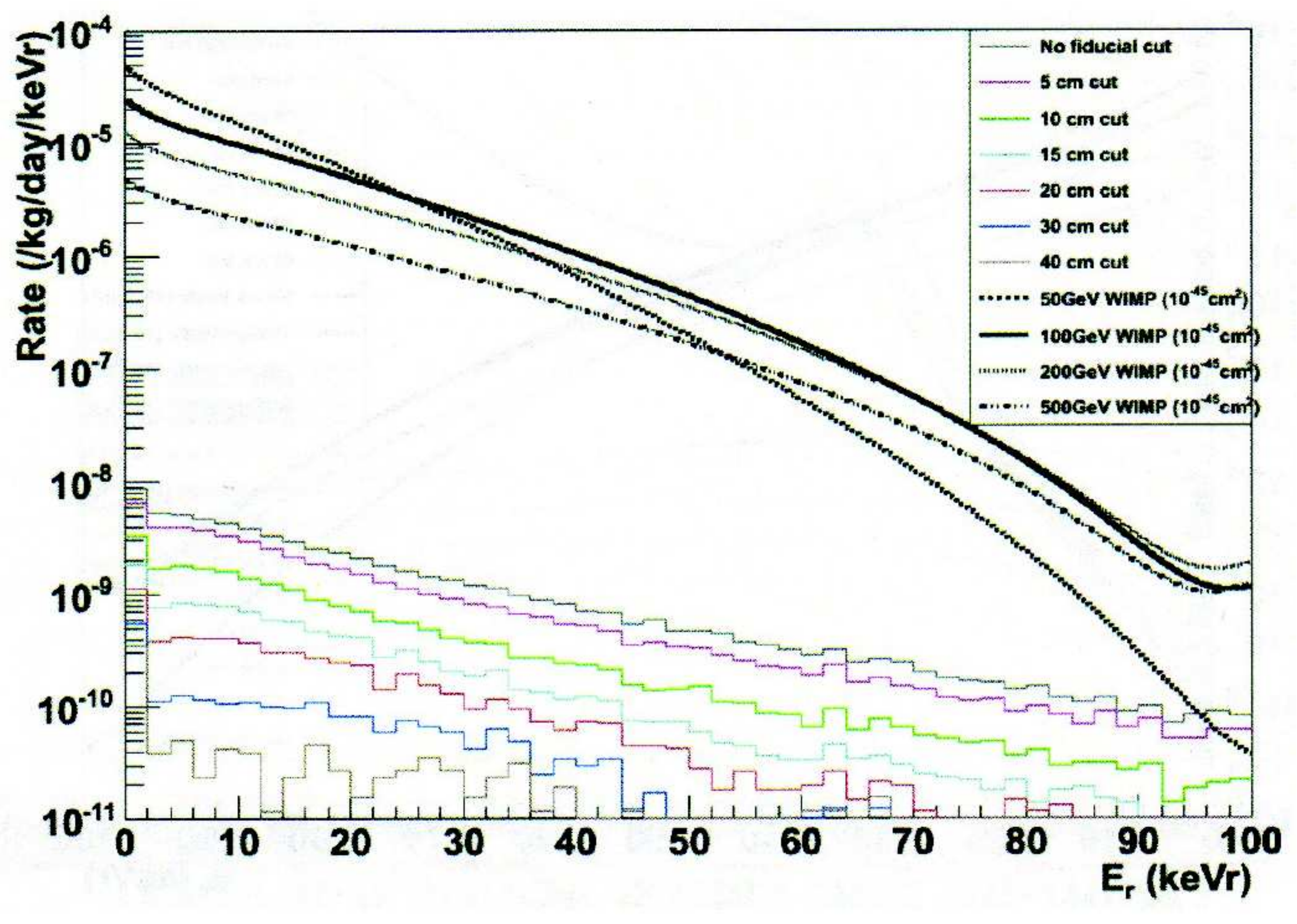}
  \caption{Neutron single scatter spectra in fiducial 10-ton Xe (versus
    keVr) for outer Xe cuts of 0, 5, 10, 15, 20, 30 cm, compared with
  dark matter spectra (black curves) for WIMP-nucleon $\sigma =
  10^{-45}~\n{cm^2}$ shown for several WIMP masses.\protect\\
  \textit{(upper plots)} No outer liquid scintillator veto\protect\\
  \textit{(lower plots)} 0.5\% Gd-loaded liquid scintillator with veto
  threshold 300 keV.}
\label{fig:neutron_single_xe10t} 
\end{figure}

\begin{figure}[!htbp]
  \centering
  \includegraphics[width=0.9 \columnwidth]{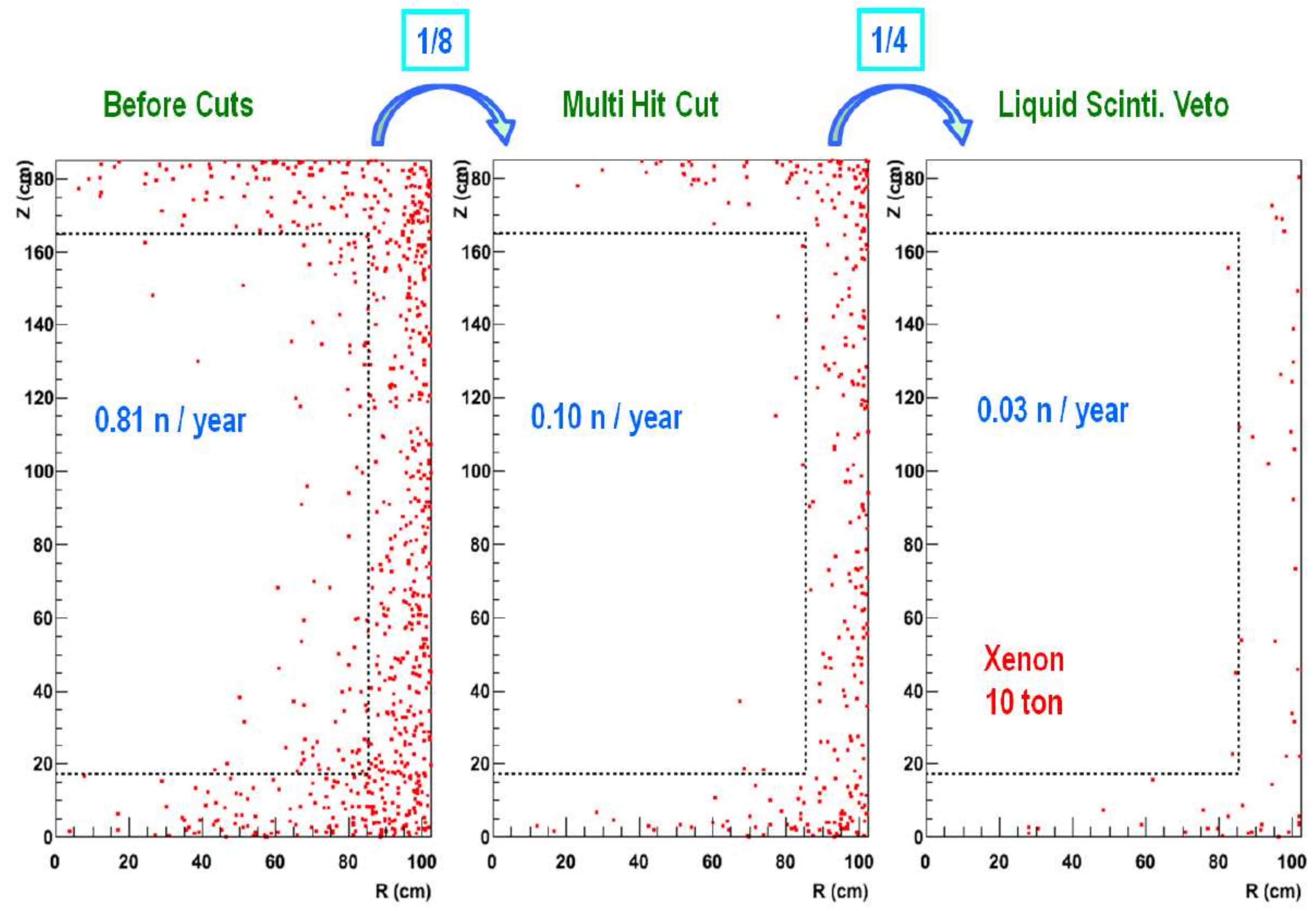}
\caption{Summary of neutron background reduction in 10-ton Xe fiducial
  region, using a 15 cm outer cut and a 100 y simulation. The multiple
  scattering cut is sufficient to reach the background level $\sim
  0.1$ events/y, with a further factor 4 gain from the veto signal.}
\label{fig:neutron_summary_xe10t} 
\end{figure}

For the neutron backgrounds, we utilize again the preceding conclusion
that the principal contribution comes from U/Th in the QUPID
photodetectors. Fig.\ref{fig:neutron_single_xe10t}(\textit{upper})
and Fig.\ref{fig:neutron_single_xe10t}(\textit{lower}) show the simulated spectra of
single-scatter events, for different outer cuts in the range $0 - 30$
cm, and with or without the liquid scintillator veto surrounding the
detector vessel. Fig.\ref{fig:neutron_summary_xe10t} is the counterpart to
Fig.\ref{fig:neutron_summary_ar5t}, showing the progressive reduction of neutron
background, firstly to $\sim 0.1$ neutrons/y by a multiple scattering
cut, and secondly to $\sim 0.03$ events/y by the liquid scintillator veto
operating at a threshold of 300 keV. The latter rejects neutrons by
simultaneous scattering in the liquid scintillator either before or
after scattering in the xenon, providing also an important diagnostic
tool in the event of unpredicted additional background in the target.

\subsection{G3 system: backgrounds for 50-ton Ar detector}
\label{subapp:bkg_ar50t}

For the 50-ton (fiducial) Ar detector, which partners the 10 ton Xe
detector, we adopt the simplifications concluded in \ref{subapp:bkg_ar5t}
for the 5-ton Ar detector:
\begin{enumerate}
  \item that the unrejected neutron background will dominate over
  the unrejected gamma background.
\item that neutrons from the QUPIDs will provide the dominant
  contribution to target background.
\end{enumerate}

\begin{figure}[!htbp]
  \centering
  \includegraphics[width=0.75 \columnwidth]{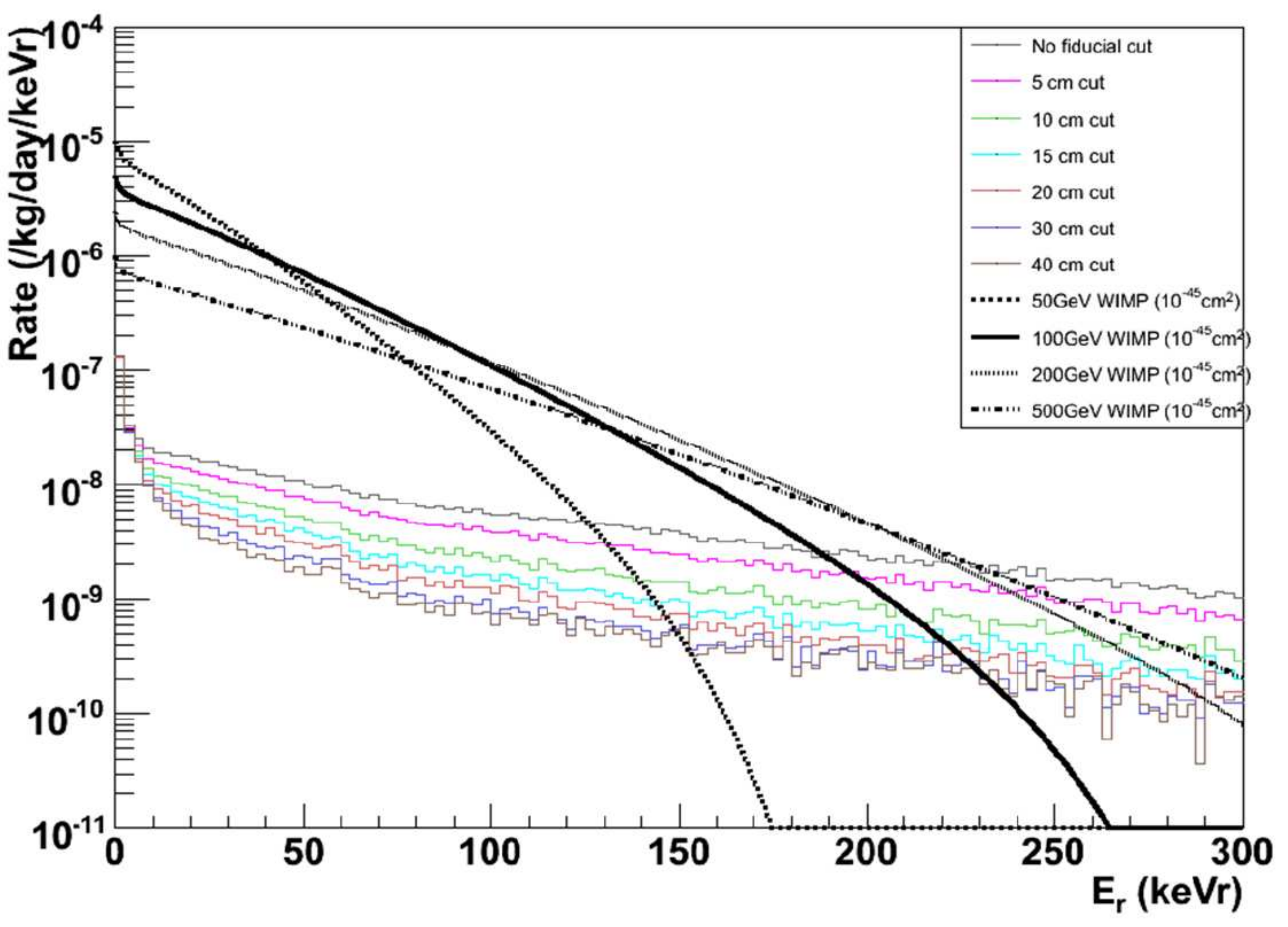}
  \includegraphics[width=0.75 \columnwidth]{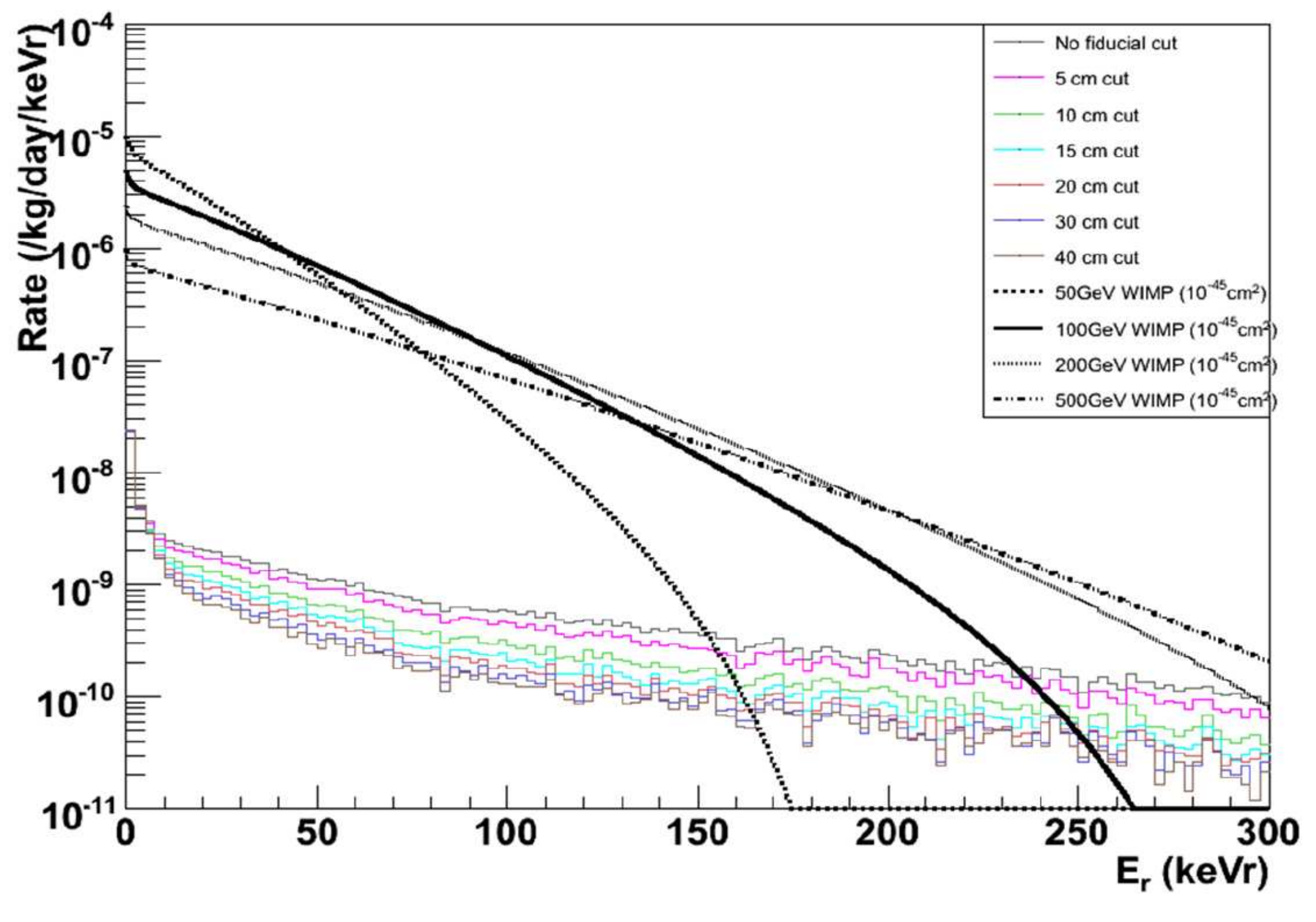}
  \caption{Neutron single scatter spectra in fiducial 50-ton Ar
    (versus keVr) for outer Ar cuts of 0, 5, 10, 15, 20, 30, 40 cm,
    compared with dark matter spectra (black curves) for WIMP-nucleon
    $\sigma = 10^{-45}~\n{cm^2}$, shown for several WIMP masses.\protect\\
    \textit{(upper plots)} No outer liquid scintillator veto\protect\\
    \textit{(lower plots)} Using veto signal with threshold 300 keV
    (spectra $\times 1.5$ for 1000 keV.)}
  \label{fig:neutron_single_ar50t} %C11
\end{figure}

Fig.\ref{fig:neutron_single_ar50t}(\textit{upper}) and
Fig.\ref{fig:neutron_single_ar50t}(\textit{lower}) show the simulated
spectrum of nuclear recoils from QUPID neutrons, for various thicknesses of outer
cut in the liquid Ar, with a multiple scattering cut, and without or
with a liquid scintillator outer veto. The comments in \ref{subapp:bkg_ar5t}
on the type of events rejected by the veto apply here also, but the
assumed veto energy threshold of 300 keV may be more difficult to
achieve in this larger detector. Repeating the simulation of
Fig.\ref{fig:neutron_single_ar50t}(\textit{lower}) with 1000 keV veto threshold increased the absolute
background numbers by a factor $\sim 1.5$.

\begin{figure}[!htbp]
  \includegraphics[width=0.9 \columnwidth]{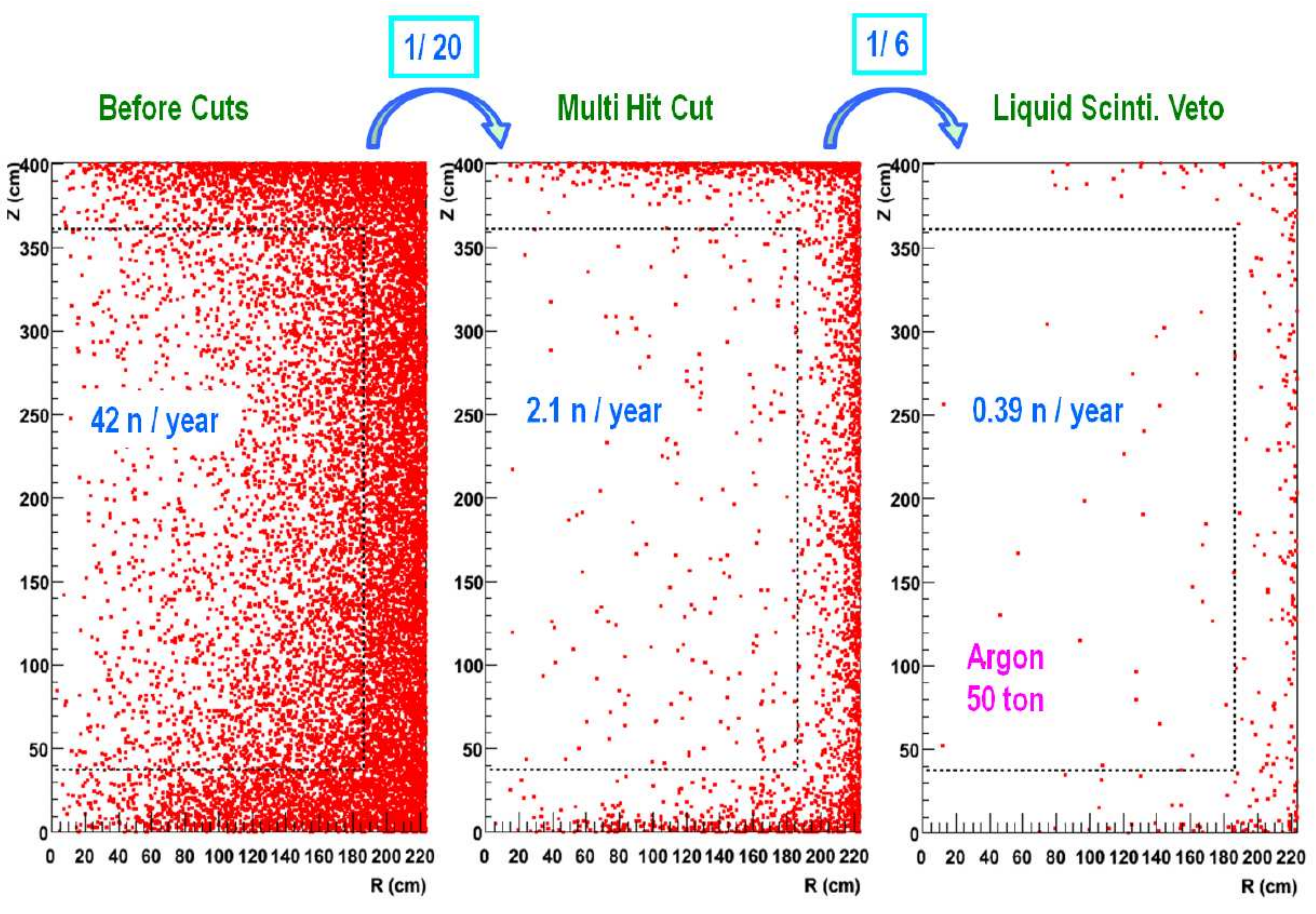}
\caption{Summary of neutron background reduction in 50-ton Ar 
fiducial region, using a 15cm outer cut. The multiple scattering cut
is sufficient to reach a background level $\sim 2$ events/y with a further
factor 5 gain from the veto signal. A larger outer cut gains
relatively slowly, reducing to 0.3 n/y for a 25 cm cut.}
\label{fig:neutron_summary_ar50t}
\end{figure}

Fig.\ref{fig:neutron_summary_ar50t} is the counterpart to Fig.\ref{fig:neutron_summary_xe10t}, again
showing two orders of magnitude gain from applying both the multiple
scattering cut and the veto cut, this time leaving a final background
of 0.4 neutrons/y for a 300 keV veto threshold, or 50\% higher at
0.6 neutrons/y for a more conservative veto threshold of 1 MeV. This
could be reduced to \textless 0.3 events/y by increasing the
self-shielding cut to 25 cm (with $\sim 7\%$ loss of fiducial
volume). 

\subsection{G4 system: backgrounds for 100-ton Xe and 500-ton Ar}
\label{subapp:bkg_g4}

To obtain a gain in detector sensitivity for G4 corresponding to the
factor 10 increase in fiducial mass, it is necessary that the
unrejected backgrounds (in absolute event numbers per year) are not
significantly greater than those in G3, i.e. remaining below the $0.5-1$
events/y level. For this, one can take advantage of the fact that
greater thicknesses of passive outer layer can be used for larger
detectors, without significant loss of the fiducial volume.

Low energy gamma and neutron backgrounds for G4 can be extrapolated
immediately from the principles established by G2 and G3 simulations
above. Total gamma emission from detector components will be
approximately proportional to detector surface area. This would have
increased the simulated absolute background numbers in the G3 Xe
detector, calculated above in C.4, by a factor 4 over those in G2, but
they were kept at a similar level ($\le $ 0.1 events/y) by increasing
the thickness of the outer layer of Xe used as passive shielding from
10cm to 15cm. 

For the G2 to G3 argon scale-up different considerations applied to
gamma and neutron rejection, as discussed in \ref{subapp:bkg_xe1t} and
\ref{subapp:bkg_xe10t}. In argon the electron/gamma pulse shape
discrimination, combined with the two phase discrimination, enabled
gamma and electron background to be reduced to a negligible level in
both G2 and G3, and a 15 cm passive layer is needed to achieve a level
0.1 neutrons/y in the G2 argon detector. A similar 15 cm
passive thickness then gives the expected area factor 4 higher
background in G3 Ar, of 0.4 events/y. As seen in
Fig.\ref{fig:neutron_summary_ar50t}, this can be reduced to $0.2-0.3$ events/y by
increasing the passive thickness to 25 cm.

Applying these principles to the G3 to G4 Xe scale up the factor 4
increase in gamma and neutron background, arising from the increased
surface area, can again be offset by an increase in the
(software-adjustable) outer passive shielding thickness from 15 cm to
20 cm, retaining 100-ton fiducial mass while reducing both backgrounds
to $\sim 0.2$ events/y. In the case of the G3 to G4 argon scale up,
the unrejected gamma background is again negligible (owing to the
additional pulse shape discrimination discussed above) and for neutron
events the expected area scaling factor $\sim 4$ increase in the
absolute number per year can again be offset by an increase in the
non-fiducial passive argon thickness, in this case from 25 cm to $\sim
35$ cm, keeping the fiducial neutron background events to below 1
event/y for the loss of only 8\% of the fiducial mass. 

\setcounter{equation}{0}
\setcounter{figure}{0}
\setcounter{table}{0}
\section{Galactic supernovae}
\label{app:nugalsupern}

\begin{figure}[!htbp]
  \centering
  \includegraphics[width=0.75 \columnwidth]{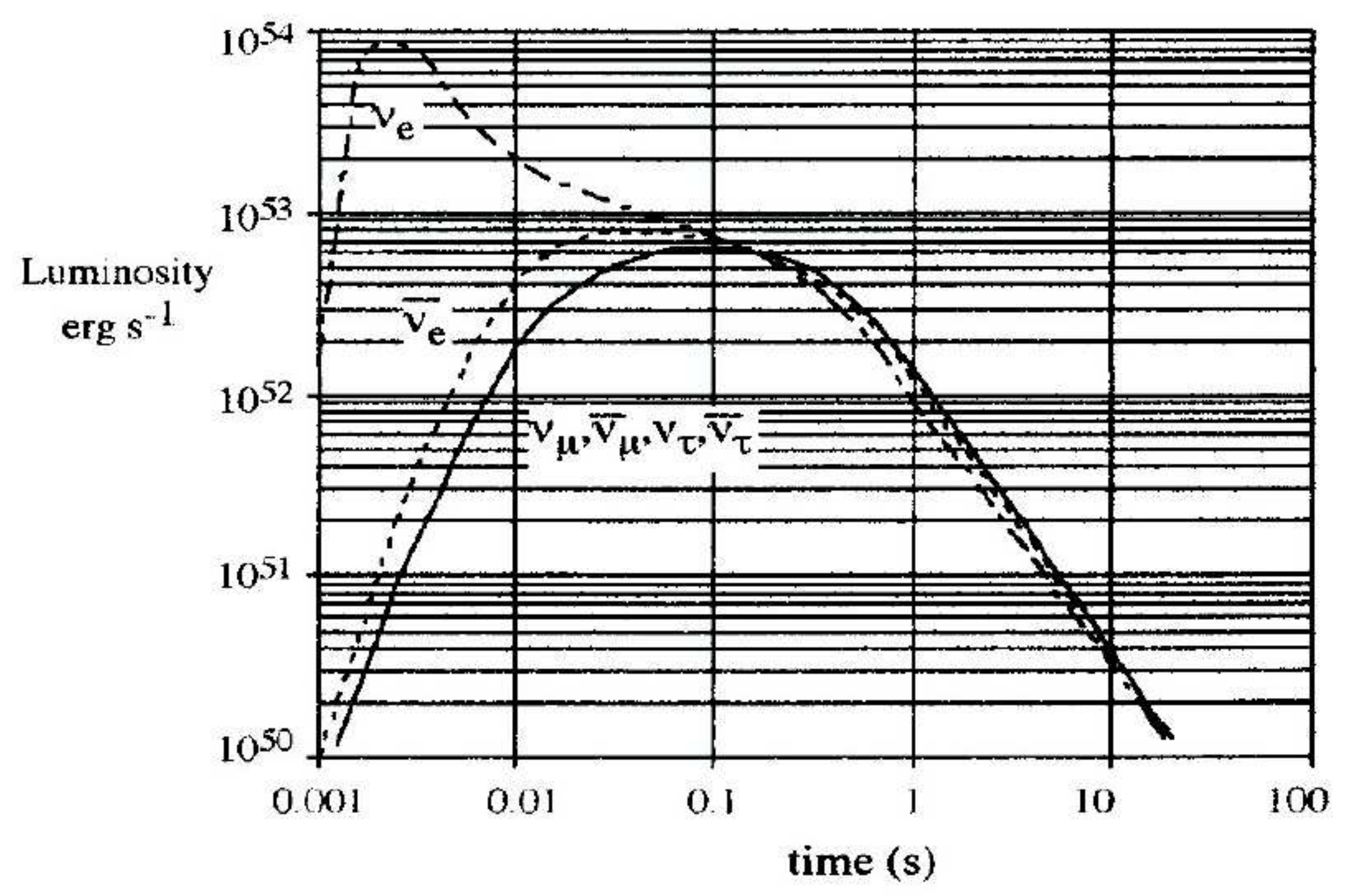}
  \caption{Generic form of supernova neutrino burst, showing
    luminosity versus time for different neutrino and antineutrino
    flavours (from~\cite{Suzuki:2000}).}
  \label{fig:nusupernovaburst}
\end{figure}

\begin{figure}[!htbp]
  \centering
  \includegraphics[width=0.5 \columnwidth]{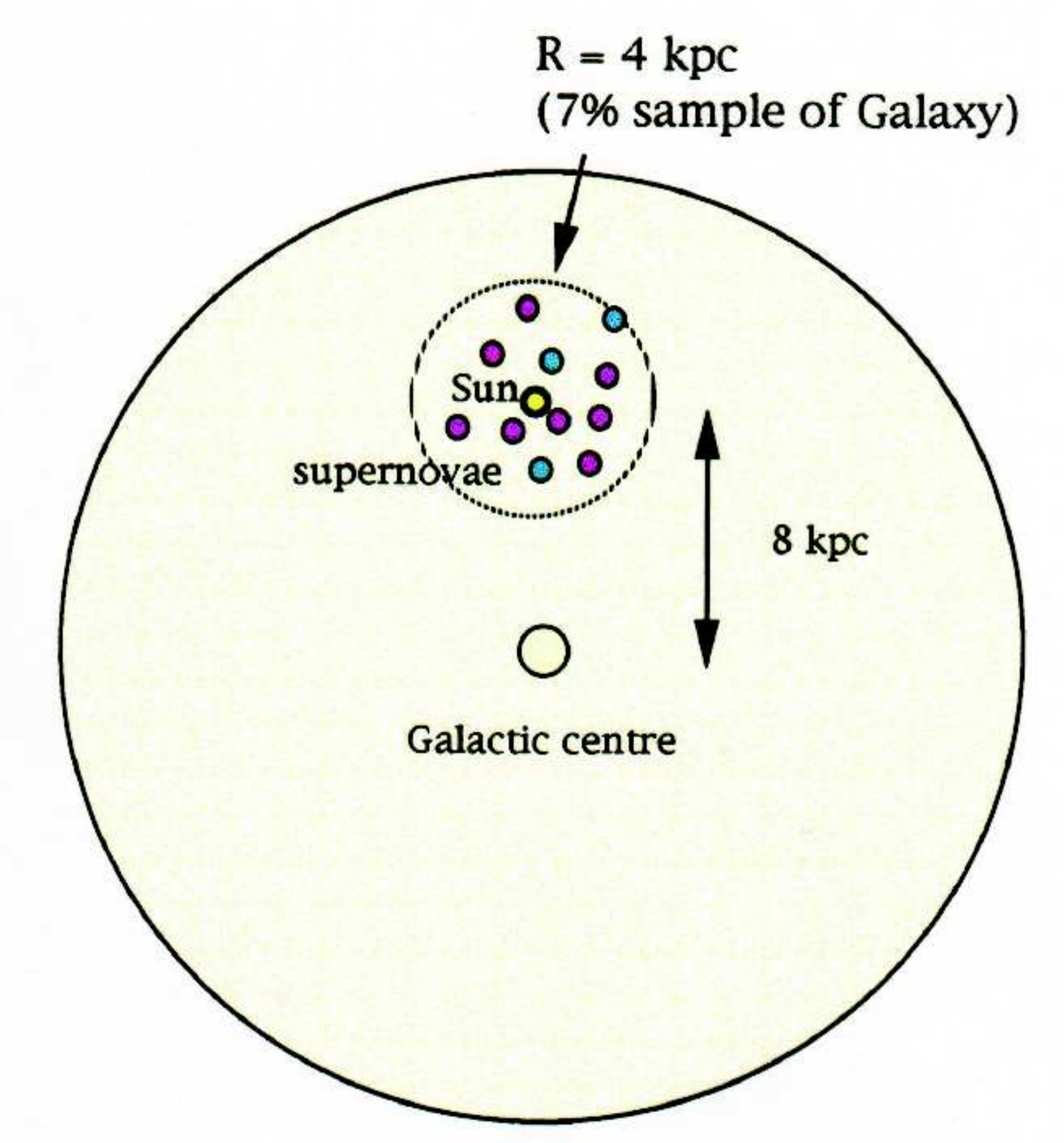}
  \caption{Sketch of Milky Way Galaxy, showing positions of observed
    Galactic supernovae from the historical visible record for the
    past 2000 years (omitting 1987a, in the Large Magellanic Cloud at
    51 kpc from Sun).\protect\\
\textit{(blue points)} SN type Ia, \textit{(purple points}) SN type II
and Ib, \textit{(yellow point) }Sun.}
\label{fig:milkyway}
\end{figure}

A type II or 1b supernova explosion releases typically $3\times
10^{53}$ ergs of energy, mostly as a burst of neutrinos and
antineutrinos of all flavours, lasting $10-20$ seconds and with the
neutrino time profiles shown in Fig.\ref{fig:nusupernovaburst}~\cite{Raffelt:1996}. With
detectors of currently foreseeable size, these neutrinos are detectable only
for supernovae in our own Galaxy. The frequency of supernovae in our
Galaxy is uncertain, a common estimate being $3\pm1$ /century. A
slightly higher rate $4\pm1$ events/century can be estimated from the
historical record of visible supernovae in our Galaxy~\cite{Fang:1990}, shown
for the past 2000 years in Fig.\ref{fig:milkyway}.

It is evident from these recorded SN locations that supernovae are not
usually seen by visible light beyond a distance from the Earth of
$4-5$ kpc, due to obscuration by material in the Galactic plane, so
the visible record covers less than 10\% of the whole Galaxy. Hence
using neutrino detection we would be able to see at least an order of
magnitude more than the number observed visibly. Allowing for some
identification uncertainties, there have been $7\pm1$ type II/Ib
supernova in 2000 years in a region $8\pm2\%$ of the Galaxy, or $\sim
4\pm1$ /century, on the assumption that the local 5 kpc (which in fact
crosses two spiral arms) represents a random sample of stellar SN
progenitors. From this it appears likely that another Galactic
supernova, visible as a neutrino burst, will occur within the lifetime
of current and foreseeable detectors, which should therefore all be
designed with data acquisition buffers able to collect and record a
high flux of incident neutrinos from this source.

\begin{table}[!htbp]
  \begin{center}
    \begin{tabular}{|c|c|c|c|c|c|c|}
      \hline
      & $\nu _{e}$ & $\bar{\nu} _{e}$ & $\nu _{\mu}$ & $\bar{\nu}
      _{\mu}$ & $\nu _{\tau}$ & $\bar{\nu} _{\tau}$ \\
      \hline
      Total energy (ergs) & $5.10^{52}$ & $5.10^{52}$ & $5.10^{52}$
      & $5.10^{52}$ & $5.10^{52}$ & $5.10^{52}$ \\
      \hline
      Mean energy $<\n{E}>$ (MeV) & $10 -12$ & $14-17$ & $24-27$ &
      $24-27$ & $24-27$ & $24-27$ \\
      \hline
      Temperature (MeV) &$ 3 - 4$ & $4 - 6$ & $7 - 9$ & $7 - 9$ & $7
      - 9$ & $7 - 9$ \\
      \hline
      Time-integrated flux ($\nu/\n{cm}^2$) & $2.4 \times 10^{11}$ &
      $1.6 \times 10^{11}$ & $1.0 \times 10^{11}$ & $1.0 \times
      10^{11}$ & $1.0 \times 10^{11}$ & $1.0 \times 10^{11}$ \\
      \hline
    \end{tabular}
    \caption{Mean energy and time-integrated flux for the three
      neutrino flavours from a Galactic SN at a distance 10 kpc from
      Earth.}
    \label{tab:nu_ene_flux}
  \end{center}
\end{table}
Tab.\ref{tab:nu_ene_flux} summarises the total energy, mean energy and
time-integrated flux for the three neutrino types, for a supernova at
10 kpc from Earth~\cite{Raffelt:1996}. For a Maxwell-Boltzman spectrum
the temperature is 1/3 of the mean energy $<E>$,
and for a Fermi-Dirac spectrum the temperature is $0.83\times (1/3)
<\n{E}>$ giving the temperature ranges shown in Tab.\ref{tab:nu_ene_flux}.

%% References without bibTeX database:

\end{document}